\documentclass[12pt]{article}
\usepackage{amsmath}
\usepackage{graphicx}
\usepackage{enumerate}
\usepackage{natbib}
\usepackage{url} 

\usepackage{chngcntr}

\usepackage[T1]{fontenc}
\usepackage[latin9]{inputenc}
\usepackage{subfigure}
\usepackage{multirow}
\usepackage{tikz}
\usepackage{amsmath}
\usepackage{amsthm}
\usepackage{amssymb}
\usepackage{amstext}
\usepackage{bm}
\usepackage{booktabs} 
\usepackage{color}
\usepackage{enumerate} 
\usepackage[shortlabels]{enumitem}
\usepackage{epsfig,epsf,psfrag}
\usepackage{epstopdf}
\usepackage{float}
\usepackage{graphics}
\usepackage{graphicx}
\usepackage{latexsym}
\usepackage{lscape}
\usepackage{mathabx} 
\usepackage{mathtools}
\usepackage{mathrsfs}
\usepackage{multirow}
\usepackage{natbib}
\usepackage{rotate}
\usepackage{setspace}
\usepackage{algorithm}
\usepackage{algorithmic}
\usepackage{sectsty}

\usepackage{xargs}[2008/03/08]
\usepackage{xcolor}

\usepackage{ifthen}
\usepackage[hidelinks]{hyperref}
\hypersetup{
	colorlinks,
	linkcolor={red!50!black},
	citecolor={blue!50!black},
	urlcolor={blue!80!black}
}

\renewcommand\thesection{\arabic{section}}

\sectionfont{\centering\bf\large}
\subsectionfont{\normalsize}

\usepackage{titlesec}

\newtheorem{prop}{Proposition}

\newtheorem{thm}{Theorem}
\newtheorem{lem}{Lemma}
\newtheorem{cor}{Corollary}
{
	\theoremstyle{remark}

	\newtheorem{assumption}{Assumption}
}

\newtheorem*{lem*}{Lemma}

\newcommand{\bea}{\begin{eqnarray*}}
\newcommand{\eea}{\end{eqnarray*}}
\newcommand{\be}{\begin{eqnarray}}
\newcommand{\ee}{\end{eqnarray}}
\newcommand{\beq}{\begin{equation}}
\newcommand{\eeq}{\end{equation}}

\newcommand{\bal}{\begin{equation}\aligned}
\newcommand{\eal}{\endaligned\end{equation}}
\newcommand{\bgt}{\begin{equation}\begin{gathered}}
\newcommand{\egt}{\end{gathered}\end{equation}}
\newcommand{\ed}{\end{document}}

\newcommand{\btab}{\begin{tabular}}
\newcommand{\etab}{\end{tabular}}

\newcommand{\bi}{\begin{itemize}}
\newcommand{\ei}{\end{itemize}}
\newcommand{\bfi}{\begin{figure}}
\newcommand{\efi}{\end{figure}}
\newcommand{\ben}{\begin{enumerate}}
\newcommand{\een}{\end{enumerate}}
\newcommand{\bay}{\begin{array}}
\newcommand{\eay}{\end{array}}

\newcommand{\F}{Fr\'echet }
\newcommand{\M}{\mathcal{M}}
\newcommand{\mP}{\mathcal{P}}
\newcommand{\mA}{\mathcal{A}}
\newcommand{\sub}{\subset}
\def\real{\mathbb{R}}

\def\bco{\iffalse}

\newcommand{\no}{\noindent}
\newcommand{\bc}{\begin{center}}
\newcommand{\ec}{\end{center}}

\definecolor{DarkBlue}{rgb}{0,.08,.45}
\definecolor{DarkRed}{rgb}{.7,0,.4}
\definecolor{DarkGreen}{rgb}{0,0.4,0}

\newcommand{\bpmt}{\begin{pmatrix}}
	\newcommand{\epmt}{\end{pmatrix}}

\newcommand{\bsp}{\begin{split}}
	\newcommand{\esp}{\end{split}}

\newcommand{\bdes}{\begin{description}}
	\newcommand{\edes}{\end{description}}

\newcommand{\bass}{\begin{assumption}}
	\newcommand{\eass}{\end{assumption}}
\newcommand{\bthm}{\begin{thm}}
	\newcommand{\ethm}{\end{thm}}
\newcommand{\blem}{\begin{lem}}
	\newcommand{\elem}{\end{lem}}
\newcommand{\bcor}{\begin{cor}}
	\newcommand{\ecor}{\end{cor}}
\newcommand{\bpf}{\begin{proof}}
	\newcommand{\epf}{\end{proof}}

\def\i1{_{i1}}

\def\cov{{\rm cov}}

\def\var{{\rm var}}

\def\trace{{\rm trace}}

\def\cp{\citep}

\def \mbb {\mathbb}

\DeclareMathOperator*{\argmin}{argmin}

\def\real{\mbb{R}}

\def \M{\mathcal{M}}

\newcommand{\ltwoNorm}[2][]{%
	\ifthenelse{ \equal{#1}{} }
	{\ensuremath{\|#2\|}}
	{\ensuremath{\|#2\|_{#1}}}
} 
\usepackage{authblk}
\def\bco{\iffalse}
\newcommand{\blind}{1}

\usepackage{xr}

\begin{document}

	\def\spacingset#1{\renewcommand{\baselinestretch}%
		{#1}\small\normalsize} \spacingset{1}

%

\if1\blind
{
		

\title{\bf Inference for Dispersion and Curvature of Random Objects}
\author{Wookyeong Song$^\ast$  and  Hans-Georg M\"uller$^\ast$ \\
	$^\ast$Department of Statistics, University of California, Davis}
\maketitle
} \fi

\if0\blind
{
\bigskip
\bigskip
\bigskip
\begin{center}
{\LARGE\bf Inference for Dispersion and Curvature of Random Objects}
\end{center}
\medskip
} \fi

\bigskip
\begin{abstract}
There are many open questions pertaining to the statistical analysis of random objects, which are increasingly encountered.  A major challenge is the absence of linear operations  in such spaces. A basic statistical task is to quantify statistical dispersion or spread. For two measures of dispersion for data objects in geodesic metric spaces,  \F variance and metric variance, we derive a central limit theorem (CLT) for their joint distribution. This analysis reveals that the Alexandrov curvature of the geodesic space determines the relationship between these two dispersion measures. This suggests a novel test for inferring the curvature of a space based on the asymptotic distribution of the dispersion measures. We demonstrate how this test can be employed to detect the intrinsic curvature of an unknown underlying space, which emerges as a joint property of the space and the underlying probability measure that generates the random objects. We investigate the asymptotic properties of the test and its finite-sample behavior for various data types, including distributional data and point cloud data. We illustrate the proposed inference for intrinsic curvature of random objects using gait synchronization data represented as symmetric positive definite matrices and energy compositional data on the sphere.
\end{abstract}

\noindent%
{\it Keywords:} Alexandrov curvature; Metric statistics; \F variance; Metric variance; Bures-Wasserstein geometry

\newpage
\spacingset{1.9} 

\section{Introduction}
        \label{chap1}
	
Complex random objects are increasingly prevalent  in  data analysis and may be viewed as data residing in a  metric space. Examples of such object data include probability distributions with the Wasserstein geometry \citep{mull:23:2}, covariance matrices with various metrics on the spaces of symmetric positive definite (SPD) matrices \citep{arsigny2007geometric,lin2019riemannian}, distributional data with the Fisher-Rao metric on the Hilbert sphere \citep{dai2022statistical}, data on Riemannian symmetric spaces \citep{cornea2017regression}, finite-dimensional Riemannian manifolds \citep{eltz:18}, phylogenetic trees \citep{billera2001geometry} and  functional data \citep{wang2016functional}, among other data types \citep{marr:21,huck:21}. 

We consider here data that are situated in a separable geodesic metric space $(\M,d)$ with a distance $d: \M \times \M \rightarrow \real$ and random objects $X$ with a Borel probability measure $\mP$ on $\M;$ any elements in the space can be connected by a unique geodesic, which is a minimal length curve in the space. {The absence of algebraic structure in a general metric space $\M$ motivates a framework of metric statistics for the analysis of such data \citep{mull:24, wang2024nonparametric}, as well as new methodologies for statistical inference \citep{song2022new, chu2024manifold,  jiang2024two, dubey2023change}.} Specifically, alternative approaches are necessary for defining the expectation and dispersion for random objects $X \in \M.$ The \F mean \citep{frechet1948elements}, a generalization of the notion of expected value for random objects, is a natural measure of location,
    \begin{equation}
        \mu_{\oplus} = \operatorname*{argmin}_{\xi \in \M} \mathbb{E}d^{2}(\xi,X).
        \label{Fmean_pop}
    \end{equation}
    Note that here and in the following, expectations  $\mathbb{E}$ are with respect to the probability measure $\mP$ of $X.$ 
   
    An obvious measure of dispersion for random objects is  then the \F variance,
    \begin{equation}
        V_{F} = \mathbb{E}d^{2}(\mu_{\oplus},X).
        \label{Vf_pop}
    \end{equation}
    An empirical estimator of $V_F$ was studied in  \cite{dubey2019frechet}, who derived consistency and a central limit theorem (CLT) for  this estimator  under suitable assumptions, including  the existence and uniqueness of the \F mean $\mu_{\oplus}$ and entropy conditions; 
    this CLT also led to a test for comparing $k$ populations of metric space-valued data objects.

The metric variance $V_M$  was introduced in \cite{dubey2020functional} as an alternative to quantify  the dispersion of random objects,
    \begin{equation}
        V_{M} = \frac{1}{2}\mathbb{E}d^{2}(X,X'),
        \label{Vm_pop}
    \end{equation}
    where $X'$ represents an independent and identically distributed (i.i.d.) copy of $X.$  {Metric variance is well-defined under weak conditions, as it does not require the uniqueness of a \F mean $\mu_{\oplus}$ in \eqref{Fmean_pop}.} In Euclidean spaces, one has $V_F=V_M,$ which can be understood through the well-known identity $\frac{1}{n-1}\sum_{i=1}^n  (X_i-\bar{X})^2=\frac{1}{2n(n-1)} \sum_{i,j=1}^n (X_i-X_j)^2.$ This identity holds even in infinite-dimensional Hilbert spaces. However, in general metric spaces, $V_F \ne V_M,$ and it raises the question of what drives the difference and whether it can be exploited to learn features of the space or underlying probability measure.

    Assume that the probability measure  $\mP$ on the geodesic space $(\M,d)$ that determines the distribution of $X$  is strictly positive, i.e., for any $\delta > 0$ and any $\xi \in \M,$ one has  
    $\mP(B_{\delta}(\xi)) > 0,$ where $B_{\delta}(\xi)$ is an open ball of radius $\delta$ centered at $\xi$ with respect to the metric $d$. Under this condition, we show that the sign of curvature (in the sense of Alexandrov) of the geodesic space $\M$ can be inferred by comparing the two dispersion measures $V_F$ and $V_M$.  Determining the curvature of the space is important because it has a bearing on the asymptotic properties of estimators, including convergence rates for empirical barycenters \citep{ahidar2020convergence, hundrieser2024finite, pennec2019curvature}. 
    
    {The study of intrinsic curvature is particularly relevant when the probability measure $\mP$ of random objects $X$ is not strictly positive but instead concentrates on a simply-connected subset $\mA \subset \M$. 
    This arises because, once the ambient geodesic space $(\M, d)$ is determined, the Alexandrov curvature of $(\M, d)$ can be mathematically derived by comparing geodesic triangles in $\M$ with comparison triangles in the model space, as introduced in Section \ref{chap31}, and the underlying probability measure $\mP$ has no bearing. However, the intrinsic space $\mA,$ equipped with the intrinsic geodesic distance $d_{I},$ is derived from samples of random objects $X_{1}, \ldots X_{n},$ and is thus affected by both the space $(\M, d)$  and the measure $\mP$; a rigorous definition of $d_{I}$ is provided in Section \ref{chap4.2}.  
    Then the geometric structure, including curvature, of the space where the random objects $X$ actually reside 
    corresponds to the intrinsic space $(\mA, d_I)$ that reflects the measure $\mP$  rather than the ambient space $(\M, d).$}

    For instance, the space of multivariate distributional data equipped with the 2-Wasserstein distance, denoted as $\mathcal{W}_2(\real^p)$, $p \geq 2,$ is known to exhibit positive Alexandrov curvature \citep{otto2001geometry}. However, the intrinsic curvature of a subspace $\mathcal{D} \sub \mathcal{W}_2(\mathbb{R}^p)$ may exhibit flat geometry. 
    We provide an explicit example for this in Section \ref{WassSimul}, Example 1.
    
    To the best of our knowledge, there is currently no test available to determine the presence or nature of curvature.
To arrive at such a test, our starting point is a CLT for the joint distribution of estimators for the metric and \F variance. It turns out that the relationship between these two measures of variance depends on the presence and sign of underlying curvature. We then harness this relationship to obtain inference on curvature  for both the original space $\M$ and also for the intrinsic curvature of a set $\mA$ on which the probability measure $\mP$ concentrates, i.e., on which the measure is positive with  $\mP(\mA)=1.$ As a consequence of the joint CLT for the two variance measures, we also obtain a confidence region for the true curvature. 

    Our main contributions in this paper are as follows:
        
        {\it Joint inference for \F variance and metric variance (Section \ref{chap2}).} 
        We derive a CLT for the joint distribution of estimates $(\hat{V}_{M}, \hat{V}_{F})^{T}$ for metric variance $V_M$ and \F variance $V_F$ in Theorem \ref{thm1} and provide a consistent estimator of the asymptotic variance of the limiting distribution in Proposition \ref{prop3}. This leads to a joint confidence region for these two dispersion measures. 
        
        {\it Impact of curvature on measures of dispersion  (Section \ref{chap3}).} In Theorems \ref{thm2} and \ref{thm3}, we show that the curvature of a geodesic metric space determines the relationship between the \F variance $V_{F}$ and metric variance $V_{M}$. 
        
        {\it Inference for curvature (Section \ref{chap41}).} We propose a test statistic for detecting curvature of the original space $(\M, d)$ based on the two dispersion measures 
        and develop a consistent test for the null hypothesis that the space is flat, for contiguous alternatives that the curvature of the space is either strictly positive or negative. 
    
        {\it Detecting intrinsic curvature (Section \ref{chap4.2}).} Intrinsic curvature of a latent space $\mA$ on which the   underlying probability measure $\mP$ concentrates 
        can be inferred from observed  samples  
        by combining  Dijkstra's algorithm \citep{dijkstra1959note} and the proposed curvature test. Simulation studies for bivariate distributional data with Wasserstein geometry and 3-dimensional point cloud data are presented in Sections \ref{WassSimul} and \ref{chap63}, followed by data illustrations to detect intrinsic curvature for gait synchronization data and energy compositional data in Section \ref{chap7} and   discussion in Section \ref{chap8}. 

    \section{Inference for dispersion measures of random objects}\label{chap2}

    Consider a separable metric space $(\M,d)$ endowed with a probability measure $\mP$ and  random objects $X$, by which we mean random variables  that take values in the metric space $\M$. We observe a sample of random objects $X_{1}, \ldots, X_{n}$ with the same distribution as $X.$ The sample estimates for the \F mean $\mu_{\oplus}$ in \eqref{Fmean_pop}, which may be a set of minimizers,   and for the \F variance $V_{F}$ in \eqref{Vf_pop} are given by, respectively, 
    \begin{equation}
        \hat{\mu}_{\oplus} = \operatorname*{argmin}_{\xi \in \M} \frac{1}{n}\sum_{i=1}^{n}d^{2}(\xi,X_{i}),\quad \hat{V}_{F} = \frac{1}{n}\sum_{i=1}^{n}d^{2}(\hat{\mu}_{\oplus},X_{i}).
        \label{Vf_sam}
    \end{equation}

    The CLT for \F variance follows  from the theory of M-estimators \citep{van2023weak}, under the following conditions \citep{dubey2019frechet}:
    \vspace{2mm}
    
    \textit{(F0)} The space $\M$ is totally bounded. 
    
    \textit{(F1)} {For all $\epsilon > 0,$ $\inf\limits_{d(\xi,\mu_{\oplus})>\epsilon}\mathbb{E}d^{2}(\xi,X) > \mathbb{E}d^{2}(\mu_{\oplus},X).$} 
    
    \textit{(F2)} {$\delta \int_{0}^{1}\left[1 + \log N(\delta\epsilon/2,B_{\delta}(\xi),d) \right]^{\frac{1}{2}} d\epsilon \rightarrow 0$ as $\delta \rightarrow 0,$ for all $\xi \in \mathcal{M},$ where $B_{\delta}(\xi)$ is an open ball of radius $\delta$ with the metric $d$ centered at $\xi,$ and $N(\zeta,\mathcal{S},d)$ denotes the smallest cardinality among all $\zeta$-covers for the set $\mathcal{S}$ with the metric $d.$}  
    \vspace{2mm}
    
    Condition \textit{(F0)} is needed for weak convergence and \textit{(F1)} corresponds to the uniqueness of the population \F mean $\mu_{\oplus},$ which implies the consistency of any sample \F mean $\hat{\mu}_{\oplus}.$  Condition \textit{(F2)}  is an entropy condition that limits the complexity of the space $(\M,d).$  
    The sample estimate for the metric variance $V_{M}$ in \eqref{Vm_pop} is 
    \begin{equation}
        \hat{V}_{M} = \frac{1}{n(n-1)}\sum_{1\leq i < j \leq n}d^{2}(X_{i},X_{j}).
        \label{Vm_sam}
    \end{equation}
 To derive the CLT for metric variance, we require {the following} conditions.
    \vspace{2mm}
    
    \textit{(M0)} $0<\mathbb{E}d^{2}\left(X,X'\right)< \infty,$ for i.i.d. random objects $X, X' \in \mathcal{M}.$  
    
    \textit{(M1)} {$\mathbb{E}d^{4}(X,X') < \infty,$ for i.i.d. random objects $X, X' \in \mathcal{M}.$} 
    
    \textit{(M2)} $\text{Var}_{X}\left[\mathbb{E}_{X'|X}\left\{d^{2}\left(X,X'\right)|X\right\}\right] > 0.$
    \vspace{2mm}
    
    Here  \textit{(M0)} is needed for the metric variance to be  well-defined, while \textit{(M1)} is a weak requirement 
    to bound the central moment of the metric variance estimates $\hat{V}_{M}$ and \textit{(M2)} ensures that the metric variance estimate $\hat{V}_{M}$ is meaningful.
    
    
    The metric variance estimate $\hat{V}_{M}$ takes the form of a U-statistics, making it possible to leverage 
    martingale properties \citep{hoeffding1961strong,berk1966limiting}. While U-statistics are traditionally discussed in the context of Euclidean spaces 
    \citep{serfling2009approximation,lee2019u}, the underlying concept can be extended to more general metric spaces $\mathcal{M}$ \citep{alberink2001berry}. 
    The following Proposition is a 
    consequence of these martingale properties. All proofs are in the Supplement. 
    \begin{prop}\label{prop1}
        Under (M0), $\hat{V}_{M} \rightarrow V_{M}$ almost surely. Furthermore, assume that conditions (M0) and (M1) hold. Then, $\mathbb{E}\lVert \hat{V}_{M} - V_{M} \rVert^{2} = O(n^{-1}).$
    \end{prop}
    Considering the joint distribution of estimates of metric variance and \F variance,  their theoretical foundations
    differ substantially. While in the CLT for \F variance \cite{dubey2019frechet} primarily draw upon M-estimation theory, the  uniqueness of the \F mean $\mu_{\oplus}$ \textit{(F1)} and the  entropy condition \textit{(F2)}, the  CLT for metric variance does not require either of these assumptions. Instead, one can more directly utilize 
    properties of U-statistics  \citep{hoeffding1948class} and \textit{(M0)-(M2)}. {Note that total boundedness \textit{(F0)} implies the moment condition \textit{(M1)}. We now present the following joint central limit theorem.}

    \begin{thm}\label{thm1}
        Under  (F0), (F1), (F2), (M0) and (M2), 
        \begin{equation*}
        \sqrt{n}\left((\hat{V}_{M},\hat{V}_{F})^{T}
        - (V_{M}, V_{F})^{T} \right) \rightarrow N\left(\mathbf{0},\Sigma\right) \quad \text{in distribution,}
    \end{equation*} where $\Sigma = \begin{pmatrix}
            \sigma^{2}_{M} & \sigma_{FM}\\
            \sigma_{FM} & \sigma^{2}_{F}
        \end{pmatrix},$ $\sigma^{2}_{M} = \var_{X}\left[\mathbb{E}_{X'|X}\left\{d^{2}\left(X,X'\right)|X\right\}\right], \sigma^{2}_{F} = \var\left(d^{2}(\mu_{\oplus},X)\right),$ and $\sigma_{FM} = \cov\left( \mathbb{E}_{X'|X}\left\{d^{2}\left(X,X'\right)|X\right\}, d^{2}(\mu_{\oplus},X)\right)$.
    \end{thm}


    The convergence rate towards the limiting distribution for the metric variance estimates $\hat{V}_{M}$ is based on the Berry-Essen theorem for U-statistics \citep{alberink2001berry}, an extension of  previous bounds of   \cite{callaert1978berry, van1984berry, friedrich1989berry}. The following additional finite moment assumption is  needed.
    \vspace{2mm}
    
    \textit{(M3)} $\mathbb{E}_{X}\lvert \mathbb{E}_{X|X'} \left\{d^{2}(X,X')|X \right\} \rvert^{3} < \infty,$ for i.i.d. random objects $X, X' \in \mathcal{M}.$ 
    
    \vspace{2mm}
    
    For a bounded metric space $\mathcal{M},$  $\mathbb{E}_{X}\lvert\mathbb{E}_{X'|X}\left\{d^{2}\left(X,X'\right)|X\right\}\rvert^{p} $ is finite for all $p > 0,$ so that \textit{(M3)} automatically holds.
    \begin{prop}\label{prop2}
        Under {(M0)-(M3)}, $\sup\limits_{x}\left|\mathbb{P}\left(\sqrt{n}(\hat{V}_{M}-V_{M})/\sigma_{M} \leq x \right) - \Phi(x)\right| = O(n^{-\frac{1}{2}})$ \newline for $n \geq 2$,
        where $\Phi(x)$ is the distribution function of the standard normal distribution.
    \end{prop}
    
    
    The proof relies on an application of the Essen  lemma for the characteristic function of  metric variance 
   and   involves decomposing  $\hat{V}_{M}$ into a projection and a residual part. The projection component is addressed similarly as in the standard proof for the Berry-Esseen theorem for i.i.d. summands, while the residual component 
    is handled by utilizing a martingale structure.
    Simultaneous inference for two dispersion measures can be obtained from  the asymptotic normality established in Theorem \ref{thm1} by supplementing sample-based estimators for the asymptotic covariance matrix \begin{equation}
        \hat{\Sigma} = \begin{pmatrix}
            \hat{\sigma}^{2}_{M} & \hat{\sigma}_{FM}\\
            \hat{\sigma}_{FM} & \hat{\sigma}^{2}_{F}
        \end{pmatrix},
        \label{asym_var_sam}
    \end{equation}
    given by $\hat{\sigma}^{2}_{M} = \frac{1}{n}\sum\limits_{i=1}^{n}\{\frac{1}{n-1} \sum\limits_{j \neq i}^{n} d^{2}(X_{i},X_{j})\}^{2} - \{\frac{2}{n(n-1)}\sum\limits_{1 \leq i < j \leq n} d^{2}(X_{i},X_{j} )\}^{2},$ 
    
  \no  $\hat{\sigma}_{FM} = \frac{1}{n}\sum\limits_{i=1}^{n} \frac{d^{2}(\hat{\mu}_{\oplus},X_{i})}{n-1} \{\sum\limits_{j \neq i}^{n} d^{2}(X_{i},X_{j})\} - \{\frac{1}{n}\sum\limits_{i=1}^{n}d^{2}(\hat{\mu}_{\oplus},X_{i})\} \{\frac{2}{n(n-1)}\sum\limits_{1 \leq i < j \leq n} d^{2}(X_{i},X_{j} ) \},$ 
  
  \no and $\hat{\sigma}^{2}_{F} = \frac{1}{n}\sum\limits_{i=1}^{n}d^{4}(\hat{\mu}_{\oplus},X_{i}) - \{\frac{1}{n}\sum\limits_{i=1}^{n}d^{2}(\hat{\mu}_{\oplus},X_{i})\}^{2}.$
    
    The following Proposition establishes the consistency of these estimates. 
    
    \begin{prop}\label{prop3}
        Assume that the conditions of Theorem \ref{thm1} hold. Then \newline
    \hspace{2cm}  $ \hat{\sigma}_{FM}^2 \rightarrow \sigma_{FM}^2,$ $\hat{\sigma}^{2}_{M} \rightarrow \sigma^{2}_{M},$ $\hat{\sigma}^{2}_{F} \rightarrow \sigma^{2}_{F}$ in probability. 
    \end{prop}
      
    Setting $\hat{\eta} = \left(\hat{V}_{M}, \hat{V}_{F}\right)^{T},$ $\eta = \left(V_{M}, V_{F}\right)^{T},$ these results enable the construction of an asymptotic 
    joint $100(1-\alpha)\%$ confidence region for $\eta$ given by    \begin{equation}
        \mathcal{C}_{n}(1-\alpha) := \left\{\eta \in \mathbb{R}^{2} \mid \left(\hat{\eta} - \eta\right)^{T}\left(\hat{\Sigma}/n\right)^{-1}  \left(\hat{\eta} - \eta\right) \leq \chi^{2}_{2}(1-\alpha)\right\},
        \label{CR}
    \end{equation}
    where $\chi^2_{q}(1-\alpha)$ is the $(1-\alpha)$ quantile of the chi-square distribution with $q$ d.f. 

    \section{Alexandrov curvature and dispersion measures $V_M$ and $V_F$}
    \label{chap3}


    \subsection{Metric geometry}\label{chap31}

Consider random objects $X$ taking values in the Hilbert space $\mathcal{H}.$ In this case, the \F mean $\mu_{\oplus}$ of $X$ is equivalent to the expectation $\mathbb{E}X.$ Using the inner product of the Hilbert space, it can be readily shown that the \F variance $V_{F}$ and metric variance $V_{M}$ are identical,  $V_F=V_M$;  this identity  does not hold in general metric spaces  \cp{dubey2020functional}.
    We provide a brief overview of basic metric geometry.  Further details can be found in \cite{burago2001course, lang2012differential,lin2021total}. {For a metric space $(\mathcal{M},d)$, a path $\gamma : \left[0, T \right] \rightarrow \mathcal{M}$ is a geodesic from $X \in \mathcal{M}$ to $Y \in \mathcal{M}$ if $\gamma(0) = X,$ $\gamma(T) = Y$ and $d\left(\gamma(t),\gamma(t')\right) = \lvert t-t' \rvert$ for all $t, t' \in \left[0, T \right]$. }
    The metric space $(\mathcal{M},d)$ is a geodesic space if any pair of points $X,Y \in \mathcal{M}$ can be connected by a geodesic. The metric space $(\mathcal{M},d)$ is a unique geodesic space if geodesics are unique. Euclidean space is a unique geodesic space and geodesics are straight lines connecting any two points.
    
    In contrast to Euclidean spaces, general metric spaces can exhibit \textit{curvature}, which measures deviation from flatness. For a geodesic space $\mathcal{M}$, curvature is assessed by comparison triangles situated in model metric spaces 
    $(\mathcal{M}_{\kappa},d_{\kappa})$, where $\kappa$ is a curvature parameter.
    
    For $\kappa > 0,$ the model space is the sphere $\mathbb{S}^{2} = \left\{(x,y,z) \mid x^2 + y^2 + z^2 = 1 \right\}$ with the angular distance $d_{\kappa}(a,b) = \arccos(x_{a}x_{b} + y_{a}y_{b} + z_{a}z_{b})/\sqrt{\kappa},$ where $a = (x_{a}, y_{a}, z_{a}) $ and $b = (x_{b}, y_{b}, z_{b}).$ 
    For $\kappa <0,$ the model space is the hyperbolic space $\mathbb{H}^{2} = \{ (x,y,z) \mid x^2 + y^2 - z^2 \newline 
    = -1 \}$ with the hyperbolic distance $d_{\kappa}(a,b) = \text{arccosh}(z_{a}z_{b} - x_{a}x_{b} - y_{a}y_{b})/\sqrt{-\kappa}$. Finally, for $\kappa = 0$, the model space is the Euclidean plane $\real^{2}$ with the Euclidean distance. 
      A \textit{geodesic triangle} for vertices $a, b, c$ on the space $\mathcal{M},$ denoted by $\triangle(a, b, c),$ consists of three geodesic edges $\gamma_{ab}, \gamma_{bc},$ and $\gamma_{ca},$ connecting the corresponding vertices.  
      A \textit{comparison triangle} $\tilde{\triangle}(\tilde{a},\tilde{b},\tilde{c})$ on the model space $(\mathcal{M}_{\kappa}, d_{\kappa})$ is formed by reference vertices $\tilde{a}, \tilde{b}, \tilde{c}$ and geodesic edges $\tilde{\gamma}_{\tilde{a}\tilde{b}},\tilde{\gamma}_{\tilde{b}\tilde{c}}$ and $\tilde{\gamma}_{\tilde{c}\tilde{a}}$ 
      that have lengths identical to the lengths of $\gamma_{ab}, \gamma_{bc}$ and $\gamma_{ca}.$ 
      
\begin{figure}[h!]
		\centering
		\subfigure[]{\includegraphics[width=0.26\linewidth]{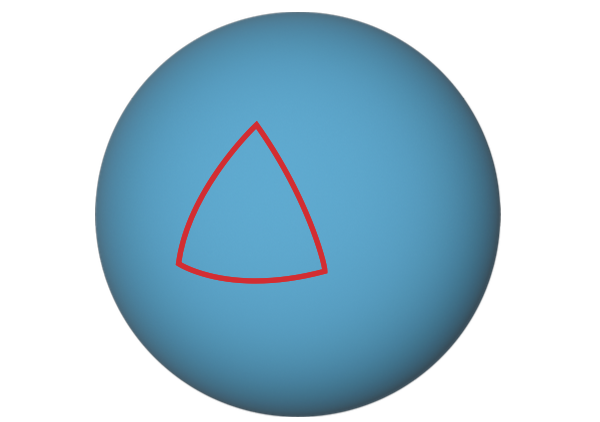}} 
		\subfigure[]{\includegraphics[width=0.26\linewidth]{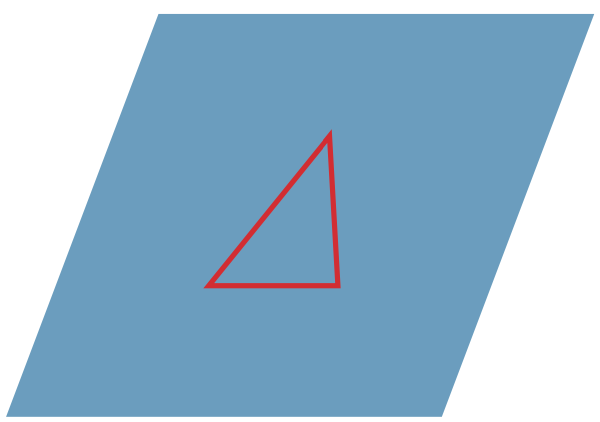}}
		\subfigure[]{\includegraphics[width=0.26\linewidth]{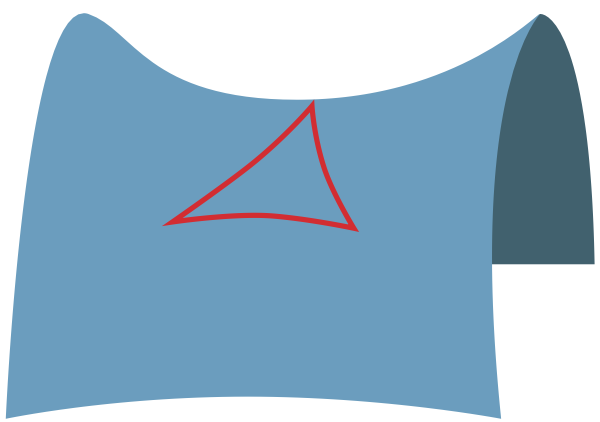}}
		\caption{Model spaces $\M_{\kappa},$ with their comparison triangles $\tilde{\triangle}$. (a) Positive curvature parameter, $\kappa > 0,$ (b) Flat, $\kappa=0,$ and (c) Negative, $\kappa < 0.$}
		\label{fig:universe}
	\end{figure}

      The formal definition of curvature involves comparing the comparison triangle $\tilde{\triangle}(\tilde{a}, \tilde{b}, \tilde{c})$ with the geodesic triangle $\triangle(a, b, c)$ of perimeter $< 2 D_{\kappa},$ where $D_{\kappa} = \infty$ for $\kappa \leq 0,$ and $D_{\kappa} = \frac{\pi}{\sqrt{\kappa}}$ for $\kappa >0.$
      For any point $x$ on segment $\gamma_{bc}$ of $\triangle(a, b, c),$ a corresponding point $\tilde{x}$ exists on segment $\tilde{\gamma}_{\tilde{b}\tilde{c}}$ of $\tilde{\triangle}(\tilde{a}, \tilde{b}, \tilde{c})$ such that $d(x, b) = d_{\kappa}(\tilde{x}, \tilde{b}).$ If $d(x, a) \leq d_{\kappa}(\tilde{x}, \tilde{a})$ (or $d(x, a) \geq d_{\kappa}(\tilde{x}, \tilde{a}),$ respectively) for all such $x,$ as well as for $x$ on $\gamma_{ab}$ or $\gamma_{ac},$ then  $(\mathcal{M},d)$ has curvature $\leq \kappa$ (or $\geq \kappa,$ respectively). Figure \ref{fig:universe} illustrates the shapes of comparison triangles $\tilde{\triangle}$ on  model spaces $(\M_{\kappa}, d_{\kappa}),$ with $\kappa >0$, $\kappa = 0$, and $\kappa <0$.
    
  \subsection{Relationship between  $V_M$ and $V_F$  reflects underlying  curvature}\label{chap32}
    
    	Let $X$ be random objects situated on a complete geodesic space $(\M,d)$, where the underlying  probability measure $\mP$ satisfies the following assumption:
	\vspace{2mm}
	
	\textit{(S1)} For all $\delta >0$ and $\xi \in \M,$ $\mP( X \in B_{\delta}(\xi) ) > 0,$ where $B_{\delta}(\xi)$ is an open ball of radius $\delta$ with the metric $d$ centered at $\xi.$
	\vspace{2mm}
	
	Condition \textit{(S1)} corresponds to the notion of a strictly positive measure \citep{kell:59} and is needed to prevent probability measures $\mP$ from being concentrated on any strict closed subspace $\mathcal{D} \sub \M,$ as the curvature of $\mathcal{D}$ might differ from that of the original space $\M.$
A complete geodesic space $(\M,d)$ featuring non-positive curvature is a \textit{Hadamard space}, or complete CAT(0) space. Geodesic triangles in a Hadamard space are thinner or equal to triangles in a Euclidean space. For random objects $X$  
    in Hadamard spaces, the function $F(\cdot) = \mathbb{E}d^{2}(\cdot,X)$ is strongly convex, ensuring a unique \F mean $\mu_{\oplus}$ if $F(\xi) < \infty$ for some $\xi$ \citep{stur:03}. Examples of Hadamard spaces include  flat spaces that are complete, such as Hilbert space, the space of univariate probability distributions $\mathcal{W}_{2}(\mathbb{R})$ with the 2-Wasserstein metric \citep{kant:06:1}, or SPD matrices with the Frobenius, power and log-Frobenius, Cholesky and log Cholesky metrics  \citep{dryden2009factored, feragen2015geodesic, pigoli2014distances,lin2019riemannian}. Non-flat Hadamard spaces include the space of SPD matrices with the affine-invariant Riemannian metric \citep{bhatia2009positive}  and the phylogenetic tree space with the BHV metric \citep{billera2001geometry}.

	
	Applying the law of cosines and properties of directional derivatives in Hadamard spaces, including Theorem 4.5.6 in \cite{burago2001course}, we find that the following relationship holds between $V_M$ and $V_F$.

      \begin{thm}\label{thm2}
        Suppose $(\M,d)$ is a Hadamard space and  random objects $X$ on $\M$ satisfy $\mathbb{E}d^{2}(\xi,X) < \infty$ for some $\xi \in \M.$ Then  
            ${V}_{M} \geq {V}_{F}.$
        Furthermore, if $(\M, d)$ has  strictly negative curvature and condition $\textit{(S1)}$ holds, then
               $ {V}_{M} > {V}_{F}.$
    \end{thm}

    Next, we establish the relationship for the case of complete geodesic spaces  $(\mathcal{M},d)$ with positive curvature. Examples include the space of SPD matrices with the Bures-Wasserstein metric \citep{dryden2009non,han2021riemannian}, finite-dimensional spheres $\mathbb{S}^{p}$, the  Hilbert sphere $\mathbb{S}^{\infty}$ with geodesic metric  \citep{dai2022statistical}  
    and the 2-Wasserstein space for distributions with Euclidean $\mathcal{W}_{2}(\mathbb{R}^{p}),~ p \geq 2$ \citep{otto2001geometry}. \bco Specifically, let $(\mathcal{X},d_{\mathcal{X}})$ be a complete separable metric space, $\mathcal{B}(\mathcal{X})$ be the Borel $\sigma$-algebra on $\mathcal{X}$, and $\mathcal{P}_{2}(\mathcal{X})$ be the set of Borel probability measures on a space $\mathcal{X}$. Define the $2$-Wasserstein space on $\mathcal{X}$ as
    \begin{equation}
        \mathcal{W}_{2}(\mathcal{X}) = \left\{ \nu \in \mathcal{P}_{2}(\mathcal{X}): \int_{\mathcal{X}} d_{\mathcal{X}}(x,y) d\nu(y) < \infty \right\},
        \label{Wass_space}
    \end{equation}
    for some $x \in \mathcal{X}.$ Consider the $2$-Wasserstein space $\mathcal{W}_{2}(\mathcal{X})$ of probability space on $(\mathcal{X},\mathcal{B}(\mathcal{X})),$ endowed with the $\mathcal{L}^{2}$-Wasserstein distance
    \begin{equation}
        d_{\mathcal{W}}(\nu_1,\nu_2) = \left( \inf_{\pi \in \Pi(\nu_1,\nu_2)} \int_{\mathcal{X}\times \mathcal{X}} \lVert x - y \rVert^{2} d\pi(x,y) \right)^{1/2},
        \label{Wass_dist}
    \end{equation}
    where $\Pi(\nu_1,\nu_2)$ is the set of probability measures $\pi$ on $\mathcal{X}\times \mathcal{X}$ such that $\nu_1$ and $\nu_2$ are the first and second marginals of $\pi,$ i.e., $\pi(A \times \mathcal{X}) = \nu_1(A)$ and $\pi(\mathcal{X} \times B) = \nu_2(B)$ for all $A,B \in \mathcal{B}(\mathcal{X}).$ \fi


    \begin{thm}\label{thm3}
        Suppose $(\M,d)$ is a complete unique geodesic space with non-negative curvature and condition \textit{(F1)} (uniqueness of the \F mean) holds.  Then 
            ${V}_{M} \leq {V}_{F}.$
        Furthermore, if $(\M, d)$ has  strictly positive curvature and condition $\textit{(S1)}$ holds, then
            ${V}_{M} < {V}_{F}.$
    \end{thm}
    
   {In spaces with positive curvature, the assumption of a unique \F mean $\mu_{\oplus}$ in Theorem \ref{thm3} is not guaranteed to be satisfied.  An example is a probability measure uniformly distributed over an entire sphere $\mathbb{S}^{p}$. Then  every point on the sphere is a  minimizer of the \F function. However, random objects $X$ on the sphere may still have a unique \F mean under certain conditions. For example, if the support of a subset $\mathcal{M} \subset \mathbb{S}^{p}$ is contained within a geodesic ball $B_{r}(\xi)$ centered at some $\xi \in \mathbb{S}^{p}$ with radius $r < \frac{\pi}{2},$ then random objects $X$ on $\M$ have a unique \F mean \citep{afsari2011riemannian}. Moreover, \cite{dai2022statistical} demonstrated that a subset of the Hilbert sphere $\M \sub \mathbb{S}^{\infty}$ with $\sup\limits_{x,y \in \mathcal{M}} d(x,y) \leq \frac{\pi}{2}$ satisfies this assumption.} 
    {For random objects $X$ in 2-Wasserstein space $\mathcal{W}_{2}(\mathbb{R}^{p})$ with $p \geq 2,$ which has a positive curvature, the uniqueness assumption is satisfied when $X$ is absolutely continuous \citep{agueh2011barycenters}. For a broader perspective, \cite{ahidar2020convergence} studied variance inequalities that guarantee the uniqueness of a population \F mean.} 
    
    Combining Theorems \ref{thm2} and \ref{thm3}, we obtain the following result. 
    
   

  \begin{cor}\label{corr1}
        Suppose that the geodesic spaces $(\M,d)$ is flat and complete, random objects $X$ on $\M$ satisfy $\mathbb{E}d^{2}(\xi,X) < \infty$ for some $\xi \in \M,$ and condition $\textit{(S1)}$ holds. Then 
        $V_{M} = V_{F}.$
    \end{cor}


\section{Inference for curvature} \label{chap4}
    
      \subsection{A test for metric curvature}\label{chap41}

    Suppose that $(\M, d)$ is complete and a totally bounded geodesic space. Consider population \F variance $V_{F}$ in \eqref{Vf_pop} and metric variance $V_{M}$ in \eqref{Vm_pop}. We define the {\it metric curvature}  $\rho$ of the space $\M$ as 
    \begin{equation}
        \rho = \frac{{V}_{F}}{{V}_{M}} - 1.
        \label{curv_alex}
    \end{equation}
	
     In the following we show that $\rho<0$  when $(\M, d)$ exhibits strictly negative curvature, as outlined in Theorem \ref{thm2}. Similarly, $\rho>0$  when $(\M, d)$ has strictly positive curvature (Theorem \ref{thm3}) and  $\rho=0$ when $(\M, d)$ is  a flat space (Corollary \ref{corr1}). 
    
    \begin{prop}\label{prop5}
    Under condition \textit{(S1)}, if $\left(\M, d\right)$ has  strictly negative curvature, then $\rho < 0.$ If  $\left(\M, d\right)$ has strictly positive curvature and condition \textit{(F1)} holds, then $\rho > 0.$ Furthermore, if $\left(\mathcal{M}, d\right)$ is a flat space, then $\rho = 0.$
    \end{prop}
	
	Consider random samples $X_{1}, \ldots, X_{n}$ with the same distribution as $X,$ and the sample \F variance $\hat{V}_{F}$ in \eqref{Vf_sam} and metric variance $\hat{V}_{M}$ in \eqref{Vm_sam}.  The sample estimator for $\rho$ in \eqref{curv_alex} is given by
    \begin{equation}
        \hat{\rho} = \frac{\hat{V}_{F}}{\hat{V}_{M}} - 1.
        \label{curv_alex_sam}
    \end{equation}
    Theorem \ref{thm1} coupled with the continuous mapping theorem yields  the asymptotic variance of $\hat{\rho}$, where $\sigma^{2} = a^{T}\Sigma a,$
        with $a = \left(-\frac{V_{F}}{V^{2}_{M}}, \frac{1}{V_{M}}\right)^{T},$ and $\Sigma = \begin{pmatrix}
            \sigma^{2}_{M} & \sigma_{FM}\\
            \sigma_{FM} & \sigma^{2}_{F}
        \end{pmatrix}.$ Proposition \ref{prop3} provides a consistent estimate of the asymptotic variance $\sigma^{2},$ given by
        \begin{equation}
            \hat{\sigma}^{2} = \hat{a}^{T} \hat{\Sigma} \hat{a},
            \label{asym_vari_rho}
        \end{equation} with $\hat{a} = \left(-\frac{\hat{V}_{F}}{\hat{V}^{2}_{M}}, \frac{1}{\hat{V}_{M}}\right)^{T}$, where the covariance matrix estimate  $\hat{\Sigma}$ is as in  \eqref{asym_var_sam}. 
    
    {The \F variance estimate $\hat{V}_{F}$ relies on the computation of the sample \F mean $\hat{\mu}_{\oplus},$ which often requires case-specific approaches depending on the underlying metric space. For Hadamard space, $\hat{\mu}_{\oplus}$ can be computed using the proximal point algorithm \citep{bacak2014computing}. For Riemannian manifolds with positive curvature, such as the sphere $\mathbb{S}^{p},$ gradient-based optimization methods \citep{fletcher2004principal} can be employed. For multivariate distributions under the 2-Wasserstein metric $\mathcal{W}_{2}(\real^{p}),$ particularly when the space does not belong to the location-scatter family, $\hat{\mu}_{\oplus}$ can be estimated using the Sinkhorn approximation \citep{cuturi2013sinkhorn} or gradient descent algorithm \citep{chewi2020gradient}.}

    
    Formally, consider the null hypothesis $H_{0}: \mP \in \Theta_{0}$ and the alternative hypothesis $H_{1}: \mP \in \Theta_{1}.$ To determine whether the space $\M$ is curved or not, we can formulate a two-sided test for $\Theta_{0} = \{ \rho \mid \rho = 0\}$ and $\Theta_{1} = \{ \rho \mid \rho \neq 0\}.$ 

    The proposed test statistic $T_{n}$ is then
    \begin{equation}
        T_{n} = \frac{\sqrt{n}\hat{\rho}}{\hat{\sigma}}.
        \label{curv_test}
    \end{equation}
    Under the null hypothesis that $\M$ is a flat space,  Propositions \ref{prop3} and  \ref{prop5} lead to  the following key property of the test statistic. 
    
    \begin{thm}\label{thm4}
        Under (F0), (F1), (F2), (M0), (M2)  and (S1),  for the curvature $\rho$ in \eqref{curv_alex},  then {
            $\sqrt{n}(\hat{\rho} - \rho)/\hat{\sigma} \rightarrow N(0, 1)$ in distribution. Thus, under $H_0: \rho = 0,$ $T_{n} \rightarrow N(0,1)$ in distribution.}
    \end{thm}
    
   
    For any level $\alpha \in (0,1),$ define a test function with size $\alpha$ as $\Phi(X_1,\dots,X_n) = \textbf{I}(R_{n,\alpha}),$ where $\textbf{I}(\cdot)$ is an indicator function and $R_{n,\alpha}$ is a rejection region.
   For a level $\alpha$ two-sided test, the rejection region is $R_{n,\alpha} =  \{  \lvert T_{n} \rvert > z_{\alpha / 2} \},$ where $z_{\eta}$ is the $\left(1-\eta \right)^{th}$ quantile of the standard normal distribution. The $100(1-\alpha)\%$ confidence interval for $\rho$ is given by
    \begin{equation}
        \mathcal{I}_{n}(1-\alpha) := \left\{\rho \in \mathbb{R}~|~\hat{\rho}-z_{{\alpha}/{2}} \frac{\hat{\sigma}}{\sqrt{n}} \leq \rho \leq  \hat{\rho}+z_{{\alpha}/{2}} \frac{\hat{\sigma}}{\sqrt{n}}\right\}.
        \label{CR_curv}
    \end{equation} 
    
    For one-sided tests, where the alternative hypothesis is either $\Theta_{1} = \{ \rho \mid \rho > 0\}$ (positive curvature) or $\Theta_{1} = \{ \rho \mid \rho < 0\}$ (negative curvature), the corresponding rejection regions are  $R_{n,\alpha} =  \{ T_{n} > z_{\alpha} \},$ or $R_{n,\alpha} =  \{ T_{n} < -z_{\alpha} \},$ respectively. Confidence regions $\mathcal{C}_{n}(1-\alpha)$ for $\eta = (V_{M}, V_{F})^{T}$ in \eqref{CR} can equally  be utilized to infer curvature.  For a level $\alpha$ test, the rejection region is  $R_{n,\alpha} = \{\eta \mid \left(\hat{\eta} - \eta\right)^{T}(\hat{\Sigma}/n)^{-1}  \left(\hat{\eta} - \eta\right) > \chi^{2}_{2}(1-\alpha)\}.$

    To evaluate the power of the proposed test, we consider alternative hypotheses defined as {$H_{1,\delta}: \rho = \rho_n,$ where $\rho_n \to 0$ and $\sqrt{n}\rho_n \to \infty$ as $n \to \infty.$ Here, $\rho_n$ is the metric curvature parameter at sample size $n$. These alternatives $\{H_{1,\delta}\}$ are contiguous, shrinking toward $H_{0}$ as $n \rightarrow \infty.$ The asymptotic normality of $\hat{\rho}$ in Theorem \ref{thm4} implies that the asymptotic power of the level $\alpha$ test under $H_{1,\delta}$ is \begin{equation*}
        \beta_{1,\delta} = \mathbb{P}(R_{n,\alpha})= \mathbb{P}(|T_{n}| > z_{\alpha/2}) \to  1 \quad \text{as} \quad n \to \infty,
    \end{equation*} demonstrating the consistency of the test. }

  \bco
  To examine the power of the test for contiguous alternatives indexed by a parameter $\delta \neq 0,$
    consider alternatives 
        $H_{a, \delta} : \mP \in \Theta_{a, \delta} := \{ \rho \mid \rho = \delta \}$
  and the power function $\beta_{a,\delta} = \inf_{X \in H_{a, \delta}} \mathbb{P}(R_{n,\alpha}).$ 
  As the asymptotic distribution of $\hat{\rho}$ converges to $\delta$ in probability,  
 the proposed test is consistent.  
 The asymptotic power $\beta_{a,\delta},$ of both two-sided and one-sided tests under $H_{a,\delta}$ converges to 1 with the rate $\frac{1}{\sqrt{n}}.$ 
 \fi

     \subsection{Detecting intrinsic curvature}\label{chap4.2}

    A practical problem of interest is to infer intrinsic curvature from data $X_{1}, \ldots, X_{n} \in \mA \subset \M,$ where the underlying probability measure $\mP$ is not strictly positive on $\M$ but is instead concentrated on a simply-connected strict subset $\mA,$ 
on which it is strictly positive with $\mP(\mA)=1.$
 Then a part of $\M$  is devoid of positive probability and the subset $\mA$ may exhibit intrinsic curvature that differs from the overall curvature of the ambient space  $\M$. 

    {This motivates extending the proposed metric curvature test from Section \ref{chap41} by incorporating the intrinsic geodesic distance $d_{I}(X, Y)$ defined on   $\mA$. Consider the set $\Gamma_{\mA}$ of paths $\gamma : [0,T] \rightarrow \mA$ that are entirely contained within the simply connected intrinsic space $\mA,$ i.e., $\gamma(t) \in \mA$ for all $t \in  [0,T]$. 
    The intrinsic distance is defined as $d_{I}(X, Y) = \operatorname*{inf}\limits_{\gamma \in \Gamma_{\mA}} \{l(\gamma) \mid \gamma(0) = X, \, \gamma(T) = Y \},$ where  $l(\gamma) := \sup\limits_{0=t_{0}<t_{1}<\ldots<t_{n}=T, n \in \mathbb{N}} \sum\limits_{i=1}^{n}d\left(\gamma(t_{i-1}),\gamma(t_{i})\right).$} A consequence of the constraint on paths routing through $\mA$ is that for all $x,y \in \mA,$ one has  $d_I(x,y) \geq d(x,y),$ where $d_I(x,y)$ is the length of the shortest path on $\mA$ connecting the two points $x,y \in \mA$.

    {Denote by $\mu_{I,\oplus},$ $V_{I, F},$ $V_{I, M},$ and $\rho_{I}$  the intrinsic \F mean, \F variance, metric variance and curvature, respectively, obtained by replacing the ambient distance $d$ in $\mu_{\oplus}$, $V_{F}$, $V_{M},$ and $\rho$ in \eqref{Fmean_pop}, \eqref{Vf_pop}, \eqref{Vm_pop}  and \eqref{curv_alex} with the intrinsic distance $d_{I}.$ For a random sample $\mathcal{X}_{n} := \{X_{1}, \ldots, X_{n} \},$ we define the intrinsic sample \F mean  as $\hat{\mu}_{I,\oplus} = \argmin\limits_{\xi \in \mathcal{X}_n} \frac{1}{n} \sum\limits_{i=1}^{n} d_{I}^{2}(\xi, X_{i}).$ Similarly, we define the corresponding estimators $\hat{V}_{I,F},$ $\hat{V}_{I,M},$ $\hat{\rho}_{I}$ and $\hat{\sigma}_{I}$ by replacing the ambient distance $d$ with the intrinsic distance $d_{I}.$ To infer whether the intrinsic space $\mA$ is curved or not, 
    we consider the test statistic
    \begin{equation}
        T_{I, n} = \frac{\sqrt{n}\hat{\rho}_{I}}{\hat{\sigma}_{I}},
        \label{test_stat}
    \end{equation}
    in analogy to \eqref{curv_test} in Section \ref{chap41}. }
     
\bco

    We define  the \textit{intrinsic curvature} 
    \begin{equation}
        \rho_{I} = \frac{V_{I, F}}{V_{I,M}} - 1,
        \label{curv_ts}
    \end{equation}
where we suppress the dependence on the set $\mA$, with corresponding sample estimators 
    \begin{equation}
        \hat{\mu}_{I,\oplus} = \operatorname*{argmin}_{\xi \in \left\{ X_{1}, \ldots, X_{n}\right\}} \frac{1}{n}\sum_{i=1}^{n} d_{I}^{2}(\xi, X_{i}), \quad \hat{V}_{I,F} = \frac{1}{n} \sum_{i=1}^{n} d_{I}^{2}(\hat{\mu}_{I,\oplus}, X_{i}),
        \label{Vf_supp_sam}
    \end{equation}
    and
    \begin{equation}
        \hat{V}_{I,M} = \frac{1}{n(n-1)}\sum_{1\leq i < j \leq n}d_{I}^{2}(X_{i},X_{j}),
        \label{Vm_supp_sam}
    \end{equation}
    motivating  the following sample estimator for the intrinsic curvature \eqref{curv_ts} 
    \begin{equation}
        \hat{\rho}_{I} = \frac{\hat{V}_{I, F}}{\hat{V}_{I,M}} - 1.
        \label{curv_ts_sam}
    \end{equation}
    In analogy to \eqref{curv_test} in Section \ref{chap41}, for testing whether the intrinsic space $\mA$ is curved or not, 
    we consider the test statistic
    \begin{equation}
        T_{I, n} = \frac{\sqrt{n}\hat{\rho}_{I}}{\hat{\sigma}_{I}},
        \label{test_stat}
    \end{equation}
    where $\hat{\sigma}^{2}_{I} = \hat{\sigma}^{2}_{I,F} - 2\hat{\sigma}^{2}_{I,FM} +\hat{\sigma}^{2}_{I, M},$
    $\hat{\sigma}^{2}_{I,F} = \frac{1}{n}\sum\limits_{i=1}^{n}d_{I}^{4}(\hat{\mu}_{I,\oplus},X_{i}) - \{\frac{1}{n}\sum\limits_{i=1}^{n}d_{I}^{2}(\hat{\mu}_{I,\oplus},X_{i})\}^{2},$
    $\hat{\sigma}^{2}_{I,M} = \frac{1}{n}\sum\limits_{i=1}^{n}\{\frac{1}{n-1} \sum\limits_{j \neq i}^{n} d_{I}^{2}(X_{i},X_{j})\}^{2} - \{\frac{2}{n(n-1)}\sum\limits_{1 \leq i < j \leq n} d_{I}^{2}(X_{i},X_{j} )\}^{2},$ and $\hat{\sigma}_{I,FM} = $
    
    \no$\frac{1}{n}\sum\limits_{i=1}^{n} \frac{d_{I}^{2}(\hat{\mu}_{I,\oplus},X_{i})}{n-1} \{\sum\limits_{j \neq i}^{n} d_{I}^{2}(X_{i},X_{j})\} - \{\frac{1}{n}\sum\limits_{i=1}^{n}d_{I}^{2}(\hat{\mu}_{I,\oplus},X_{i})\} \{\frac{2}{n(n-1)}\sum\limits_{1 \leq i < j \leq n} d_{I}^{2}(X_{i},X_{j} ) \}.$
\fi

    To study the large sample behavior of the test statistic  $T_{I,n}$ in \eqref{test_stat}, we require the following assumptions: \vspace{2mm}
    
    \textit{(A0)} {The measure  $\mP$ of $X$ is strictly positive on $\mA$ and $\mP(\mA) = 1.$} 

    \textit{(A1)} The space $(\mA,d_{I})$ is totally bounded.
    
    \textit{(A2)} {For all $\epsilon > 0,$ $\inf\limits_{d_{I}(\xi,\mu_{I, \oplus})>\epsilon}\mathbb{E}d_{I}^{2}(\xi,X) > \mathbb{E}d_{I}^{2}(\mu_{I, \oplus},X).$} 
     
    \textit{(A3)} For all $\xi \in \mA,$ $\delta \int_{0}^{1}\left[1 + \log N(\delta\epsilon/2,B_{\delta}(\xi),d_{I}) \right]^{\frac{1}{2}} d\epsilon \rightarrow 0$ as $\delta \rightarrow 0.$
    
    \textit{(A4)} $\mathbb{E}\lvert d_{I}^{2}\left(X,X'\right)\rvert > 0$ and $\text{Var}_{X}\left[\mathbb{E}_{X'|X}\left\{d_{I}^{2}\left(X,X'\right)|X\right\}\right] > 0$ for i.i.d. random objects  
    
    \hspace{.8cm} $X, X' \in \mathcal{A}.$  
    \vspace{2mm}
    
    Condition \textit{(A0)} is commonly employed in manifold learning algorithms such as Laplacian eigenmaps \citep{belkin2001laplacian}  and UMAP \citep{mcinnes2018umap}. 
    Under \textit{(A0)}, the measure is concentrated on $\mA.$ Conditions \textit{(A1)-(A3)} impose boundedness, uniqueness of the intrinsic \F mean and an entropy condition for the intrinsic space $\mA$, which are necessary for the application of empirical process theory. Condition \textit{(A4)} is required to ensure the existence of $\rho_{I}$ and prevent degeneration of the asymptotic distribution.
    {\begin{thm}\label{thm5}
        Suppose conditions (A0)-(A4) hold. {Then, $\sqrt{n}(\hat{\rho}_I - \rho_I)/\hat{\sigma}_I \rightarrow N(0, 1)$ in distribution. Thus, under $H_0: \rho_I = 0,$ $T_{I, n} \rightarrow N(0,1)$ in distribution.}
    \end{thm}}

  The intrinsic distance $d_{I}$ needs to be estimated because in  practice the subspace $\mA \sub \M$ is usually unknown. One common approach to estimate the intrinsic distance between any two observations $X_{i}, X_{j} \in \mathcal{X}_{n}$ involves constructing a weighted graph based on nearest neighbors and computing the shortest paths using Dijkstra's algorithm \citep{dijkstra1959note}, as detailed in Algorithm \ref{table:algo}. 
    For a discussion of the convergence of Dijkstra's algorithm 
we refer to Section S.3 of the Supplement.

    \begin{algorithm}[h!]
    \caption{Intrinsic distance estimation}
    \label{table:algo}
    \begin{algorithmic}
    \STATE \textbf{Input}: Observations $\mathcal{X}_{n},$ ambient distance $d$ and neighbor ball size parameter $r.$
    \STATE \textbf{Output}: Pairwise intrinsic distance estimate $d_{I}(X_{i}, X_{j}),$ for all $X_{i}, X_{j} \in \mathcal{X}_{n}.$
    \STATE \quad 1. Construct a weighted graph for each pair $(X_{i}, X_{j}) \in \mathcal{X}_{n},$ connecting if $d(X_{i}, X_{j}) \leq r.$\\
    \quad 2. Set $d_{I}(X_{i}, X_{j})$ to be the length of the shortest path calculated by Dijkstra's algorithm \citep{dijkstra1959note}.
    \end{algorithmic}
    \end{algorithm}

\vspace{2mm}

   Following the same argument as in Section \ref{chap41}, for a two-sided test at level $\alpha$, the $(1-\alpha)$ confidence intervals for the intrinsic curvature $\rho_{I}$ and for the intrinsic dispersion measures $\eta_{I} = (V_{I, M}, V_{I, F})^{T}$ are given by:
   \begin{align}
       \mathcal{I}_{I,n}(1-\alpha) &= \left\{\rho_{I} \in \mathbb{R} \mid \hat{\rho}_{I}-z_{{\alpha}/{2}} \frac{\hat{\sigma}_{I}}{\sqrt{n}} \leq \rho_{I} \leq  \hat{\rho}_{I}+z_{{\alpha}/{2}} \frac{\hat{\sigma}_{I}}{\sqrt{n}}\right\},\label{CR_curv_int} \\
       \mathcal{C}_{I, n}(1-\alpha) &= \left\{\eta_{I} \in \mathbb{R}^{2} \mid \left(\hat{\eta}_{I} - \eta_{I}\right)^{T}\left(\hat{\Sigma}_{I}/n\right)^{-1}  \left(\hat{\eta}_{I} - \eta_{I}\right) \leq \chi^{2}_{2}(1-\alpha)\right\}. \label{CR_int}
   \end{align}
    {Here  $\hat{\Sigma}_{I}$ is obtained by replacing $d$ in $\hat{\Sigma}$ \eqref{asym_var_sam} with $d_{I}$ and we plug in the estimates  $\hat{\eta}_{I} = (\hat{V}_{I, M}, \hat{V}_{I, F})^{T}.$} 

   \bco
    \begin{equation}
        \mathcal{I}_{I,n}(1-\alpha) :=\left\{\rho_{I} \in \mathbb{R} \mid \hat{\rho}_{I}-z_{{\alpha}/{2}} \frac{\hat{\sigma}_{I}}{\sqrt{n}} \leq \rho_{I} \leq  \hat{\rho}_{I}+z_{{\alpha}/{2}} \frac{\hat{\sigma}_{I}}{\sqrt{n}}\right\}.
        \label{CR_curv_int}
    \end{equation}
    One-sided tests can be constructed in the same way as in  Section \ref{chap41}.

    The  $100(1-\alpha)\%$ confidence regions for the intrinsic dispersion measures \newline $\eta_{I} = (V_{I, M}, V_{I, F})^{T}$ are the elliptical sets 
    \begin{equation}
        \mathcal{C}_{I, n}(1-\alpha) := \left\{\eta_{I} \in \mathbb{R}^{2} \mid \left(\hat{\eta}_{I} - \eta_{I}\right)^{T}\left(\hat{\Sigma}_{I}/n\right)^{-1}  \left(\hat{\eta}_{I} - \eta_{I}\right) \leq \chi^{2}_{2}(1-\alpha)\right\}.
        \label{CR_int}
    \end{equation}
   Under the null hypothesis $H_{0}: \mP \in \Theta_{0}$ and contiguous alternatives $H_{a, \delta} :  \mP \in \{ \rho_{I} \mid \rho_{I} = \delta \},$ $\delta \neq 0,$ the intrinsic curvature test of size $\alpha,$ denoted as $\Phi(X_{1}, \ldots, X_{n}) = \textbf{I}(R_{n,\alpha}),$ is consistent by the same arguments as in  Section \ref{chap41}.
   \fi

\begin{figure}[ht]
      \centering
        \includegraphics[width=0.24\textwidth]{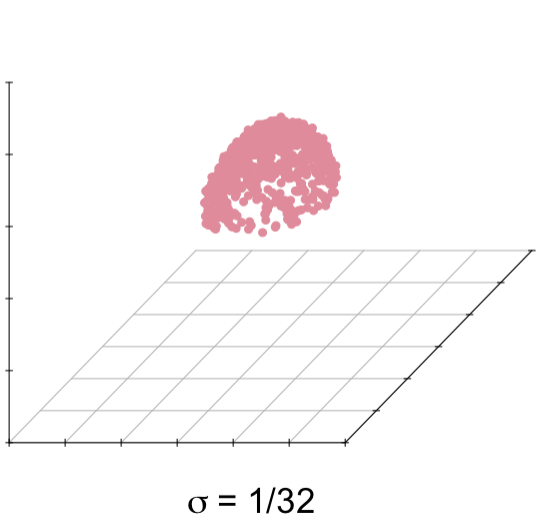}
        \includegraphics[width=0.24\textwidth]{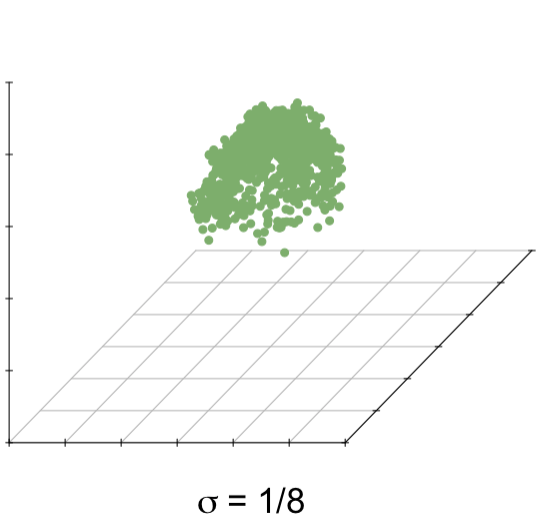}
        \includegraphics[width=0.24\textwidth]{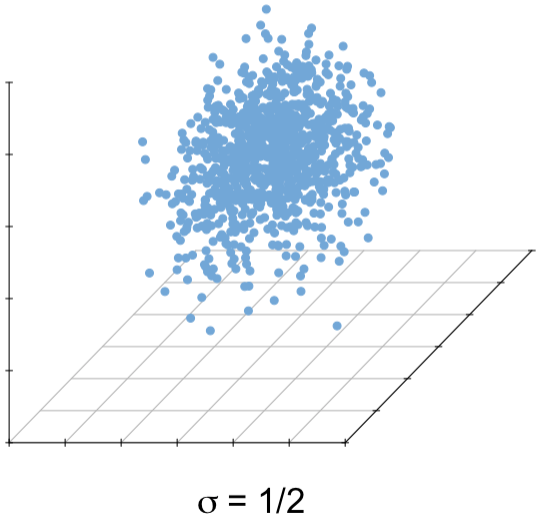}
        \includegraphics[width=0.24\textwidth]{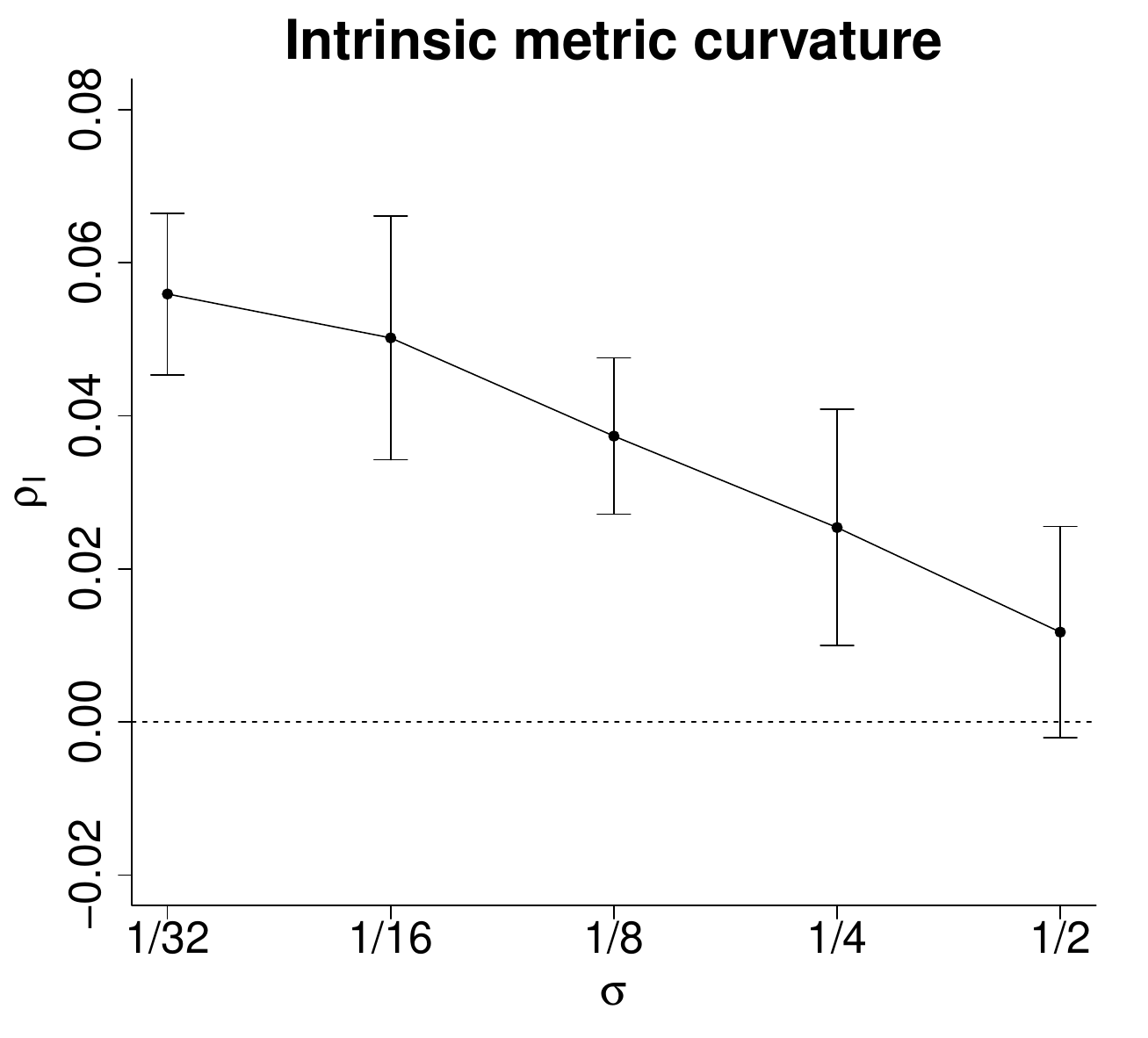}
      \caption{Random samples following a uniform distribution on $\mathbb{S}^{2}_{+}$ perturbed with truncated Gaussian error $TN_{[-1.5,1.5]^3}(\mathbf{0}, \sigma^{2} \text{I}_{3}),$ where $\sigma = 1/32$ (leftmost panel), $\sigma = 1/8,$ (left middle panel), and $\sigma = 1/2$ (right middle panel) and confidence intervals $\mathcal{I}_{I,n}(1-\alpha)$ in \eqref{CR_curv_int} for the intrinsic curvature $\rho_{I}$  with $\alpha = 0.05,$ as a function of $\sigma$ (rightmost panel).}
      \label{fig:contam_curv}
    \end{figure}

    Figure \ref{fig:contam_curv} depicts random objects $X$ following a uniform distribution on the upper hemisphere $\mathbb{S}^{2}_{+}:= \{(x,y,z) \mid x^{2} + y^{2} + z^{2} = 1,\, z \geq 0 \},$ perturbed by truncated Gaussian noise  $TN_{[-1.5,1.5]^{3}}(\mathbf{0}, \sigma^{2}\text{I}_{3}),$ restricted to the domain $[-1.5, 1.5]^{3} \subset \real^{3},$ where $\mathbf{0} = (0,0,0)^{T},$ and $\text{I}_{3}$ is the identity matrix. As the noise variance $\sigma^2$ increases, the intrinsic space $\mA$ induced by the probability measure $\mathcal{P}$  transitions from the positively curved space $\mathbb{S}^{2}_{+}$ for small $\sigma$  to the flat ambient space $\real^{3}$ for larger $\sigma$.  The rightmost panel illustrates the confidence interval $\mathcal{I}_{I,n}(1-\alpha)$ in \eqref{CR_curv_int} for the intrinsic curvature $\rho_{I}$ in dependence on  $\sigma$ for $\alpha=0.05$.   Notably, when $\sigma$ is small, the interval indicates a positive intrinsic curvature for $\mathcal{A}$ and does not include 0;  as $\sigma$ increases, the interval starts to include 0, so that the no curvature null hypothesis cannot be rejected anymore.

         \section{Simulations}\label{chap6}
 \subsection{Example 1: Distributional data with 2-Wasserstein distance}\label{WassSimul}

{Consider a measure $\mP$ on the bivariate Wasserstein space, $\mathcal{W}_2(\real^2),$ such that the measure is positive on bivariate normal distributions $\mathcal{D}$ with $\mP(\mathcal{D})=1$, for   \begin{equation}
    \mathcal{D} = \left\{ N(0, \Lambda(\theta)) \mid \Lambda(\theta) = R(\theta) \Lambda_{0} R(\theta)^{T} \in \real^{2 \times 2}, ~ \theta \in [0,1] \right\},
    \label{2dIntrinspaceD}
\end{equation}
with rotation matrix $R(\theta) = \begin{pmatrix}
    \cos(\frac{\pi}{2}\theta) & - \sin(\frac{\pi}{2}\theta) \\
    \sin(\frac{\pi}{2}\theta) & \cos(\frac{\pi}{2}\theta)
\end{pmatrix}$ and $\Lambda_{0} = \text{diag}(\lambda_1, \lambda_2),$ with $\lambda_{1} = 4$ and $\lambda_{2} = 1.$ The ambient Wasserstein space $\mathcal{W}_2(\real^2)$ is known to exhibit positive Alexandrov curvature. However, the intrinsic curvature of a subspace $\mathcal{D} \sub \mathcal{W}_2(\mathbb{R}^2)$ may exhibit flat geometry, as random samples only depend on the one-dimensional normalized rotation angle $\theta \in [0,1].$}

{To infer the intrinsic curvature of $\mathcal{D} \sub \mathcal{W}_{2}(\real^{2}),$ we generate $X_{i} = N(0, \Lambda(\theta_{i})) \in \mathcal{D},$ where $\theta_{i} \sim \text{Beta}(2, 2),$ $i = 1, \ldots, 100.$ Figure \ref{fig:sub12dwassobs} illustrates $95\%$ density contour plots of random bivariate normal distributions, $X_{i}$ $i = 1, \ldots, 100.$ We can infer the intrinsic curvature of $\mathcal{D},$ where the observed distributions are located, by comparing the intrinsic \F variance $\hat{V}_{I,F}$ and the metric variance $\hat{V}_{I,M}$. The confidence regions $\mathcal{C}_{I, n}(1-\alpha)$ in \eqref{CR_int} for $\eta_{I} = (\eta_{I,1}, \eta_{I,2}) := (V_{I, M},V_{I,F})^{T}$ are shown in Figure \ref{fig:sub22dwassobs}. Since all confidence regions intersect with $\left\{\eta_{I}~|~\eta_{I,1} = \eta_{I,2} \right\},$ they are compatible with an intrinsic flat geometry (although they do not establish this fact). 

\begin{figure}[ht]
  \centering
    \subfigure[]{\includegraphics[width=0.45\textwidth]{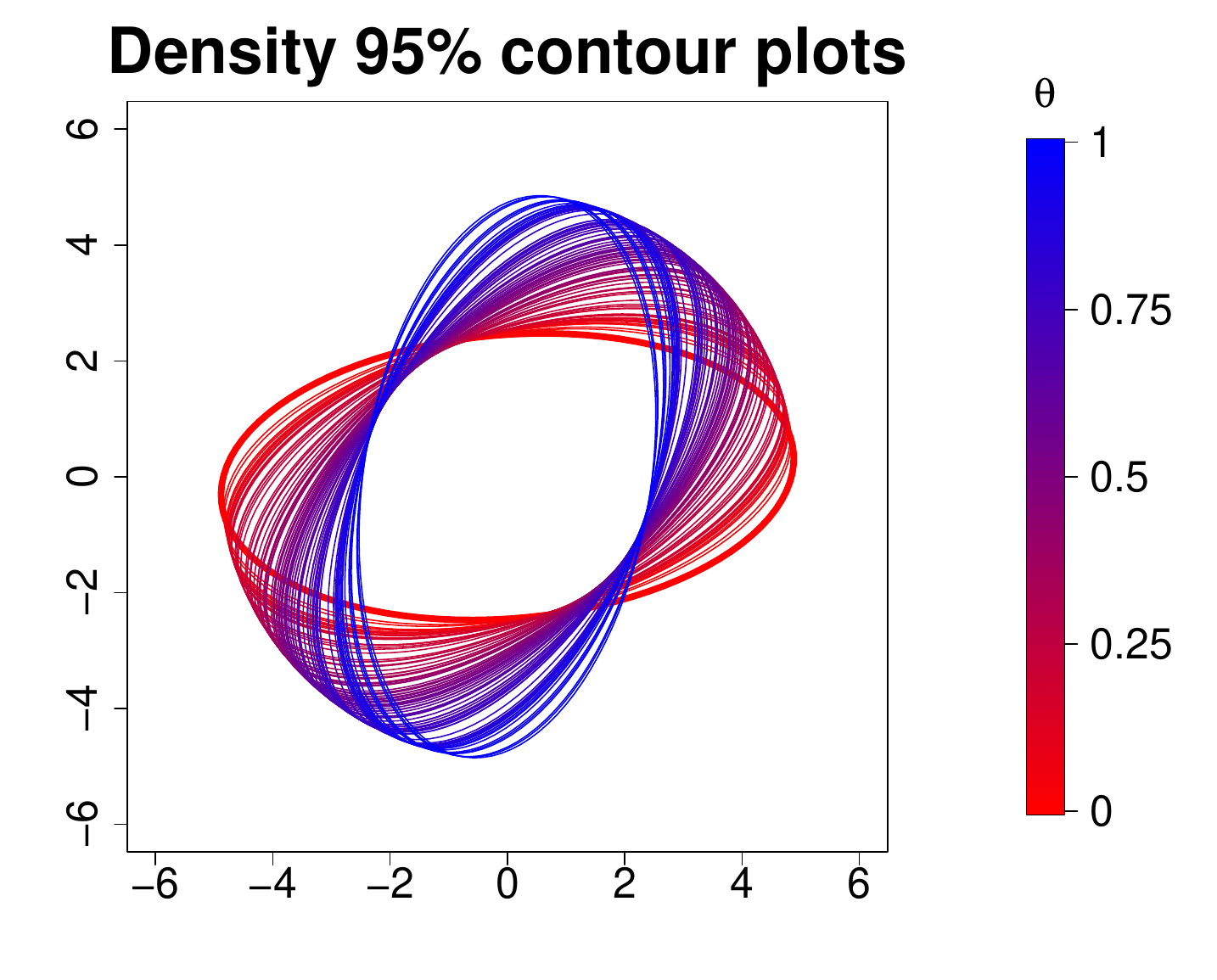}\label{fig:sub12dwassobs}}
    \subfigure[]{\includegraphics[width=0.37\textwidth]{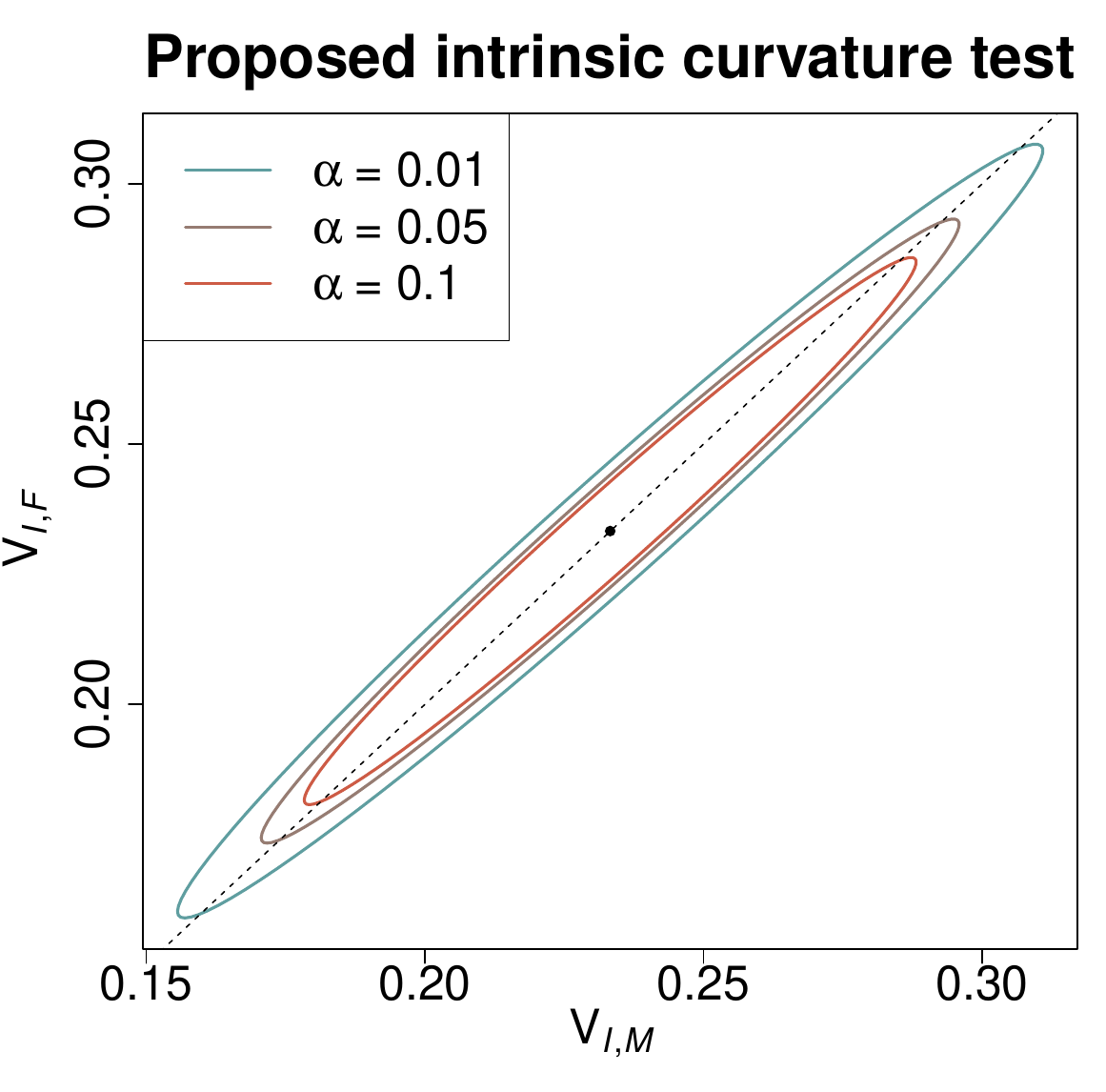}\label{fig:sub22dwassobs}}
  \caption{(a): The $95\%$ density contour plots of random bivariate normal distributions, $X_{i} = N(0, \Lambda(\theta_{i})),$ where $\theta_{i} \sim \text{Beta}(2, 2),$ $i= 1, \ldots, 100,$ generated from the intrinsic space $\mathcal{D}$ in \eqref{2dIntrinspaceD}. (b): The proposed intrinsic curvature test using the confidence regions $\mathcal{C}_{I,n}(1-\alpha)$ in \eqref{CR_int} with $\alpha = 0.01, 0.05, 0.1,$ inferring flat intrinsic curvature of $\mathcal{D}.$ }
  \label{fig:2dwassobs}
\end{figure}

Thus, even though the ambient space with Wasserstein geometry is positively curved, the flat intrinsic space $\mathcal{D}$ satisfies the assumptions of traditional manifold learning algorithms, such as ISOMAP \citep{tenenbaum2000global}, which require the space to be globally isometric to a convex subset of Euclidean space. To represent the intrinsic geodesic using ISOMAP, we follow the approach of \cite{chen2012nonlinear}.} 

Consider the intrinsic geodesic $\gamma(t) : [0,1] \rightarrow \mathcal{D}$ between two observations $x,y \in \mathcal{D}.$ 
We can then estimate the intrinsic geodesic as
\begin{equation}
    \hat{\gamma}(t) :=  \hat{\psi}^{-1}(s(t)) = \hat{\psi}^{-1}((1-t)\hat{\psi}(x) + t\hat{\psi}(y)), \quad t \in [0,1], 
    \label{intrin_isomap_geod}
\end{equation}
where $\hat{\psi} : \mathcal{D} \rightarrow \real^{q}$ is the representation map of ISOMAP, $s(t) = (1-t)\hat{\psi}(x) + t\hat{\psi}(y)$ is the ISOMAP representation interpolation, and the inverse map $\hat{\psi}^{-1} : \real^{q} \rightarrow \mathcal{D}$ is estimated by the weighted barycenter of the distributional observations $X_{1}, \ldots, X_{n}.$ For $\zeta \in \real^{q},$ we estimate $\hat{\psi}^{-1}(\zeta) = \argmin\limits_{\xi \in \mathcal{D}} \sum_{i=1}^{n} w_{i}(\zeta)d^{2}(X_{i},\xi),$
where $w_{i}(\zeta) = \frac{\kappa(H^{-1}(\hat{\psi}(X_{i}) - \zeta)) }{\sum_{i=1}^{n} \kappa(H^{-1}(\hat{\psi}(X_{i}) - \zeta))},$ and $\kappa$ is a non-negative $q$-dimensional kernel, with $H = h\textbf{I}_{q}$ for a suitably chosen bandwidth $h.$ We note that this representation map 
requires the intrinsic space to be flat and its estimation could alternatively be implemented by Fr\'echet regression \citep{pete:19}. 

\begin{figure}[ht] 
    \centering
    \subfigure[]{\includegraphics[width=0.32\linewidth]{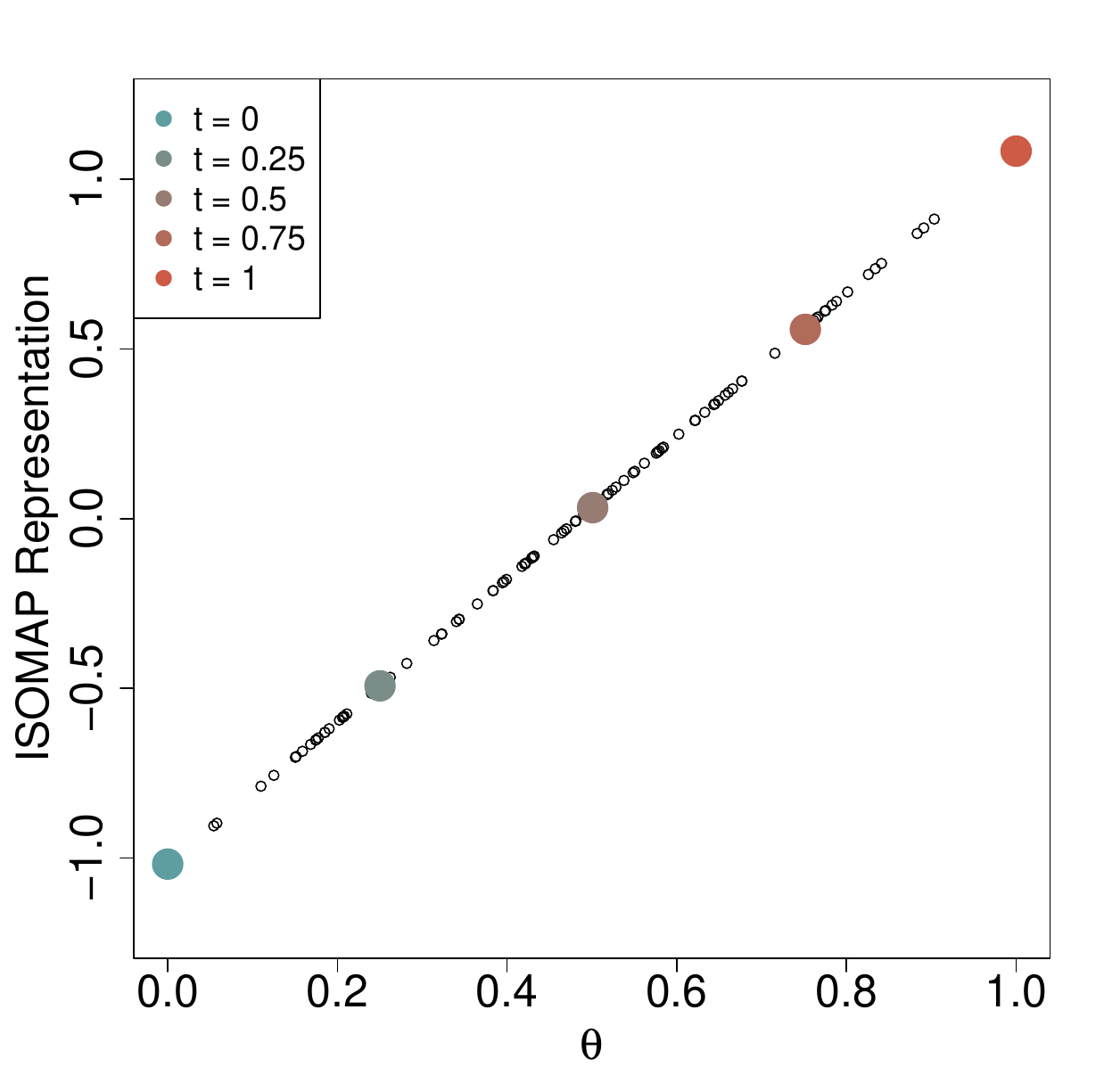}\label{fig:sub1WassSim}}
    \subfigure[]{\includegraphics[width=0.32\linewidth]{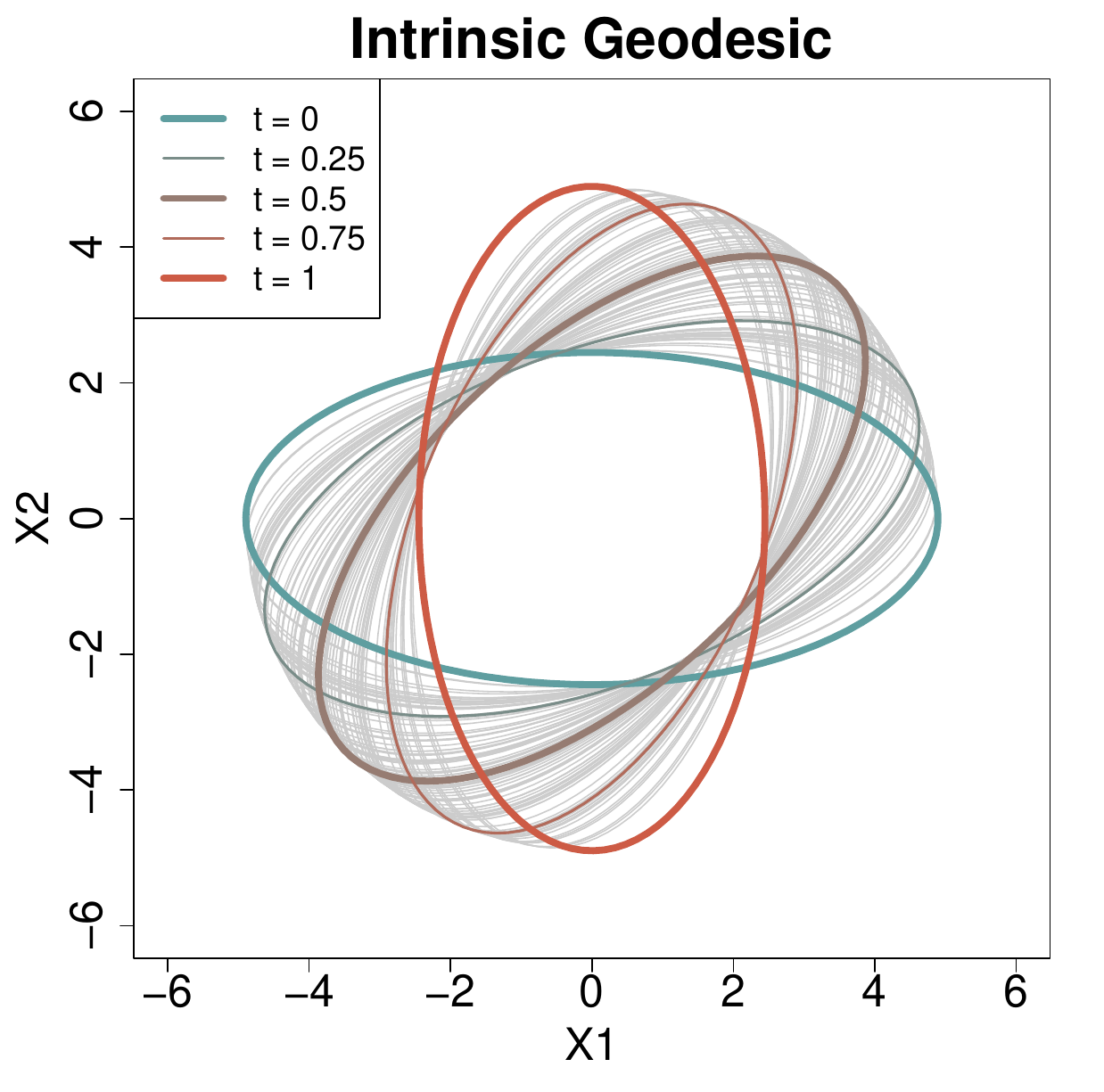}\label{fig:sub2WassSim}}
    \subfigure[]{\includegraphics[width=0.32\linewidth]{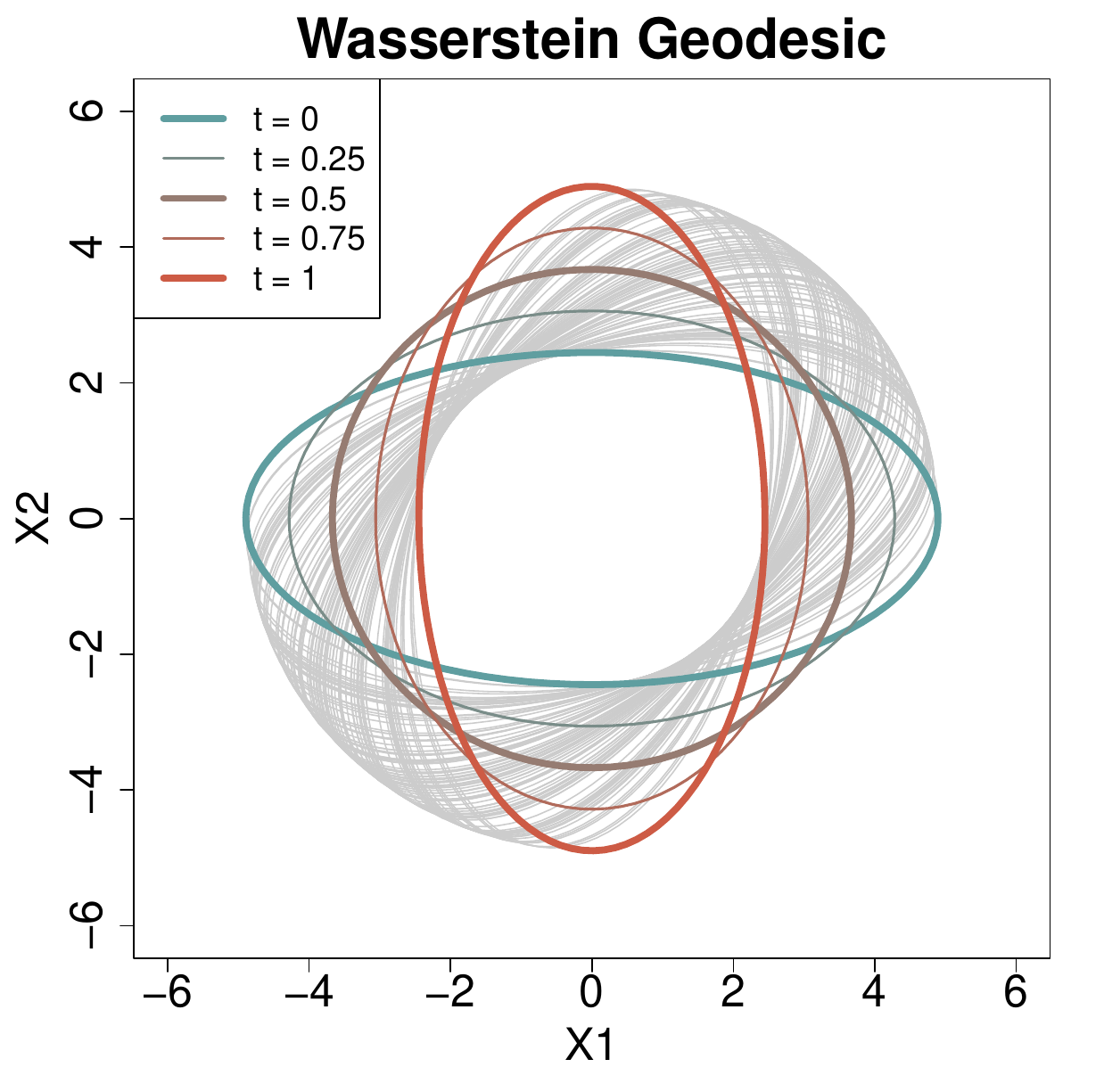}\label{fig:sub3WassSim}}
    \caption{(a) ISOMAP representation interpolation $s(t)$ for  random bivariate normal distributions, $X_{1}, \ldots, X_{100}$, generated from the intrinsic space $\mathcal{D}$ in \eqref{2dIntrinspaceD} as a function of normalized rotation angle $0 \leq \theta \leq 1.$ (b) Estimated intrinsic geodesic path $\hat{\gamma}(t)$ in \eqref{intrin_isomap_geod} and (c) Wasserstein ambient geodesic path $\gamma_{W}(t)$ in \eqref{wass_geod} between blue $(t = 0)$ and red $(t=1)$ ellipse, represented by $95\%$ density contour plots.}
    \label{fig:2dwassgeo}
\end{figure}

In our analysis, we use a Gaussian kernel with an intrinsic dimension of $k = 1.$ We demonstrate the resulting geodesic to connect the distributions  at $x = N\left(0, \Sigma(0)\right)$ and $y = N(0, \Sigma(1)) \in  \mathcal{D},$ illustrating the ISOMAP representation interpolation at $s(t),$ $t = 0, 0.25, 0.5, 0.75,$ and $1$ in Figure \ref{fig:sub1WassSim} and the intrinsic geodesic $\hat{\gamma}(t) = \hat{\psi}^{-1}\left(s(t)\right)$ in Figure \ref{fig:sub2WassSim}.  Due to the flat intrinsic curvature $\rho_{I} = 0$ of $\mathcal{D}$, as ascertained with the proposed method, the ISOMAP-based intrinsic geodesics $\hat{\gamma}(t)$ in \eqref{intrin_isomap_geod} successfully maintain the same ratio between the two eigenvalues of the covariance matrix  along the entire geodesic from $t = 0$ to $t =1,$ corresponding to an interpretable and insightful rotation of the visualized contour ellipse.

As a comparison method, we consider the Wasserstein geodesic that is defined in the ambient (rather than intrinsic) geometry. For two bivariate normal distributions $N(0, U),$ and $N(0,V),$ the Wasserstein geodesic $\gamma_{W}(t)$ is 
\begin{equation}
    \gamma_{W}(t) = N(0, W(t)), \quad W(t) = \left\{(1-t)\textbf{I}_{2} + tT  \right\}U \left\{(1-t)\textbf{I}_{2} + tT  \right\},
    \label{wass_geod}
\end{equation}
where $T = V^{\frac{1}{2}}( V^{\frac{1}{2}}U V^{\frac{1}{2}})^{-\frac{1}{2}} V^{\frac{1}{2}},$ for $t \in [0,1]$ \citep{takatsu2011wasserstein}. Figure \ref{fig:sub3WassSim} shows that the Wasserstein geodesic paths between $x = N\left(0, \Sigma(0)\right),$ and $y = N(0, \Sigma(1))$ go through a spherical shape at the halfway point and thus leave the intrinsic space. Thus they are far less intuitive and lack interpretability  (see Section \ref{rda_gait} for a data example).    

    \subsection{Example 2: Point cloud data with intrinsic distance}\label{chap63}
    Here we apply the proposed  test to ascertain  intrinsic curvature for point cloud data. Many problems in statistical machine learning involve data where each data point is drawn from an underlying probability density situated in an unknown intrinsic space \citep{lee2022statistical,chew2022manifold}. The data clouds  delineate the spatial locations of objects within an ambient space, while the curvature of the intrinsic space is of interest but remains elusive.  
    We consider a Euclidean ambient space $\mathbb{R}^3$ with strict subspaces $\mathcal{B}_{1}, \mathcal{B}_{2},\mathcal{B}_{3} \subset \mathbb{R}^3$ given by 
    \begin{align*}
        \begin{split}
            \mathcal{B}_{1} &= \left\{(x,y,z) \in \mathbb{R}^{3} ~ | ~ x^{2} + y^{2} + z^{2} = 1, ~ z\geq0 \right\},\\
            \mathcal{B}_{2} &= \left\{(x,y,z) \in \mathbb{R}^{3} ~ | ~ x^{2} - y^{2} - z^{2} = -1, ~ {0 \leq z \leq 4} \right\},\\
            \mathcal{B}_{3} &= \left\{(x,y,0) \in \mathbb{R}^{3} ~ | ~ 0 \leq x,y \leq 1 \right\},
        \end{split}
    \end{align*}
    where $\mathcal{B}_{1}$ represents the open upper hemisphere,  $\mathcal{B}_{2}$ is a hyperboloid obtained by rotating a hyperbolic graph around the $z$-axis  and $\mathcal{B}_{3}$ is a plane. These spaces are positively curved, negatively curved and flat, respectively. We generate i.i.d. point cloud data for each of these spaces  as follows: For $i = 1, \ldots, n,$
    \begin{align*}
        \begin{split}
            (x_{1i}, y_{1i}, z_{1i}) &= (\cos\theta_{i}\sin\psi_{i}, \sin\theta_{i}\sin\psi_{i}, \cos\psi_{i} ) \in \mathcal{B}_{1}, ~\psi_{i} \sim {\text{U}[0, \pi/2]},\, \theta_{i} \sim {\text{U}[0, 2\pi ]};\\
            (x_{2i}, y_{2i}, z_{2i}) &= (\upsilon_{i}, \sqrt{1+\upsilon_{i}^{2}}\cos\theta_{i}, \sqrt{1+\upsilon_{i}^{2}}\sin\theta_{i})\in \mathcal{B}_{2}, ~\upsilon_{i} \sim {TN_{[-r,r]}(0, 1)}, \, \theta_{i} \sim \text{U}[0, \pi]; \\
            (x_{3i}, y_{3i}, z_{3i}) &= \left(\mu_{i}, \nu_{i}, 0 \right) \in \mathcal{B}_{3}, ~ \mu_{i}, \nu_{i} \sim {\text{U}[0,1]}.
        \end{split}
    \end{align*}
    Here $\text{U}[a, b]$ denotes a uniform distribution over the interval $[a, b],$ and $TN_{[-r, r]}(0, 1)$ is a one-dimensional truncated normal distribution with mean 0 and variance 1, restricted to the interval $[-r, r].$  where  $r = \sqrt{15}.$

\begin{figure}[ht]
      \centering
        \includegraphics[width=0.32\textwidth]{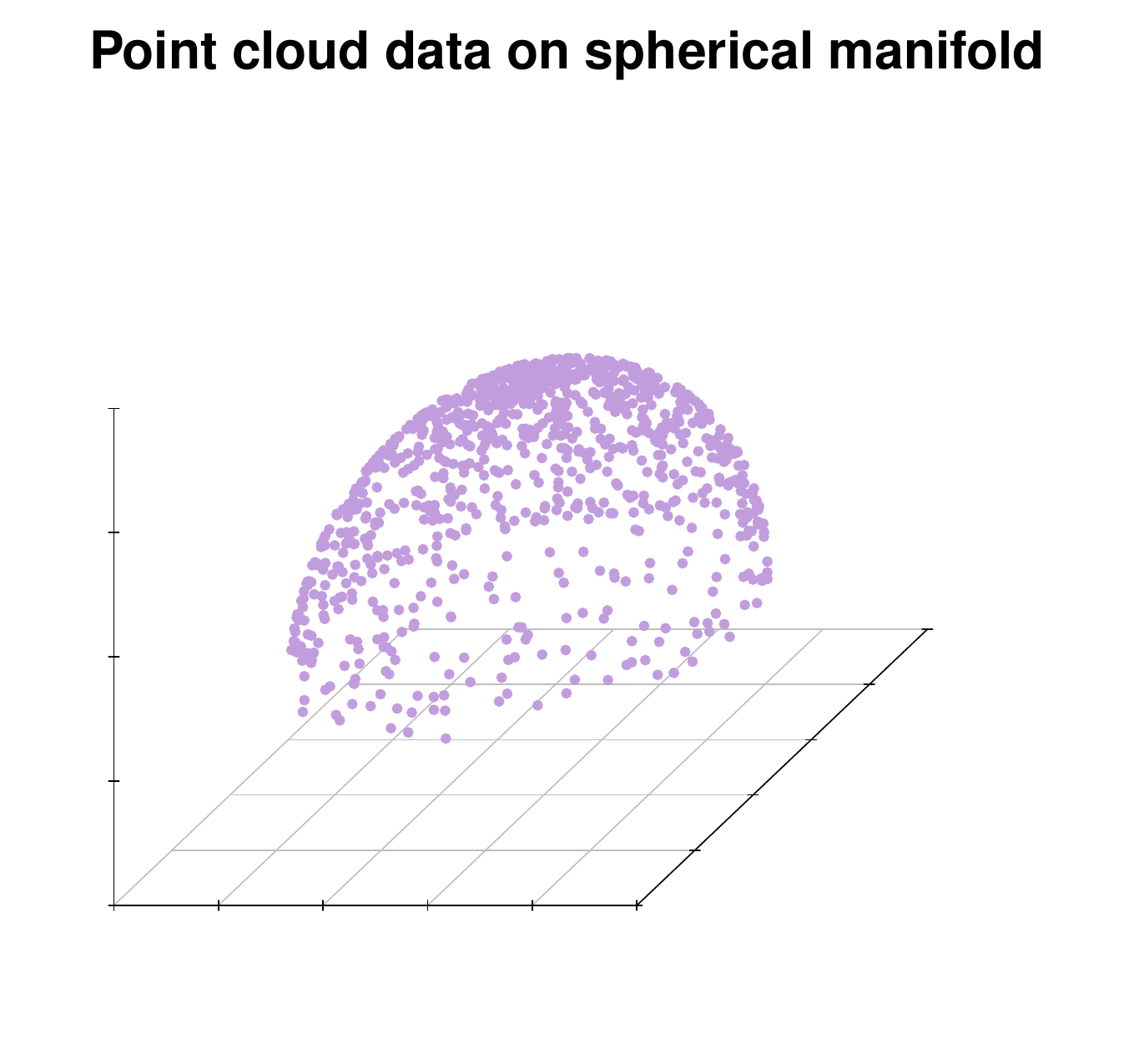}
        \includegraphics[width=0.32\textwidth]{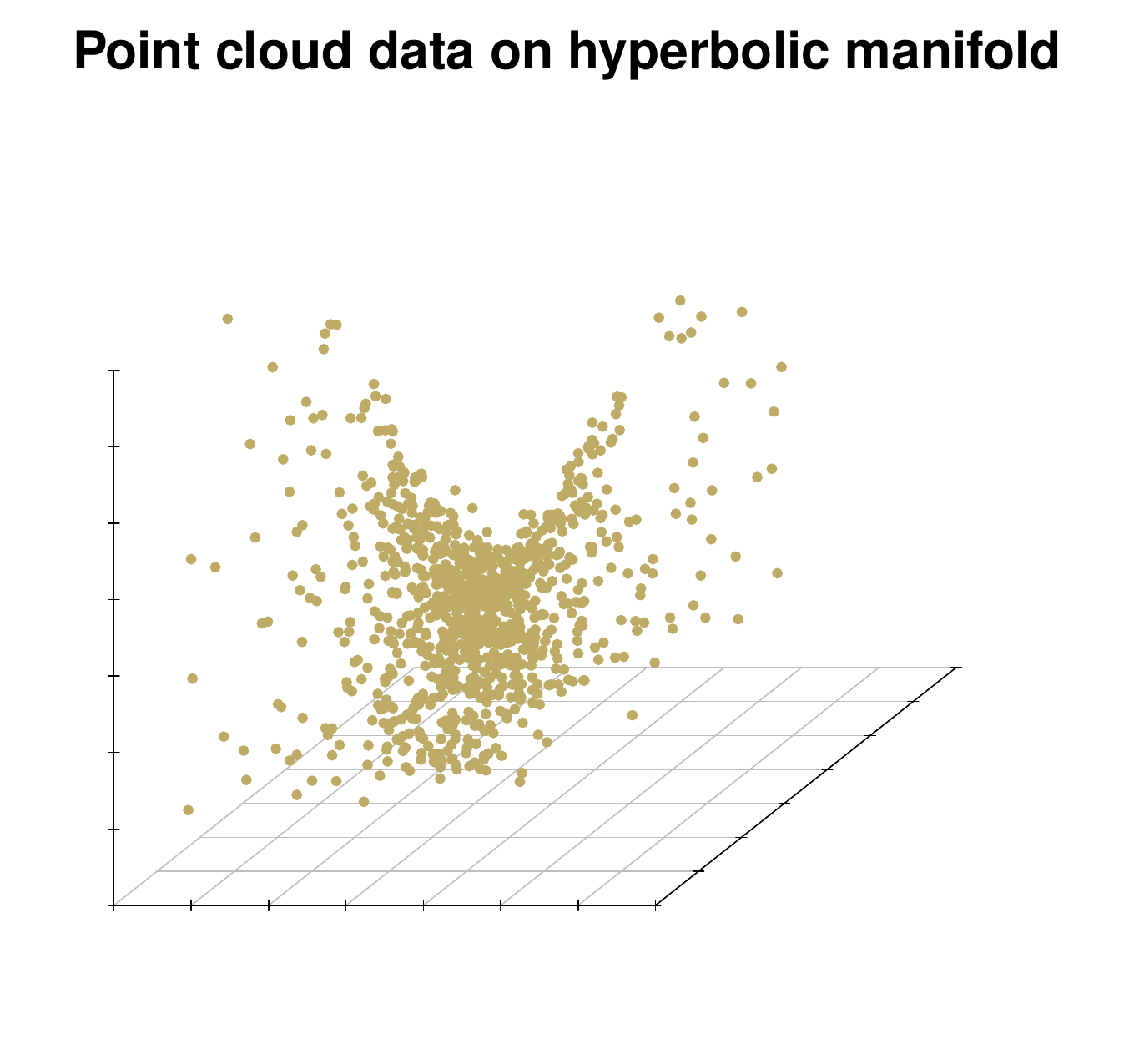}
        \includegraphics[width=0.32\textwidth]{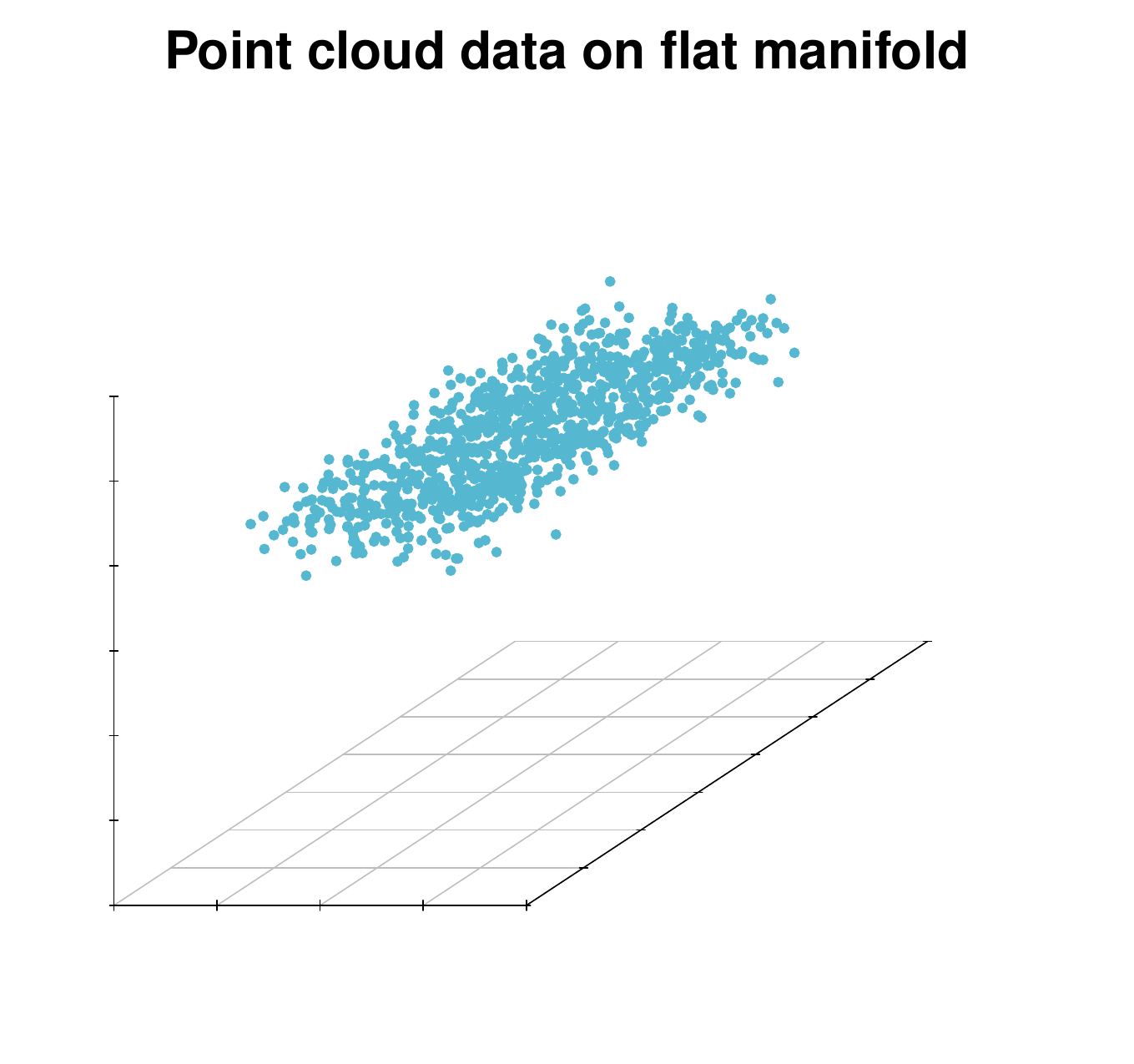}
        \includegraphics[width=0.32\textwidth]{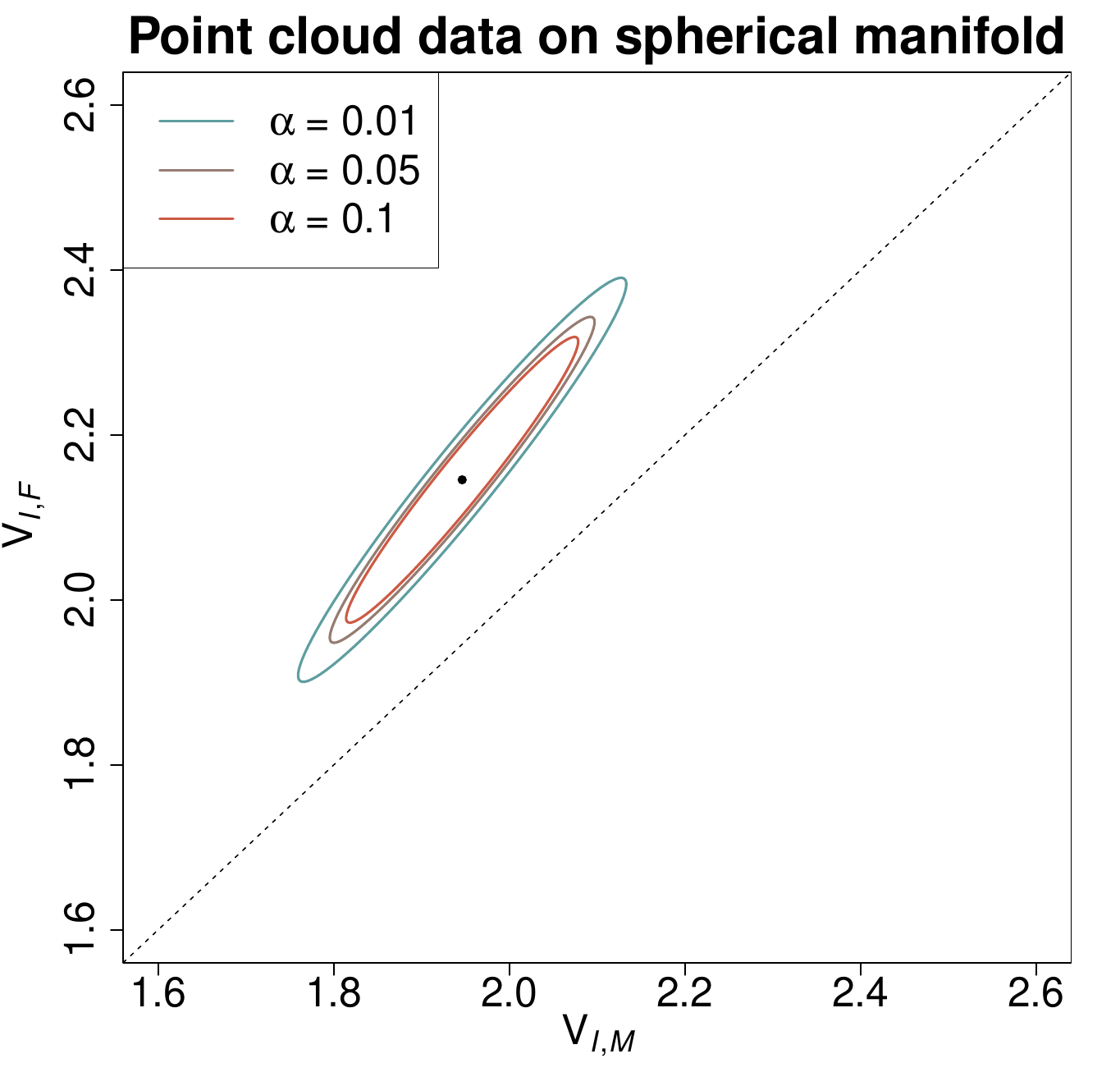}
        \includegraphics[width=0.32\textwidth]{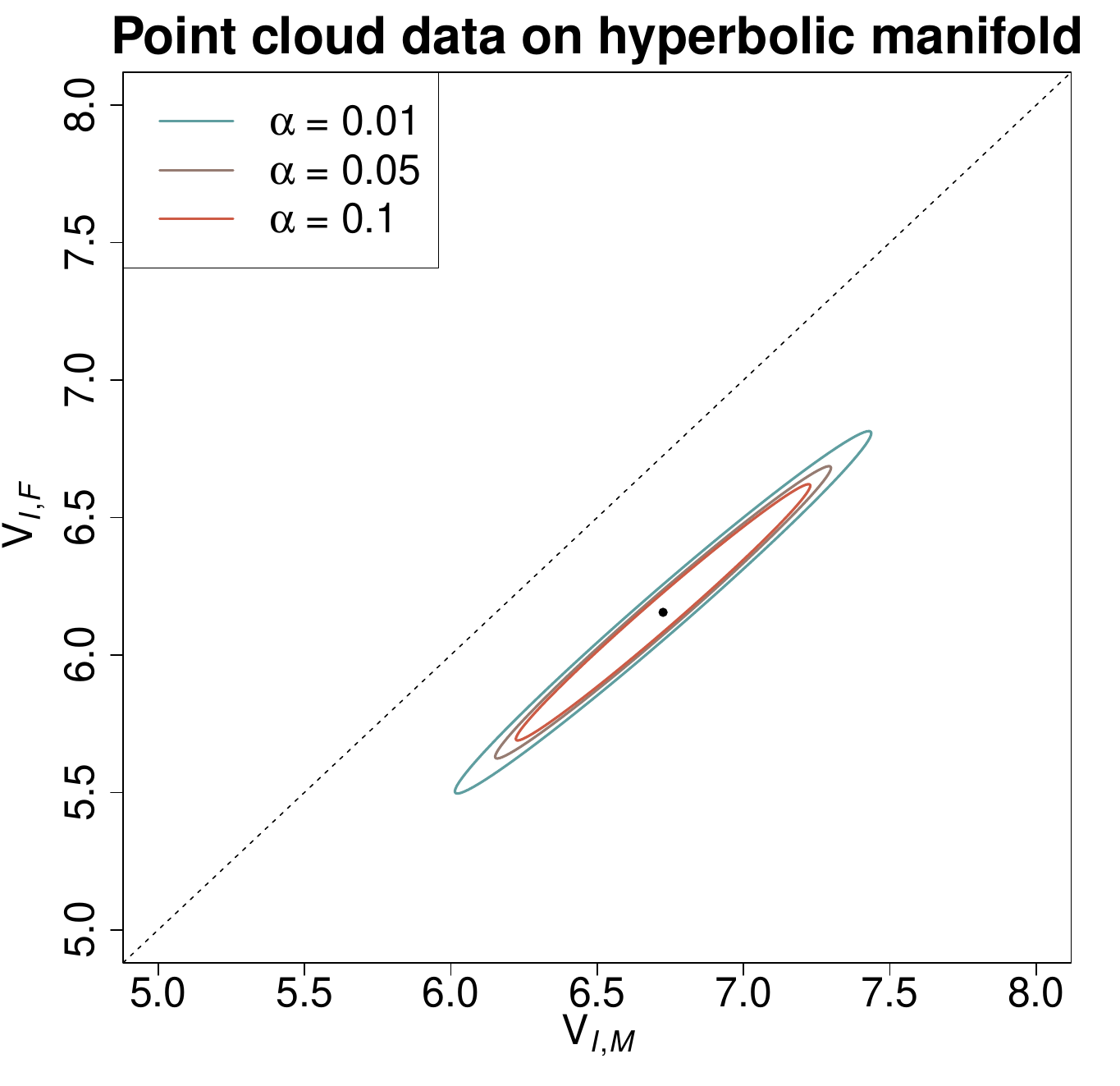}
        \includegraphics[width=0.32\textwidth]{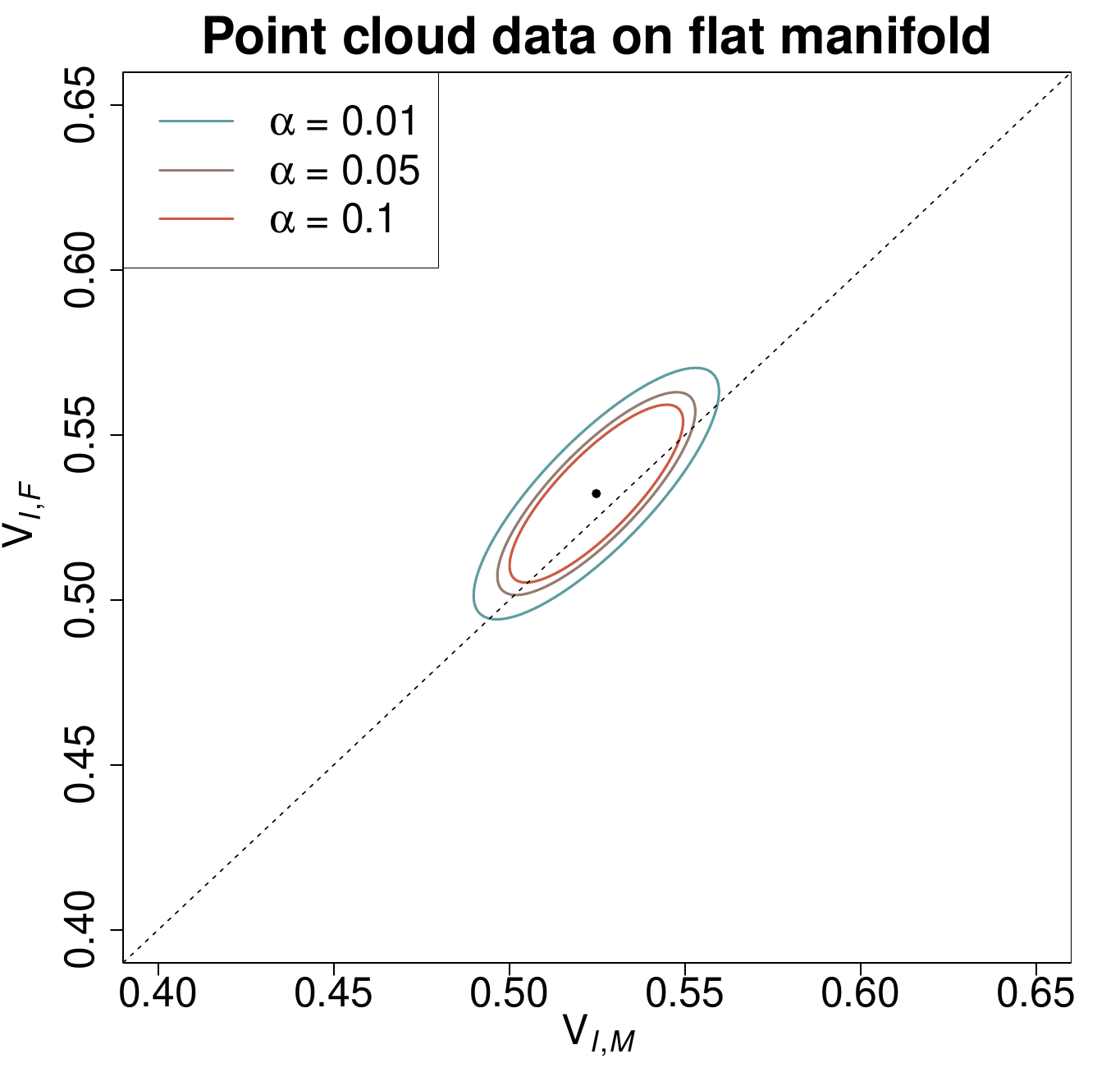}    
      \caption{(First row) Simulated data clouds $\left\{(\Tilde{x}_{li}, \Tilde{y}_{li}, \Tilde{z}_{li})\right\}_{i=1}^{n}$ located in the subspaces $\mathcal{B}_{l},$  $l = 1,2,3$ of $\mathbb{R}^3,$ with 
      some noise added. (Second row)  The proposed intrinsic curvature test utilizing confidence regions $\mathcal{C}_{I,n}(1-\alpha)$ for $\eta_{I}$ in \eqref{CR_int} that are obtained with the respective data cloud in the same column as input, for levels  $\alpha = 0.01, 0.05, 0.1.$  This provides evidence for positive, negative and flat intrinsic curvature, respectively.}
      \label{fig:cloud}
    \end{figure}

    In real world  settings, observations usually are not exactly on target,  but are corrupted by errors; we reflect this by  incorporating small perturbations $(\Tilde{x}_{li}, \Tilde{y}_{li}, \Tilde{z}_{li}) = (x_{li}, y_{li}, z_{li}) + \mathbf{\epsilon}_{li},$ $\mathbf{\epsilon}_{li} \sim N(\mathbf{0}, \sigma^{2} \mathbf{I}_{3} ),$  where $\mathbf{0} = (0,0,0)^{T}$ and $\mathbf{I}_{3}$ is a $3 \times 3$ identity matrix, $l=1,2,3,\quad i=1,\ldots, n,$ where the sample size was $n= 1000$ and  the noise level $\sigma$ was set to $\frac{1}{10}.$ The first row of Figure \ref{fig:cloud} displays the data clouds $\left\{(\Tilde{x}_{li}, \Tilde{y}_{li}, \Tilde{z}_{li})\right\}_{i=1}^{n}$ for each space $\mathcal{B}_{l},$   $l = 1,2,3.$
    
    The intrinsic geodesic distance $d_{\mathcal{B}_{l}},$ $l=1,2,3,$ can be obtained from the given Euclidean input distance using Algorithm \ref{table:algo}. We emphasize that this approach relies solely on Euclidean distance to estimate intrinsic geodesic distances and does not utilize any information about the curvature of the latent spaces. For each space $\mathcal{B}_{l},$ we estimate the sample intrinsic \F mean $\hat{\mu}_{I, \oplus}$, \F variance $\hat{V}_{I, F}$ and metric variance $\hat{V}_{I, M}$,  using error-perturbed samples $\left\{(\Tilde{x}_{li}, \Tilde{y}_{li}, \Tilde{z}_{li})\right\}_{i=1}^{n}$  as input.
    
    The second row of Figure \ref{fig:cloud} illustrates $100(1-\alpha) \%$ confidence regions $\mathcal{C}_{I, n}(1-\alpha)$ in \eqref{CR_int} for $\eta_{I} = (\eta_{I,1}, \eta_{I,2}) := (V_{I, M},V_{I,F})^{T}$ and  $\alpha = 0.01, 0.05, 0.1.$ The intrinsic space corresponding to  data cloud $\left\{ (\Tilde{x}_{1i}, \Tilde{y}_{1i}, \Tilde{z}_{1i}) \right\}_{i=1}^{n}$ in the first column is found to exhibit significant positive curvature, as the confidence regions $\mathcal{C}_{I, n}(1-\alpha)$ 
    are contained in the set $\left\{\eta_{I}~|~ \eta_{I,1} < \eta_{I, 2}\right\}$ for all three values of $\alpha.$  Analogously,  the intrinsic space corresponding to the point cloud  $\left\{ (\Tilde{x}_{2i}, \Tilde{y}_{2i}, \Tilde{z}_{2i}) \right\}_{i=1}^{n}$ in the second column is found to have  significant negative curvature, while for the point cloud $\left\{ (\Tilde{x}_{3i}, \Tilde{y}_{3i}, \Tilde{z}_{3i}) \right\}_{i=1}^{n}$ in the third column one cannot reject the null hypothesis that $H_{0}: \mP \in \Theta_{0},$ where $\Theta_{0}: \left\{\eta_{I} ~ | ~ \eta_{I, 1} = \eta_{I, 2}  \right\},$ which is compatible (but does not establish) no significant curvature. Thus the actual curvature is correctly assessed. 
    
    Table \ref{tab:cloud} shows confidence intervals $\mathcal{I}_{I,n}(1-\alpha)$ in \eqref{CR_curv_int} for the intrinsic curvature $\rho_{I}$, corresponding to each data cloud $\left\{(\Tilde{x}_{li}, \Tilde{y}_{li}, \Tilde{z}_{li})\right\}_{i=1}^{n}$ sampled from $\mathcal{B}_{l},$ $l = 1, 2, 3$ at confidence levels $\alpha = 0.01, 0.05,$ and $0.1,$ with intervals widening for smaller $\alpha$ and confirming that the proposed curvature measure  correctly detects the intrinsic curvature of these spaces.
    

     \begin{table}     
    \caption{\label{tab:cloud}Confidence intervals $\mathcal{I}_{I,n}(1-\alpha)$ in \eqref{CR_curv_int} for the intrinsic curvatures $\rho_{I}$ of point cloud data generated from the underlying manifolds $\mathcal{B}_{1}, \mathcal{B}_{2},$ and $\mathcal{B}_{3}$.}
        \centering
        \begin{tabular}{lccc}\toprule
             & $\mathcal{B}_{1}$ & $\mathcal{B}_{2}$ & $\mathcal{B}_{3}$\\
             \midrule\midrule
        $\alpha = 0.1$ & $\left(0.083,0.122\right)$ & $\left(-0.093, -0.077\right)$ & $\left(-0.009, 0.038\right)$\\
        \midrule
        $\alpha = 0.05$ & $\left(0.080,0.125\right)$ & $\left(-0.095, -0.075\right)$ & $\left(-0.014, 0.042\right)$ \\
        \midrule
        $\alpha = 0.01$ & $\left(0.072,0.132\right)$ & $\left(-0.098, -0.072\right)$ & $\left(-0.023, 0.051\right)$\\
        \bottomrule
        \end{tabular}
    \end{table} 

       \section{Data illustrations}\label{chap7}

    \subsection{Gait synchronization analysis}\label{rda_gait}

    Human gait analysis is critical in medical research due to its potential to reveal the impact of various pathologies, such as Parkinson's disease and arthritis, which can lead to reduced mobility and increased fall risk \citep{bachlin2009wearable, barrois2015quantify}. For this analysis, we use gait time series data collected between April 2014 and October 2015. For detailed information about  the collection process and a comprehensive description of the data we refer to  \cite{truo:19}. The data are available at \url{https://github.com/deepcharles/gait-data}. 

    We used measurements from Inertial Measurement Units (IMUs) attached on the foot of $52$ participants in the healthy group with no known medical issues, and another $52$ participants in the orthopedic disorders group, who had lower limb osteoarthritis or cruciate ligament injuries. We considered the $\real^{2}$-valued time signal corresponding to the angular velocity (in $deg/s$) along the vertical axis perpendicular to the ground (referred to as the V-axis) and along the axis perpendicular to the sensor placed on a tilted dorsum of the foot for each participant (referred to as the Z-axis). Each recorded signal corresponds to a gait cycle in which participants stand for 6 seconds, walk 10 meters at their preferred walking speed on a level surface to a designated turn point, turn around, walk back to the starting point, and finally stand for 2 seconds.
    
    For each participant, we computed a $2 \times 2$ SPD sample covariance matrix to measure the degree of synchronization between the  V- and Z- axes from the bivariate signal data vectors.  Consider $X_{1}, \ldots, X_{n_{1}},$ as the SPD matrices for $n_{1} = 52$ participants in the healthy group and $Y_{1}, \ldots, Y_{n_{2}},$ as the SPD matrices for $n_{2} = 52$ participants in the orthopedic disorders group. We consider the SPD matrices of size $2 \times 2$, denoted by $\mathcal{S}^{2}_{++} = \left\{X \in \mathbb{R}^{2 \times 2}: X^{T} = X, X \succ 0 \right\}$ with the Bures-Wasserstein metric in the  ambient space of SPD matrices, $d_{BW}(U,V) = \left[ \trace(U) + \trace(V) - 2 \trace(U^{\frac{1}{2}}VU^{\frac{1}{2}})^{\frac{1}{2}} \right]^{\frac{1}{2}}.$ This metric is also known as the Bures distance in quantum information and is closely related to the  Wasserstein metric for Gaussian distributions \citep{bhatia2019bures}. We provide additional sensitivity analysis results on the choice of ambient metrics $d$ for the space of SPD matrices in Section S.6 of the Supplement.

\begin{figure}[ht]
      \centering
        \subfigure[]{\includegraphics[width=0.24\linewidth]{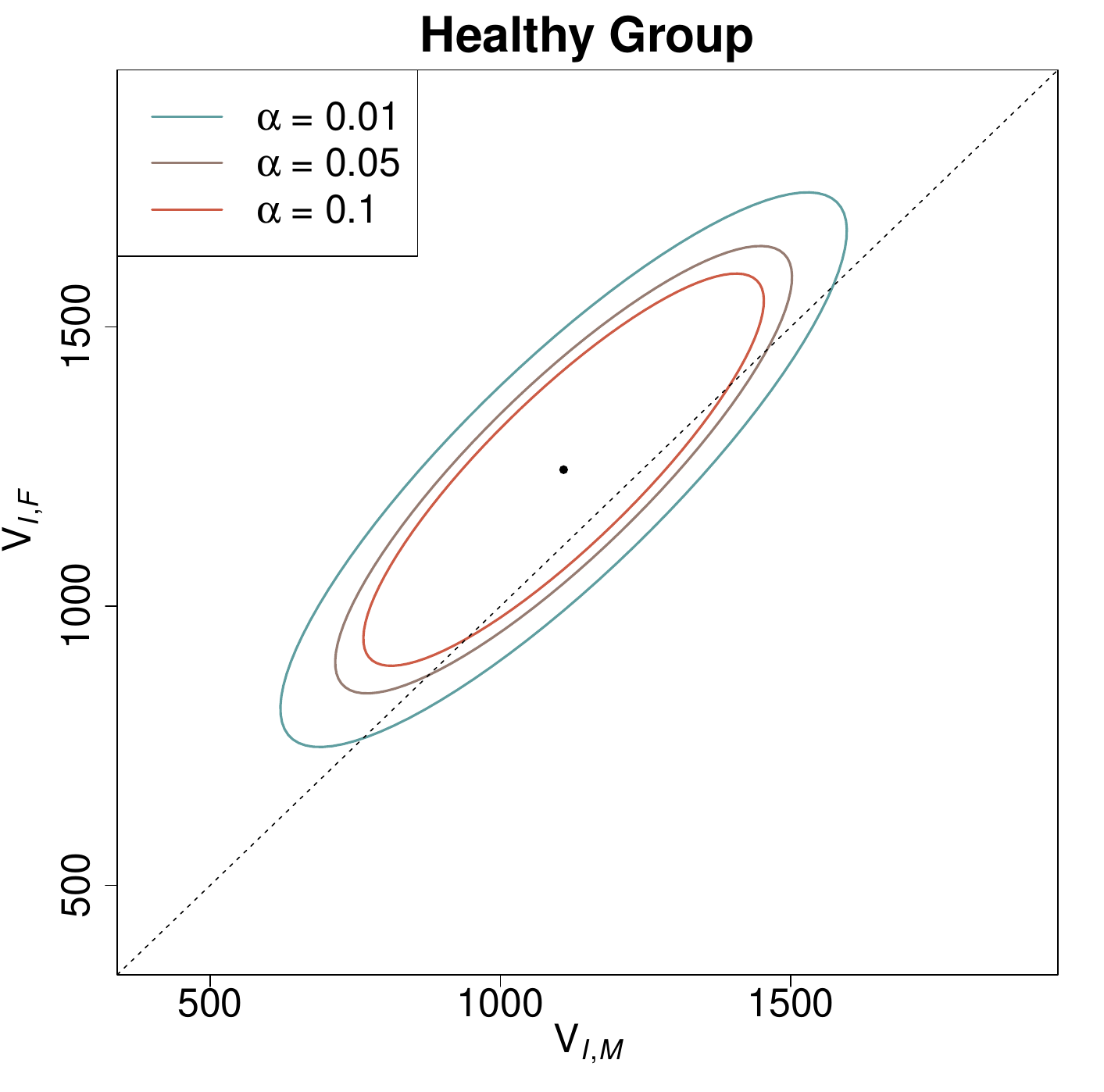}\label{fig6:sub1}}
        \subfigure[]{\includegraphics[width=0.24\linewidth]{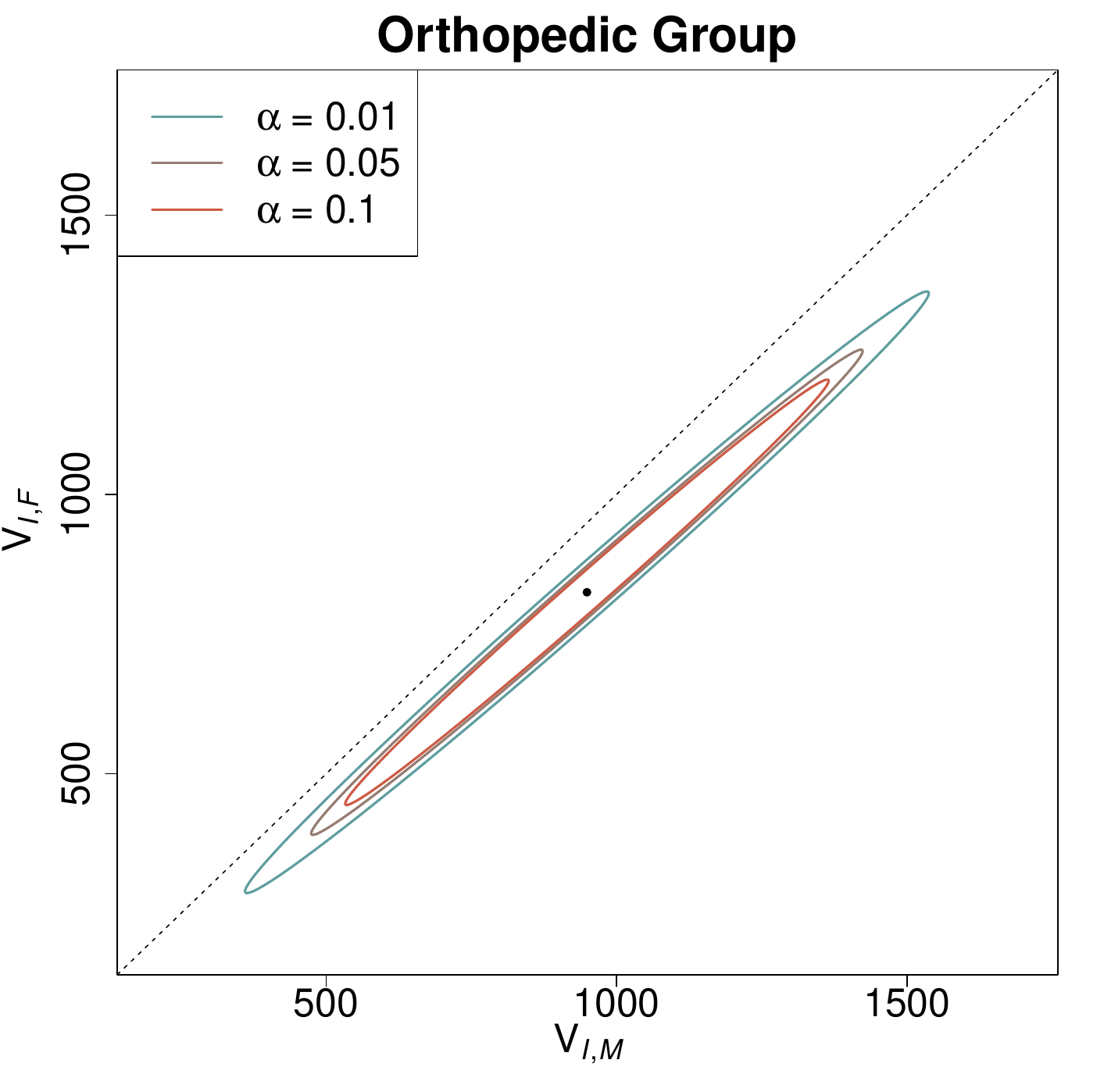}\label{fig6:sub2}}
        \subfigure[]{\includegraphics[width=0.24\linewidth]{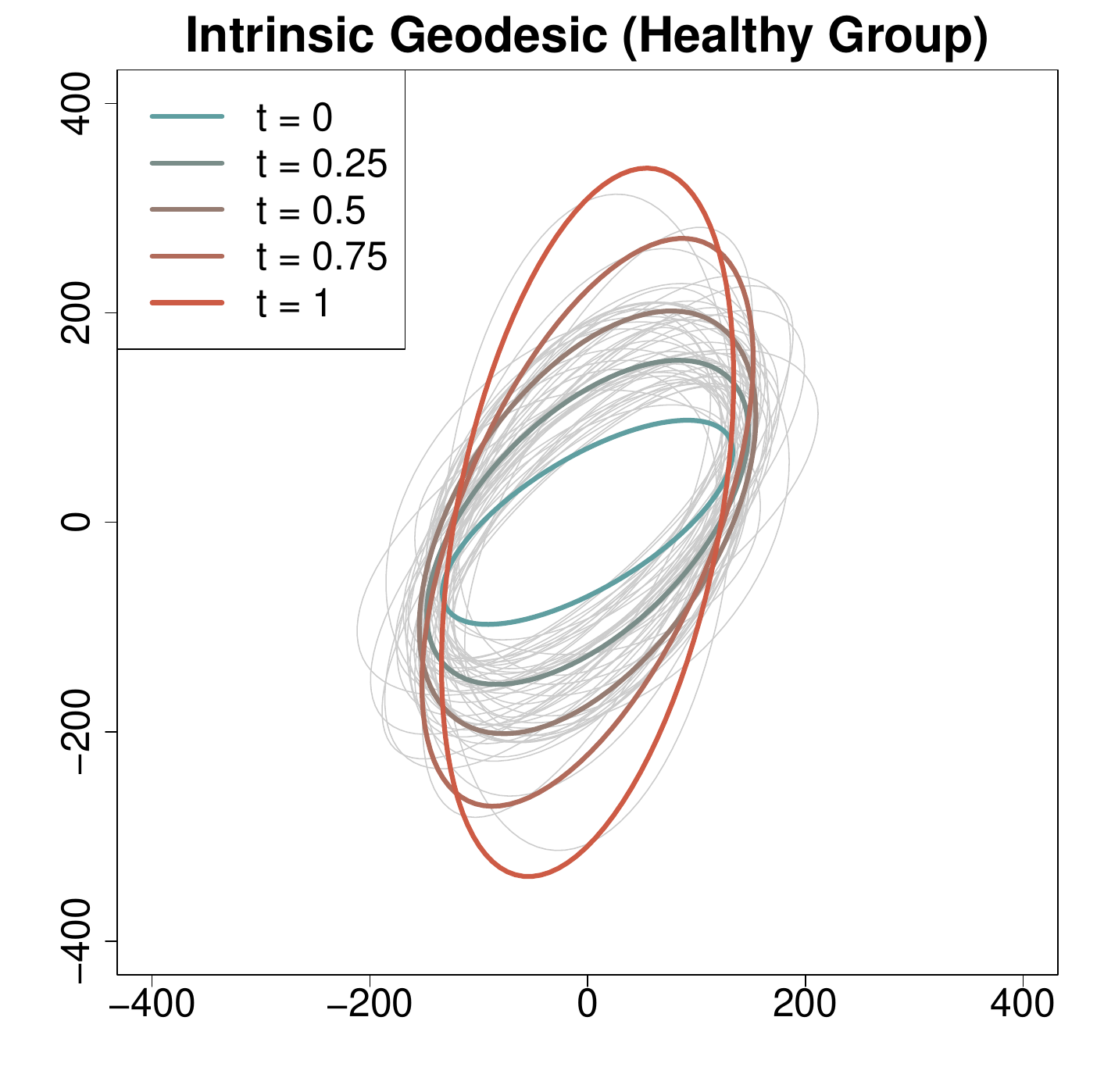}\label{fig6:sub3}}
        \subfigure[]{\includegraphics[width=0.24\linewidth]{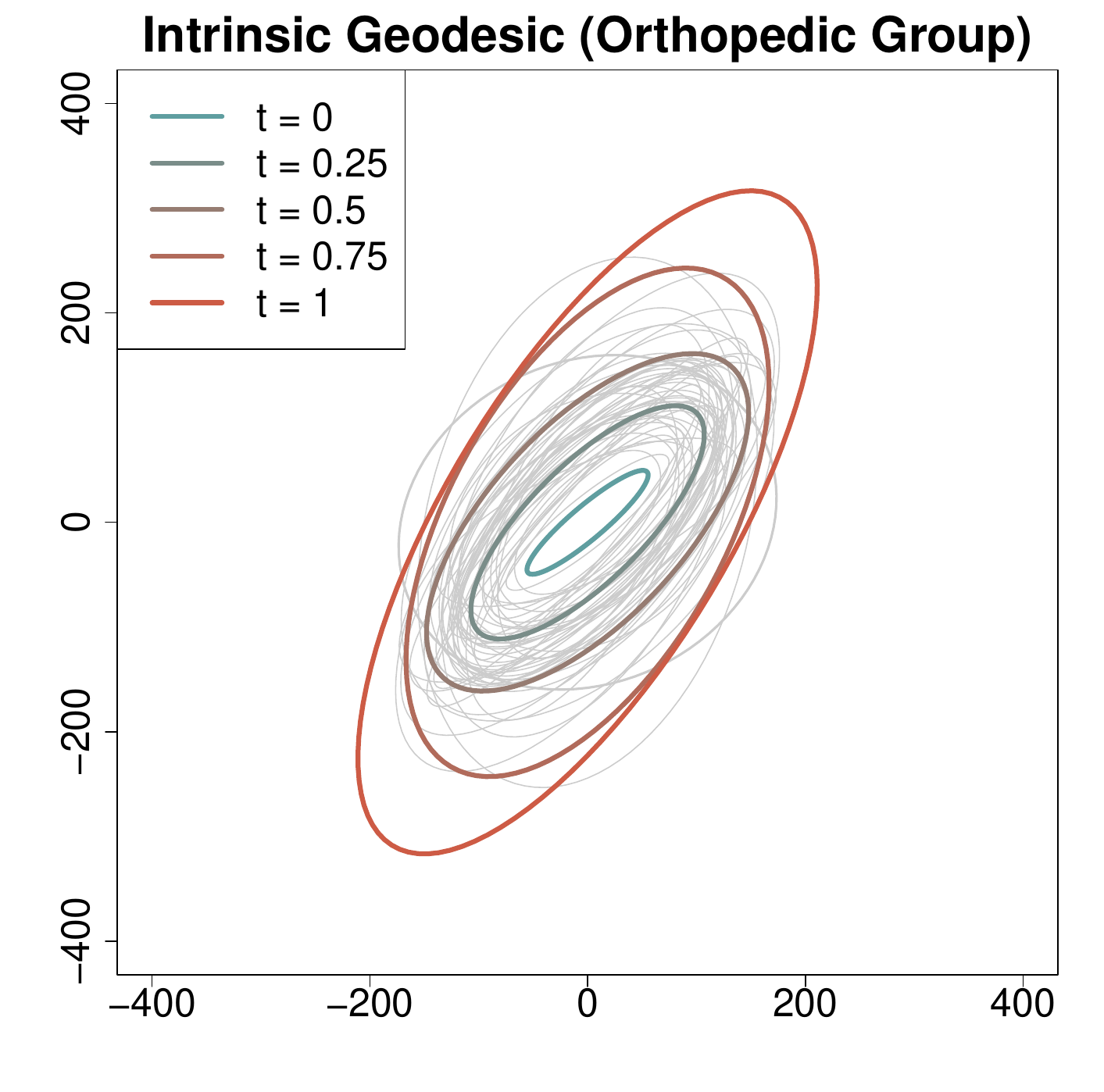}\label{fig6:sub4}}
      \caption{(a), (b): Confidence regions $\mathcal{C}_{I,n}(1-\alpha)$ in \eqref{CR_int} for $\eta_{I}$ representing gait synchronization SPD matrices for healthy and orthopedic disorder groups, with $\alpha = 0.01, 0.05, 0.1$. (c): Intrinsic geodesic $\hat{\gamma}(t)$ in \eqref{intrin_isomap_geod} from the blue observation ($t = 0$) to red ($t = 1$), for the healthy group. (d): Intrinsic geodesic $\hat{\gamma}(t)$ in \eqref{intrin_isomap_geod} from the blue observation ($t = 0$) to red ($t = 1$), for the orthopedic disorder group.}
      \label{fig:gait}
    \end{figure}

   To infer the curvature of the underlying subset of $\mathcal{S}^{2}_{++}$ in which the observed gait synchronizations are located, we compare the intrinsic \F variance $\hat{V}_{I,F}$ and metric variance $\hat{V}_{I,M}$. The confidence regions $\mathcal{C}_{I, n}(1-\alpha)$ in \eqref{CR_int} for $\eta_{I} = (\eta_{I,1}, \eta_{I,2}) := (V_{I, M},V_{I,F})^{T}$ and $\alpha=0.1,  0.05, 0.01$ for healthy and orthopedic disorders group are illustrated in Figure \ref{fig6:sub1} and \ref{fig6:sub2}. Specifically, for the healthy group ($X_{1}, \ldots, X_{n_{1}}$), the confidence regions all intersect with  $\left\{\eta_{I}~|~\eta_{I,1} = \eta_{I,2} \right\}$ so that there is no evidence for the presence of curvature. For the participants of orthopedic group ($Y_{1}, \ldots, Y_{n_{2}}$), the confidence region is contained in the set $\left\{\eta_{I}~|~\eta_{I,1} \neq \eta_{I,2} \right\}$ at level $\alpha=0.01$, indicating the presence of negative curvature. 
As  the observed matrices for  the healthy group are compatible with an intrinsic flat curvature, we estimate the intrinsic geodesic $\hat{\gamma}(t)$ defined in \eqref{intrin_isomap_geod} with  ISOMAP. 

In Figure \ref{fig6:sub3}, the geodesic from $t = 0$ to $t = 1$ moves smoothly in a counter-clockwise direction at a constant speed for the healthy group. However, as shown in Figure \ref{fig6:sub4}, for the group with orthopedic disorders, the negative curvature implies that the ISOMAP representation of the intrinsic geodesic works less well. 
For instance, from $t = 0$ to $t = 0.5$, the path expands at a constant speed while maintaining the same eigenvectors. However, at $t = 0.75$, the eigenvector deviates significantly from this trajectory,  but subsequently returns to the original direction at $t = 1$. We also provide the confidence intervals $\mathcal{I}_{I, n}(1-\alpha)$ in \eqref{CR_curv_int} for $\rho_{I}$ in Table \ref{tab:gait}.


 \subsection{Energy source data}\label{rda_energy}

Data on energy sources for electricity generation across the United States, expressed as fractions or percentages, are available at \url{https://www.eia.gov/electricity/data/state/} and represent compositional data \citep{zhu2024spherical}. In our analysis, we consider three categories of energy sources: 1) $U_{1}$: Proportion of Coal or Petroleum for electricity generation, 2) $U_{2}$: Proportion of Natural Gas, and 3) $U_{3}$: Proportion of other sources, including Nuclear, Geothermal, Hydroelectric and Solar Thermal.

\begin{table}[h!]
    \caption{Confidence intervals $\mathcal{I}_{I,n}(1-\alpha)$ in \eqref{CR_curv_int} for the intrinsic curvature $\rho_{I}$ for the gait synchronization data.}
        \centering
        \begin{tabular}{lcc}\toprule
             & Healthy Group & Orthopedic Disorders Group \\
             \midrule\midrule
        $\alpha = 0.1$ & $\left(-0.014, 0.259\right)$ & $\left(-0.167, -0.095\right)$ \\
        \midrule
        $\alpha = 0.05$ & $\left(-0.040, 0.285\right)$ & $\left(-0.174, -0.088\right)$ \\
        \midrule
        $\alpha = 0.01$ & $\left(-0.092, 0.336\right)$ & $\left(-0.187, -0.074\right)$ \\
        \bottomrule
        \end{tabular}
        \label{tab:gait}
    \end{table}

Given the compositional nature of the data, we represent each data point as a point on the unit sphere $S^2$ by applying a square root transformation \citep{dai2022statistical,scea:14}. Specifically, denoting the monthly energy composition data from January 2012 to December 2021 by  $X_{i} =(\sqrt{U_{i1}}, \sqrt{U_{i2}}, \sqrt{U_{i3}}),$ where $U_{i1} , U_{i2} , U_{i3} \geq 0$ and $U_{i1} + U_{i2} + U_{i3} = 1$ for all $i = 1, \ldots,120,$ we consider the positive quadrant of the sphere $X_{i} \in \mathbb{S}_{+}^{2},$ where $\mathbb{S}_{+}^{2} = \{ (x,y,z) \in \real^{3} \mid x^2 + y^2 + z^2 = 1, x, y, z \geq 0 \},$ with geodesic distance as the ambient space. If the intrinsic space of energy fractions samples corresponds to the entire ambient space $\mathbb{S}_{+}^{2},$ the intrinsic curvature $\rho_{I}$ should be  positive, since the ambient space is the positively curved sphere.  

\begin{figure}[ht]
  \centering
    \subfigure[]{\includegraphics[width=0.26\linewidth]{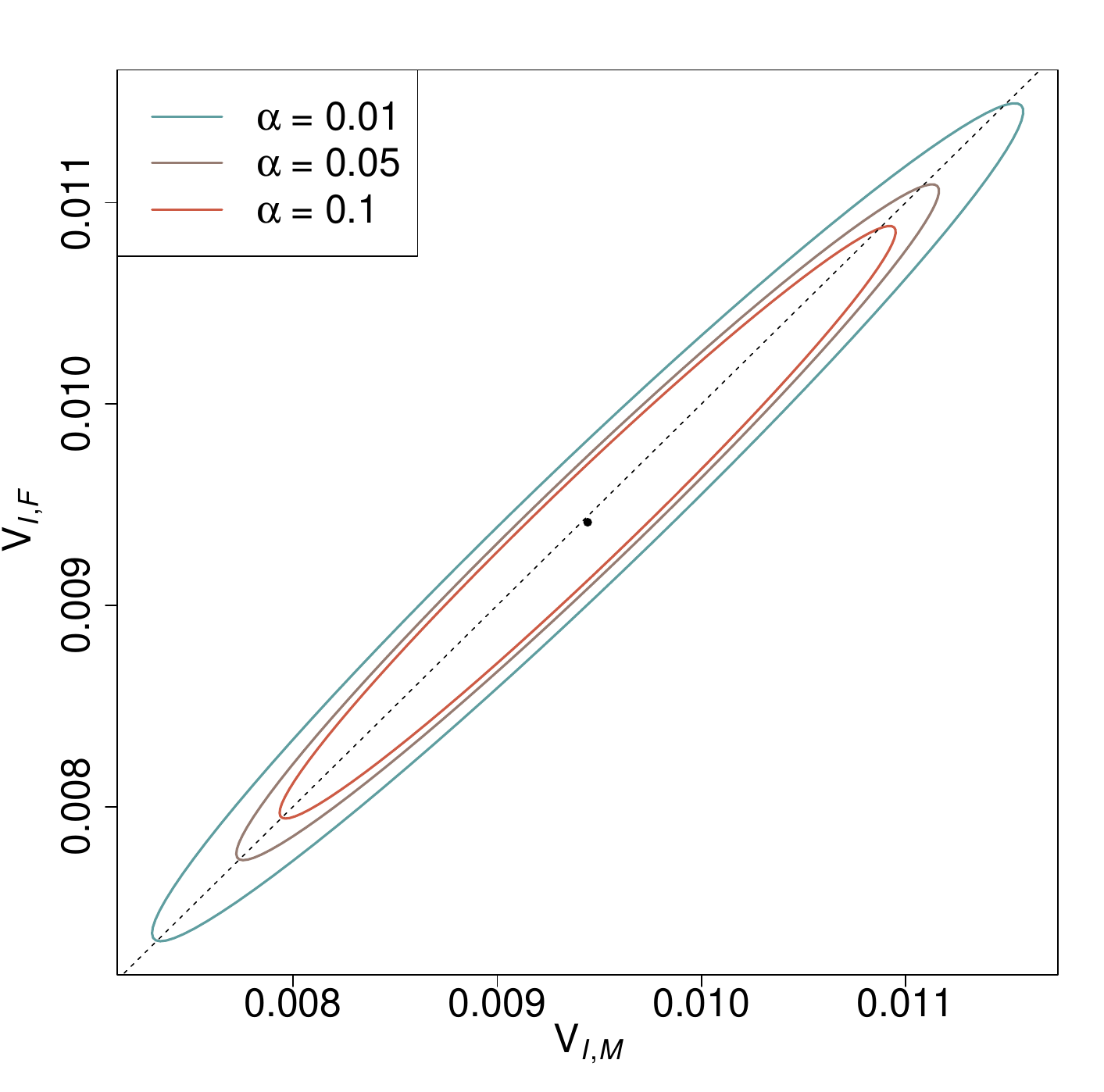}\label{fig9:sub1}}
    \subfigure[]{\includegraphics[width=0.36\linewidth]{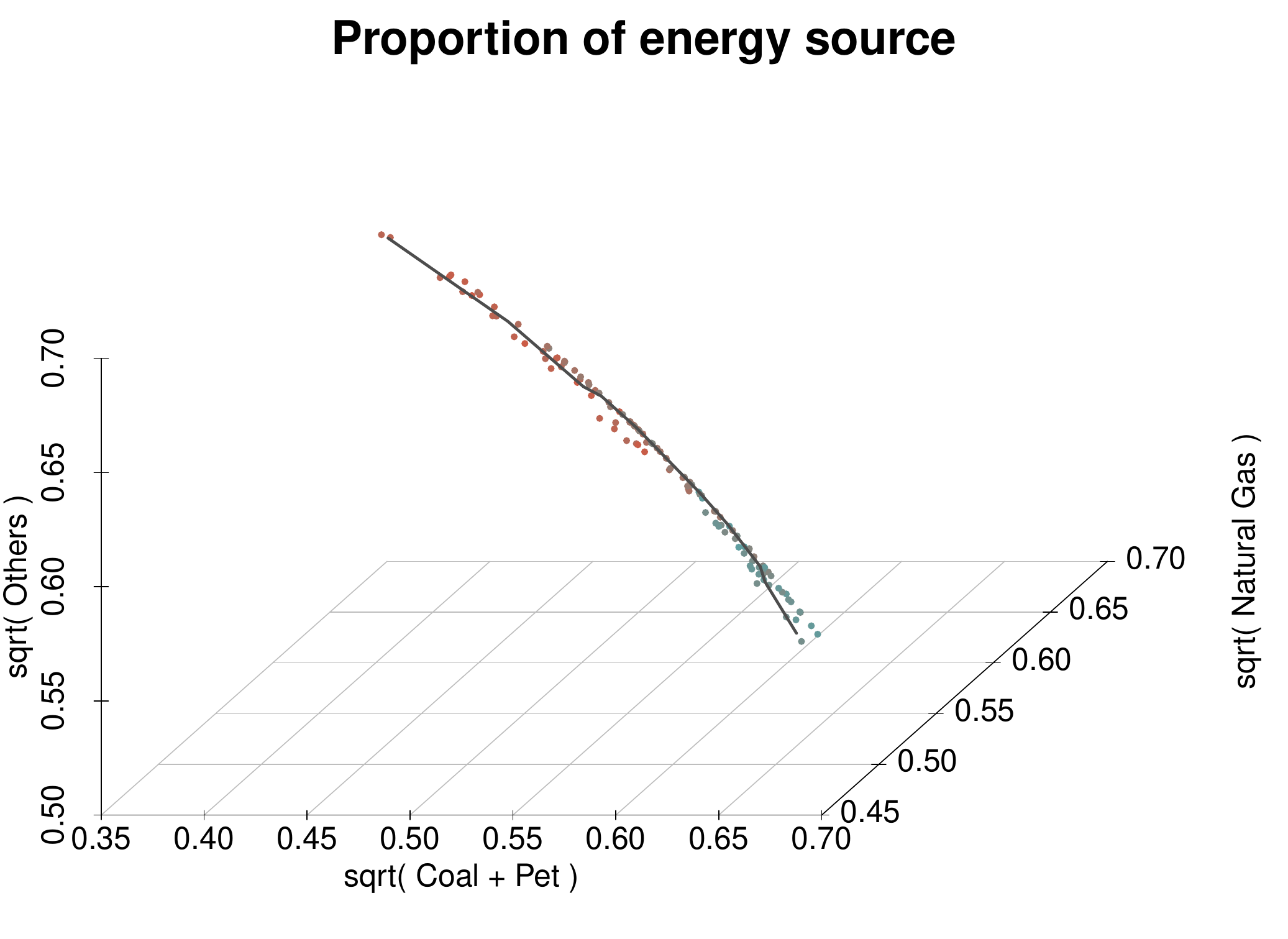}\label{fig9:sub2}}
    \subfigure[]{\includegraphics[width=0.36\linewidth]{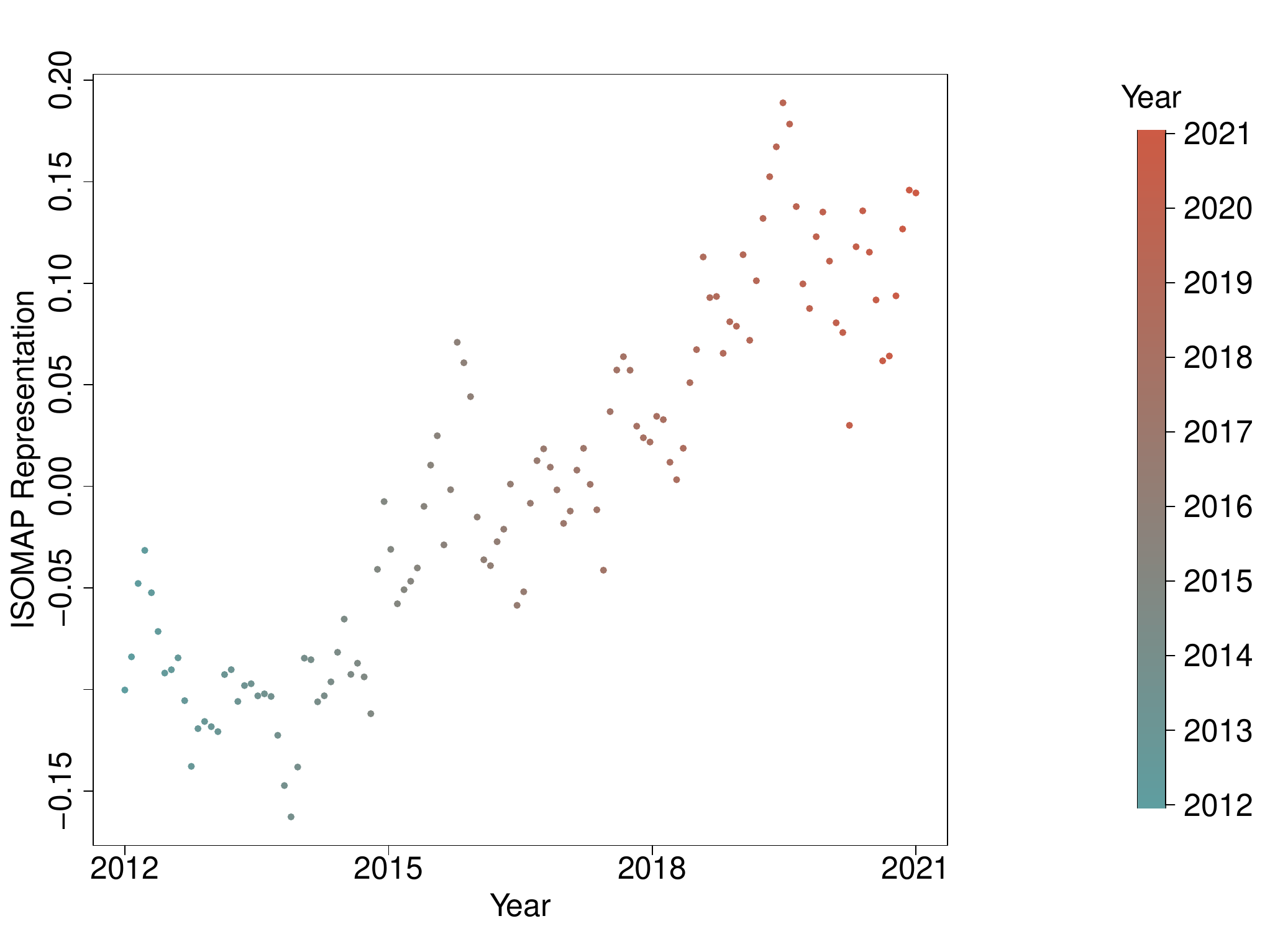}\label{fig9:sub3}}
    \caption{(a) Confidence regions $\mathcal{C}_{I,n}(1-\alpha)$ in \eqref{CR_int} for energy compositional data with $\alpha = 0.01, 0.05, 0.1.$ (b) Energy proportional data trajectory from January 2012 (blue) to December 2021 (red) with its intrinsic geodesic $\hat{\gamma}(t)$ in \eqref{intrin_isomap_geod} (black solid line). (c) Corresponding ISOMAP representation interpolation $s(t)$ 
    as a function of year.}
  \label{fig:energy}
\end{figure}

Figure \ref{fig9:sub1} demonstrates that the intrinsic curvature $\rho_{I}$ is actually not that far from being flat, indicating that the curvature of the intrinsic space, where the observations are actually located, differs from that of the ambient space $\mathbb{S}_{+}^{2}$. This then justifies to utilize the intrinsic geodesic $\hat{\gamma}(t)$ in \eqref{intrin_isomap_geod} as obtained by  ISOMAP, which a priori would not be possible for data located on the sphere.   Figure \ref{fig9:sub2} indicates that the estimated intrinsic geodesics $\hat{\gamma}(t)$ effectively capture the trajectories of these compositional  data, and as per Figure \ref{fig9:sub3}  energy sourcing moves in a linear direction from past to present. 

\section{Discussion}\label{chap8}

The study of dispersion measures, specifically metric variance and \F variance, is of general interest for the emerging field of random objects in general metric spaces. The CLT for the joint distribution of these dispersion estimates enables more detailed comparisons and  we discovered that the relation between these two measures contains information about the curvature of the metric space. Tools from metric geometry and empirical processes make it possible to obtain theory-supported inference for the underlying curvature. 

Another important application of these results is the inference of intrinsic curvature in a sample of random objects when the underlying probability measure is concentrated on a potentially curved subspace, such as a curved manifold embedded in an ambient metric space, which may itself be either flat or curved. Determining intrinsic curvature has high relevance for real data analysis as we demonstrate with SPD matrices corresponding to gait synchronization and energy compositional data located on the sphere. {While we do not make use of the ISOMAP algorithm of \cite{tenenbaum2000global}   for the proposed intrinsic curvature test, if it is discovered by applying the  test that  the curvature of $\mathcal{A}$ is compatible with that of a flat space, one can confidently use ISOMAP to represent $\mathcal{A}$ as the image of a map from $\mathbb{R}^p$ to $\mathcal{A}$. So one application of the proposed test for curvature is  to determine whether the assumptions of ISOMAP are satisfied.} 
    
{Several avenues for future research present themselves.  These include a rigorous convergence analysis of the estimated intrinsic distance for general subspaces $\mA \sub \M$  and also the choice of an appropriate metric $d$. Investigating the impact of the metric choice on statistical inference and geometric analysis remains an important open problem.}

\section*{Supplementary Materials}

The supplementary materials include all proofs, additional simulations for power analysis, further illustrations of curvature inference for the MNIST image and temperature distribution data, convergence analysis of intrinsic distance estimation, sensitivity analysis for the choice of input distance $d$ and verification of required conditions for both simulations and real data analysis.

\section*{Funding}

This research was supported in part by NSF grant DMS-2310450.

\clearpage

\setcounter{section}{0}
\setcounter{figure}{0}
\setcounter{table}{0}

\renewcommand{\thesection}{S.\arabic{section}}
\renewcommand{\thetable}{S.\arabic{table}}  
\renewcommand{\thefigure}{S.\arabic{figure}} 
\renewcommand{\thefigure}{S.\arabic{figure}} 
\renewcommand{\theequation}{S.\arabic{equation}}
\newtheorem{Theorem}{Theorem S.\ignorespaces}
\newtheorem{Lemma}{Lemma S.\ignorespaces}
\newtheorem{Corollary}{Corollary S.\ignorespaces}

{\large\bf Supplement to ``Inference for Dispersion and Curvature of Random Objects"}

\begin{abstract}
Section \ref{sec:appenex} presents additional simulation results for spaces of symmetric positive definite (SPD) matrices using six different distance metrics, as well as for spherical data and high-dimensional Euclidean random objects with low intrinsic dimension. We also provide a power and Type I error analysis. Additional real data analyses can be found in Section \ref{sec:appen2} for the MNIST image dataset and a temperature data set. Section \ref{suppl:sec:conv_intrin} discusses the convergence results of intrinsic distance estimation obtained by Algorithm 1 in Section 4.2. Section \ref{suppl:sec:prf} provides the proofs of the main results in the paper. Section \ref{suppl:sec:veri_cond} verifies the conditions assumed in our theorems for both simulations and real data analyses in the main manuscript. Section \ref{suppl:sec:sens_anal} conducts a sensitivity analysis for the choice of input distance $d$ in our proposed intrinsic curvature test introduced in Section 4.2. Section \ref{suppl:sec:rel_curv} examines the relationship between Alexandrov curvature and our proposed metric curvature.

\end{abstract}
\pagenumbering{arabic} \setcounter{page}{1}
\spacingset{1.9} 

\newpage
\section{Additional simulation results}
\label{sec:appenex}

    This section provides simulations, illustrating estimation and inference for the curvature and also how curvature may vary within the same space when employing different metrics. Specifically, this section provides additional examples, including SPD matrices with 6 different distance metrics, spherical data with geodesic distance, and high-dimensional Euclidean random objects with low intrinsic dimension. We also provide a power and Type I error analysis.  
 
   \subsection{Spaces of symmetric positive definite matrices}\label{sec:appenex:spd}

    The study of SPD matrices is relevant for various machine learning domains, including medical imaging, natural language processing, computer vision, and modeling time-varying data. The space of SPD matrices of size $p \times p$, denoted by $\mathcal{S}^{p}_{++},$ form a smooth manifold and is defined as 
    \begin{equation*}
        \mathcal{S}^{p}_{++} = \left\{X \in \mathbb{R}^{p \times p}: X^{T} = X, X \succ 0 \right\}.
    \end{equation*}  
    
\begin{figure}[h]
        \centering
        \subfigure[]{\includegraphics[width=0.32\linewidth]{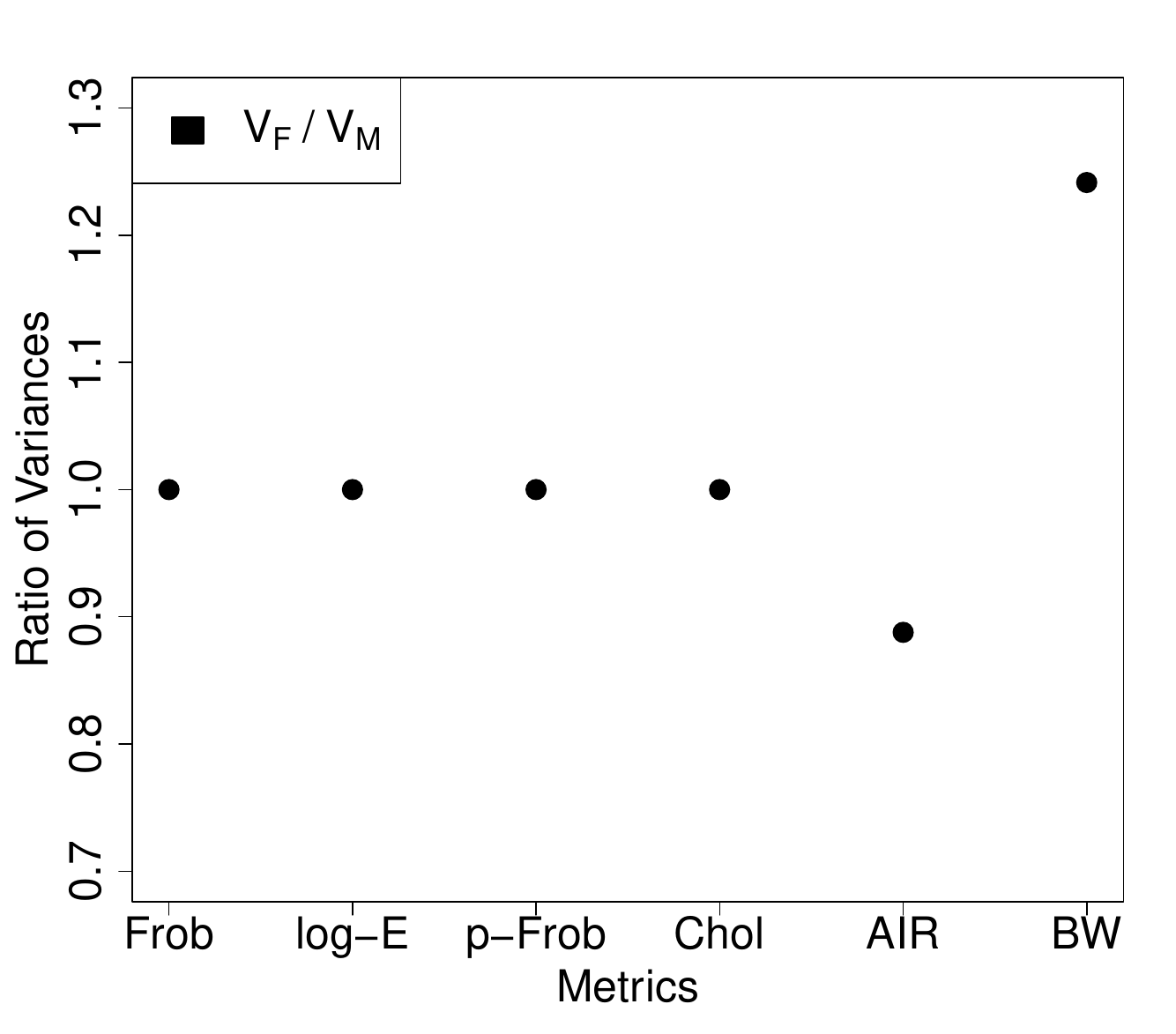}\label{fig2:sub1}}
        \subfigure[]{\includegraphics[width=0.32\linewidth]{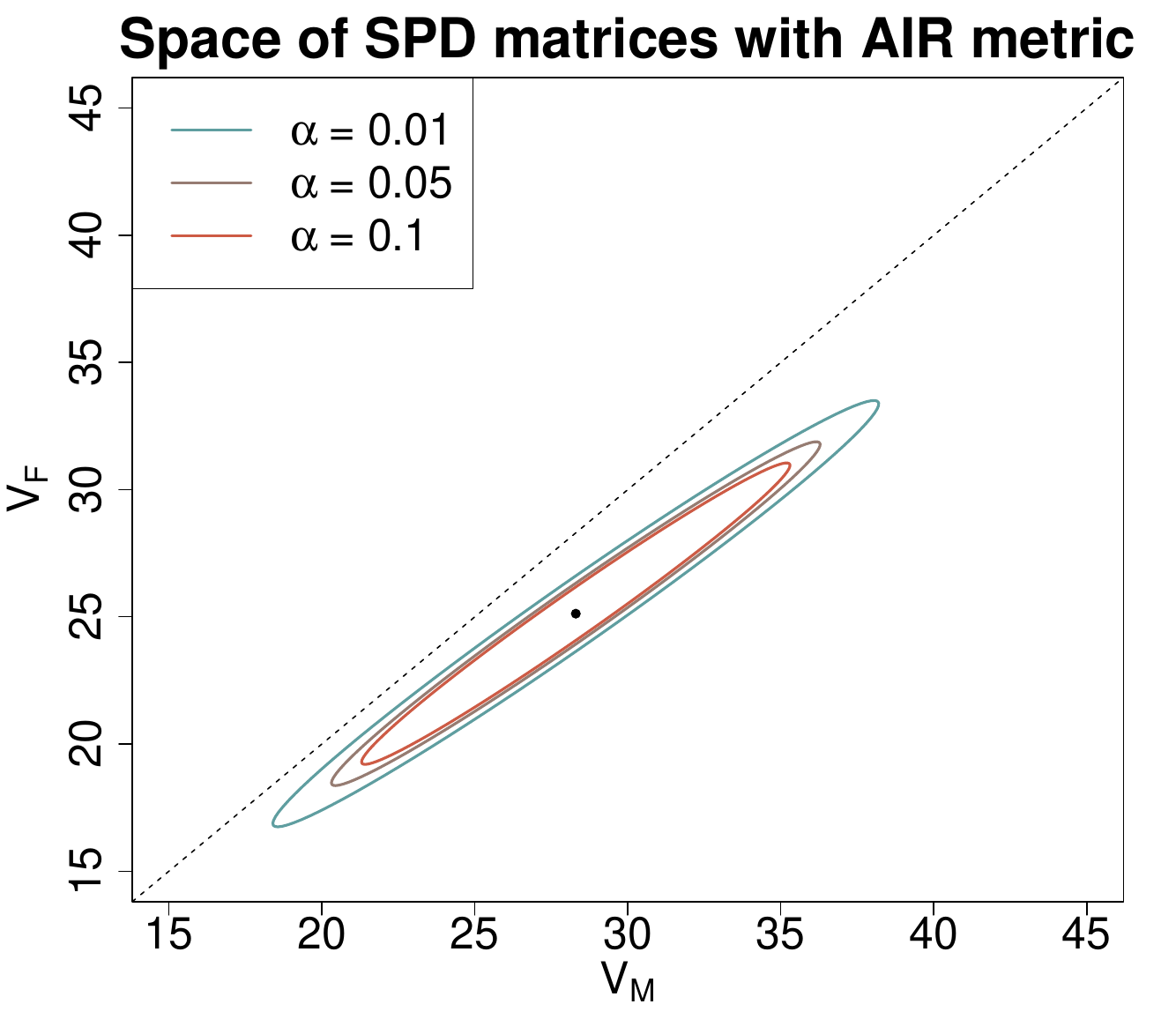}\label{fig2:sub2}}
        \subfigure[]{\includegraphics[width=0.32\linewidth]{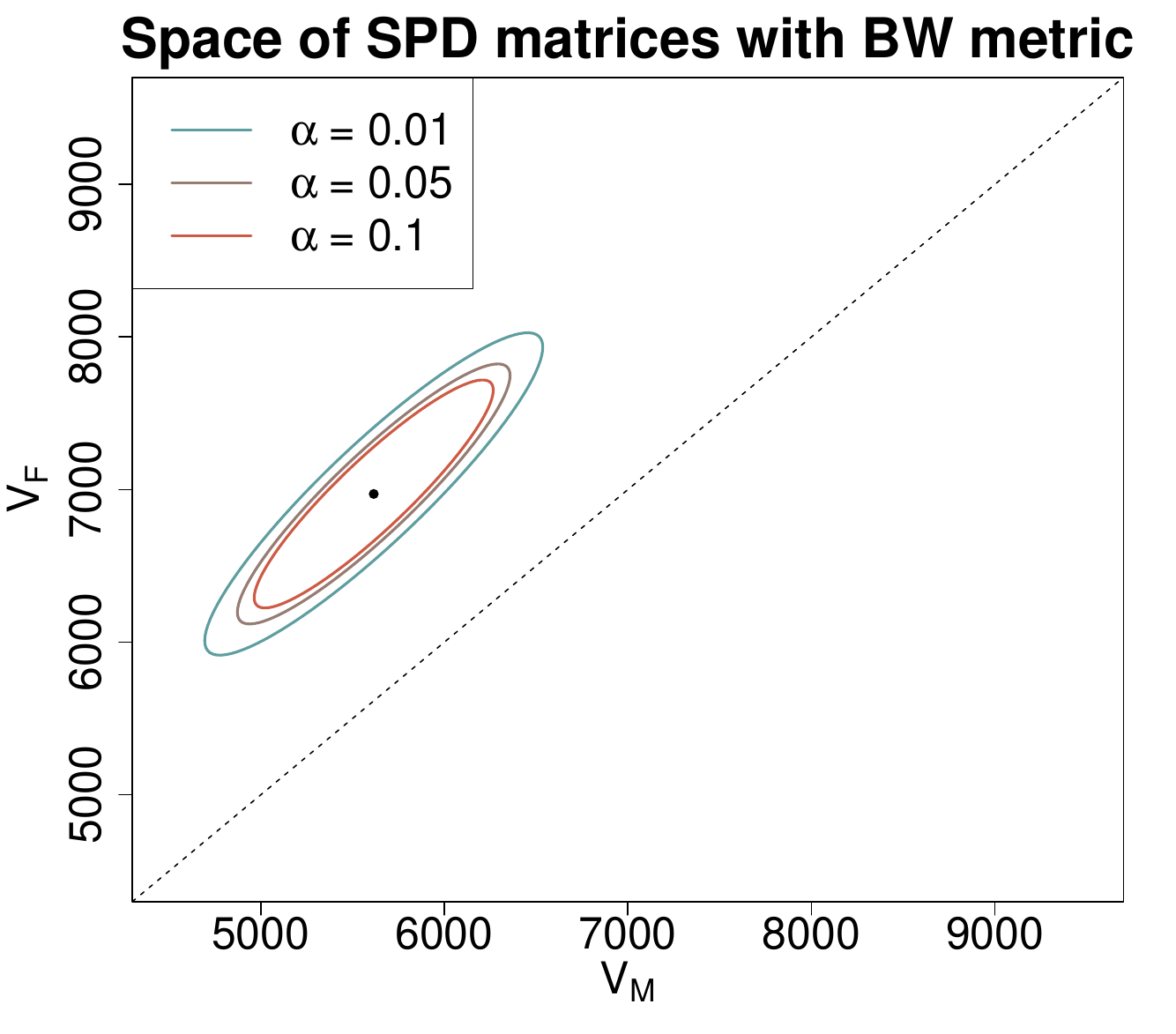}\label{fig2:sub3}}
        \caption{(a) Ratio of the two dispersion measures $\eta = (V_{M}, V_{F})^{T}$ for a sample of size 100 for random SPD matrices $\mathcal{S}^{p}_{++}$ using 6 different metrics (Frob, log-E, p-Frob, Chol, AIR, and BW). (b), (c) Corresponding confidence regions $\mathcal{C}_{n}(1-\alpha)$ in (7) for the SPD spaces with the AIR and BW metrics, with $\alpha = 0.01, 0.05,$ and $0.1,$ empirically inferring the underlying curvature of the space.}
        \label{fig:spd}
    \end{figure}

    We generated random SPD matrices  $X = V \Lambda V^{T},$ where $V = \left(\Vec{v}_{1}/\lVert \Vec{v}_{1}\rVert, \cdots ,\Vec{v}_{p}/\lVert \Vec{v}_{p}\rVert \right),$ with $\Vec{v}_{l}$ sampled independently and identically distributed (i.i.d.) from the Gaussian distribution ${N}_{p}\left(0, I_{p} \right),$ and setting $\Lambda = \text{diag}\left(\lambda_{1},\ldots, \lambda_{p}\right),$ $\lambda_{l} \sim \nu \times \text{Beta}\left(\beta, \gamma \right),$ $l= 1,\ldots, p,$ with $\beta = 3,$ $\gamma= 5,$ $\nu = 100,$ and $p = 3.$ 
    We studied six different metrics: (a) Frobenius (Frob), (b) log-Euclidean (log-E), (c) power Frobenius with power $\frac{1}{2}$ (p-Frob), (d) Cholesky (Chol), (e) Affine-Invariant Riemannian (AIR), and (f) Bures-Wasserstein (BW) metric. Figure S.\ref{fig2:sub1} depicts the ratios of metric variance $V_{F}$ over \F variance $V_{M}$ for samples each with $n =100$ SPD matrices for these six metrics.  It is known that the  spaces of SPD matrices $\mathcal{S}^{p}_{++}$ with the Cholesky and Frobenius metrics (including the log-Frobenius and power Frobenius metrics) are flat, and accordingly the \F variance $V_{F}$ and the metric variance $V_{M}$ are equal. 
    
    Figure S.\ref{fig2:sub1} aligns with established findings that the curvature of the Riemannian manifold with the affine-invariant metric is a non-positively curved space \citep{bhatia2009positive}, whereas the curvature of the SPD manifold with the BW metric has non-negative curvature \citep{han2021riemannian}, as outlined in Section 3. Figure S.\ref{fig2:sub2}-S.\ref{fig2:sub3} illustrate the confidence regions $\mathcal{C}_{n}(1-\alpha)$ in (7) for $\eta = (\eta_{1}, \eta_{2})^{T} = (V_{M}, V_{F})^{T}$, computed for the sample of SPD matrices using the AIR and BW metrics, with $\alpha = 0.01, 0.05,$ and $0.1.$ These confidence regions establish  the significance of the negative and positive curvature as determined from the sample for the AIR and BW metric, respectively, as these confidence regions are contained within the sets $\left\{\eta ~ | ~ \eta_{1} > \eta_{2} \right\},$ and $\left\{\eta ~ | ~ \eta_{1} < \eta_{2} \right\},$ respectively. These results thus confirm Theorems 2 and 3.

    \subsection{Spherical data with geodesic distance}\label{sec:appenex:sph}
    
    This simulation is designed for comparing two dispersion measures and illustrating the confidence region $\mathcal{C}_{n}(1-\alpha)$ in (7). Spherical data are common in applications, including geographical data on $\mathbb{S}^{2}$ \citep{zheng2015trajectory} and distributional data on the Hilbert sphere $\mathbb{S}^{\infty}$ \citep{dai2022statistical}.   In positively curved spaces such as  the space of spherical data, quantifying the dispersion of data using the metric variance $V_M$ 
    is often preferable, as obtaining the \F mean and the traditional \F variance $V_F$ can be numerically challenging.

       We consider here random objects taking values on the upper hemisphere $\mathbb{S}^{2}_{+} \subset \mathbb{S}^{2}$,  \begin{align*}
        \mathbb{S}^{2}_{+} = \left\{\left(x, y, z \right) \in \mathbb{R}^3 ~|~ x^2 + y^2 + z^2 = 1,~ z\geq0 \right\},
    \end{align*}
equipped with the geodesic distance.
   In this scenario, the  \F mean $\mu_{\oplus}$ of $X$ is unique with $\mu_{\oplus}=\left( 0,0, 1 \right).$ We generated $n = 50$ spherical data points, $X_{1}, \ldots, X_{n}$ from a uniform distribution on $ \mathbb{S}^{2}_{+}$ and obtained estimates of the \F variance $V_F$ and the metric variance $V_M$.  
   
\begin{figure}[!ht]
        \centering
        \includegraphics[width=0.5\linewidth]{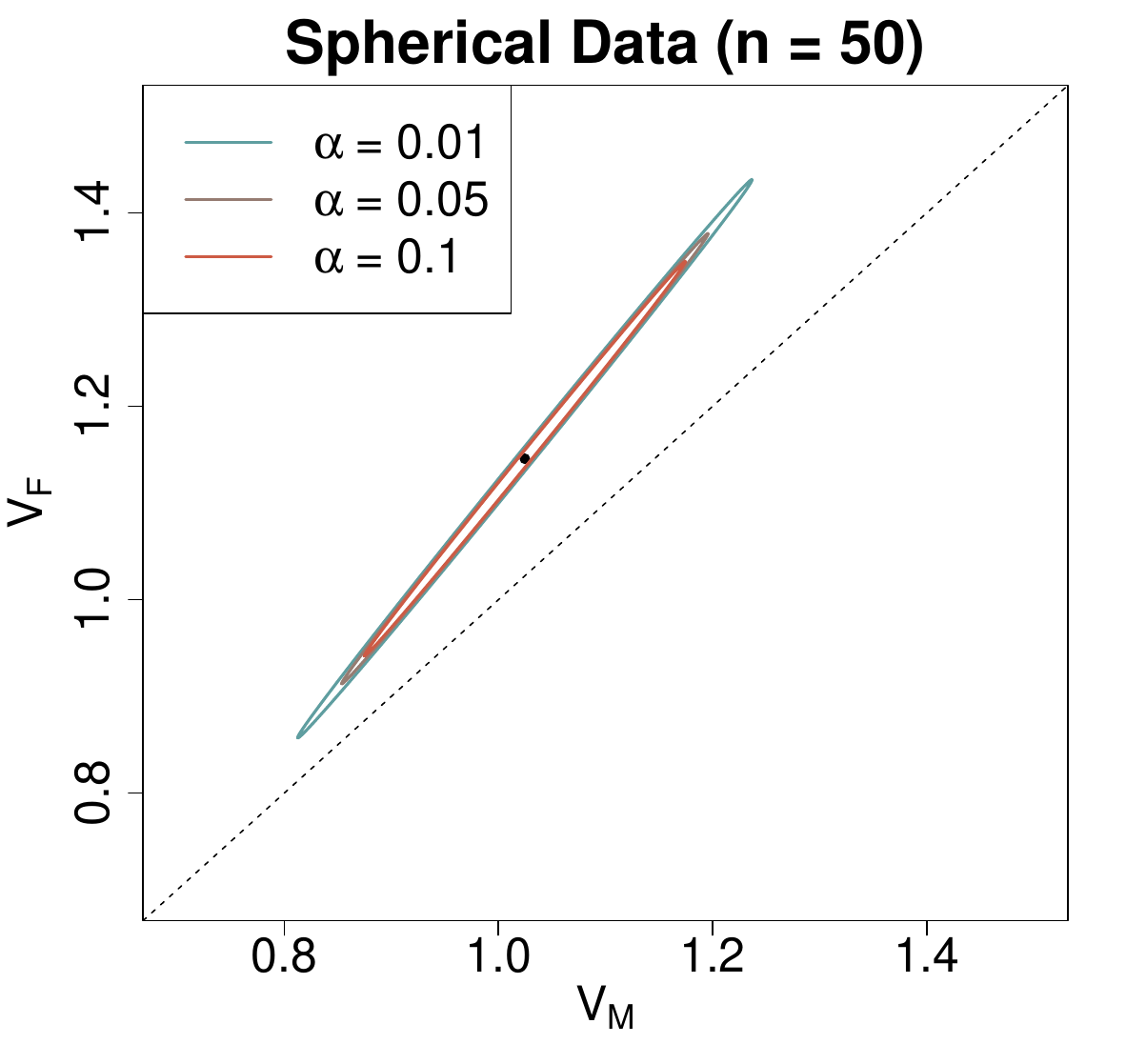}
        \caption{Confidence regions $\mathcal{C}_{n}(1-\alpha)$ for $\eta$ in (7) using $n=50$ uniform samples on the open half sphere of 
        $\mathbb{S}^{2}_{+}$ with $\alpha = 0.01, 0.05, 0.1,$ showing that the proposed curvature measure correctly detects the positive curvature. }
        \label{fig:sphere}
    \end{figure}

   Figure \ref{fig:sphere} illustrates the $100(1-\alpha) \%$ confidence region of the joint distribution for $\eta = (\eta_{1}, \eta_{2}) = (V_{M}, V_{F})^{T},$ denoted as $\mathcal{C}_{n}(1-\alpha)$ in (7), with $\alpha = 0.01, 0.05,$ and $0.1.$ We infer that the space $ \mathbb{S}^{2}_{+}$ with geodesic distance exhibits positive curvature with a confidence level exceeding $99\%$ as $\mathcal{C}_{n}(1-\alpha) \subset \Theta_{1},$ where $\Theta_{1} = \left\{\eta ~ | ~ \eta_{1} < \eta_{2} \right\}$ under  $H_{1}: \mP \in \Theta_{1}.$
   
\clearpage
\subsection{Power and Type-I error analysis}

To illustrate the finite-sample performance of the proposed intrinsic curvature test in Section 4.2, we conducted power and Type-I error analyses. Specifically, we consider the three-dimensional Euclidean space $(\M, d)$ as the ambient space and define the intrinsic space $\mathcal{D}_{\kappa}$ as 
    \begin{equation*}
        \mathcal{D}_{\kappa} = \left\{(x,y,z) \in \mathbb{R}^{3} ~\Bigg|~x^2 + y^2 + z^2 = \left(\frac{1}{\sqrt{\kappa}}\right)^2,~ \frac{1}{\sqrt{\kappa}}\cos\left(\frac{\pi\sqrt{\kappa}}{4}\right) \leq z \leq \frac{1}{\sqrt{\kappa}}  \right\},
    \end{equation*}
    for $0 < \kappa \leq 1.$ This space corresponds to a sphere of radius of $\frac{1}{\sqrt{\kappa}},$ which is known to have Alexandrov curvature $\kappa.$ The constraint on the $z$-axis $ \frac{1}{\sqrt{\kappa}}\cos\left(\frac{\pi\sqrt{\kappa}}{4}\right) \leq z \leq \frac{1}{\sqrt{\kappa}}$ ensures that the diameters of $\mathcal{D}_{\kappa}$ are identical and equal to $\frac{\pi}{2}$ under the spherical geodesic distance for all  $\kappa > 0$.
    
    For the Type-I error analysis, we also consider the following  flat null space:
    \begin{equation*}
        \mathcal{D}_{0} = \left\{(x,y,0) \in \mathbb{R}^{3} ~|~ 0 \leq x,~ y \leq \frac{\pi}{2\sqrt{2}}\right\}.
    \end{equation*}

    In the first scenario, we generate random objects $X_{1}, \ldots, X_{n}$ of size $n = 200, 500, 1000,$ following a uniform distribution on $\mathcal{D}_{\kappa},$ for all $\kappa \geq 0.$ We performed 500 Monte Carlo simulations to construct empirical power functions, where the empirical power was assessed as the proportion of test rejections at a significance level of $\alpha = 0.05$ over the $500$ runs. 
    
\begin{figure}[h!]
        \centering
        \includegraphics[width=0.5\linewidth]{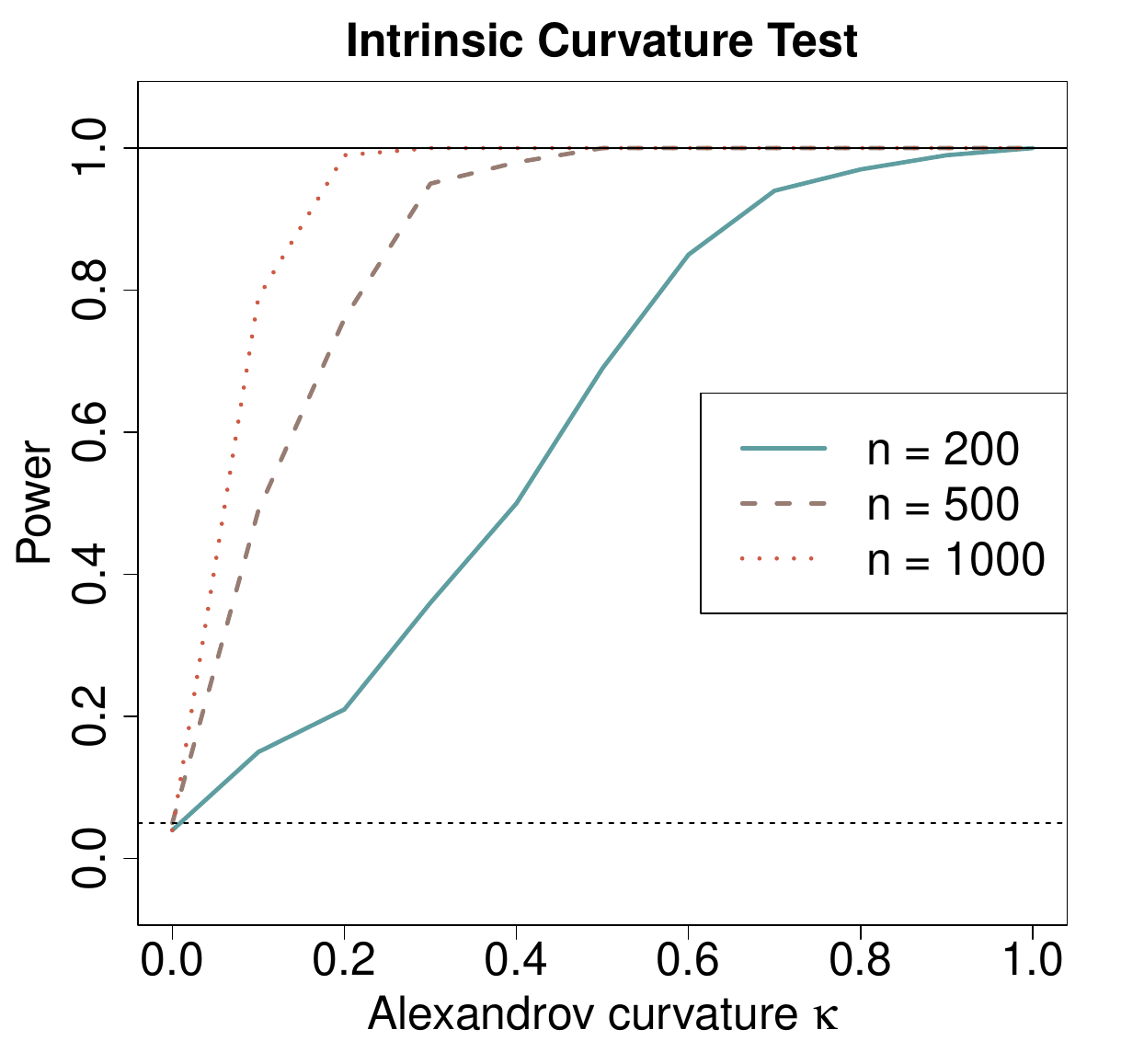}
        \caption{Power and Type-I error analyses for increasing Alexandrov curvature $\kappa,$ based on random objects  $X_{1}, \ldots, X_{n}$ with sample sizes $n = 200, 500, 1000.$ Each observation $X_{i}$ is uniformly distributed on $\mathcal{D}_{\kappa},$ for $\kappa \geq 0.$ The empirical power is computed as the proportion of test rejections at a significance level of $\alpha = 0.05$ over the $500$ Monte Carlo simulations.}
        \label{fig:ref2:power_anal}
    \end{figure}

    Figure \ref{fig:ref2:power_anal} presents the results. When the curvature parameter is  $\kappa = 0,$ the power functions remain close to $0.05$, indicating appropriate Type-I error control. For a fixed sample size $n$, the power increases as the curvature $\kappa$ grows. Similarly, for a fixed curvature  $\kappa,$ the power increases as the sample size $n$ increases.
    
\subsection{High-dimensional ambient Euclidean random objects with low intrinsic dimension}

We consider a setting where the ambient dimensionality significantly exceeds the intrinsic dimension of the manifold. Specifically, we define the following sparse $p$-dimensional Euclidean space:
    \begin{equation*}
        \mathcal{E}_{p} = \left\{(x,y,z, 0,0, \ldots, 0) \in \mathbb{R}^{p} ~\Bigg|~x^2 + y^2 + z^2 = 1,~ 0 \leq z \leq 1 \right\},
    \end{equation*}
    which represents an intrinsically two-dimensional space with positive curvature, where we vary $p$ from $p=3$ to $p=100$. 

    When the random objects lie exactly on the intrinsic manifold $\mathcal{E}_{p},$ (i.e., in the absence of noise contamination), our curvature test results remain unchanged regardless of the ambient space dimension $\real^{p}.$ This invariance holds because, for any two points  $\xi_{1} = (x_{1}, y_{1}, z_{1}, 0, \ldots, 0),$ and $\xi_{2} = (x_{2}, y_{2}, z_{2}, 0, \ldots, 0),$ the input distance is given by  $d(\xi_{1}, \xi_{2}) = (x_{1} - x_{2})^{2} + (y_{1} - y_{2})^{2} + (z_{1} - z_{2})^{2},$ which does not depend on the ambient dimension $p.$
    
\begin{figure}[h!]
        \centering
        \includegraphics[width=0.45\linewidth]{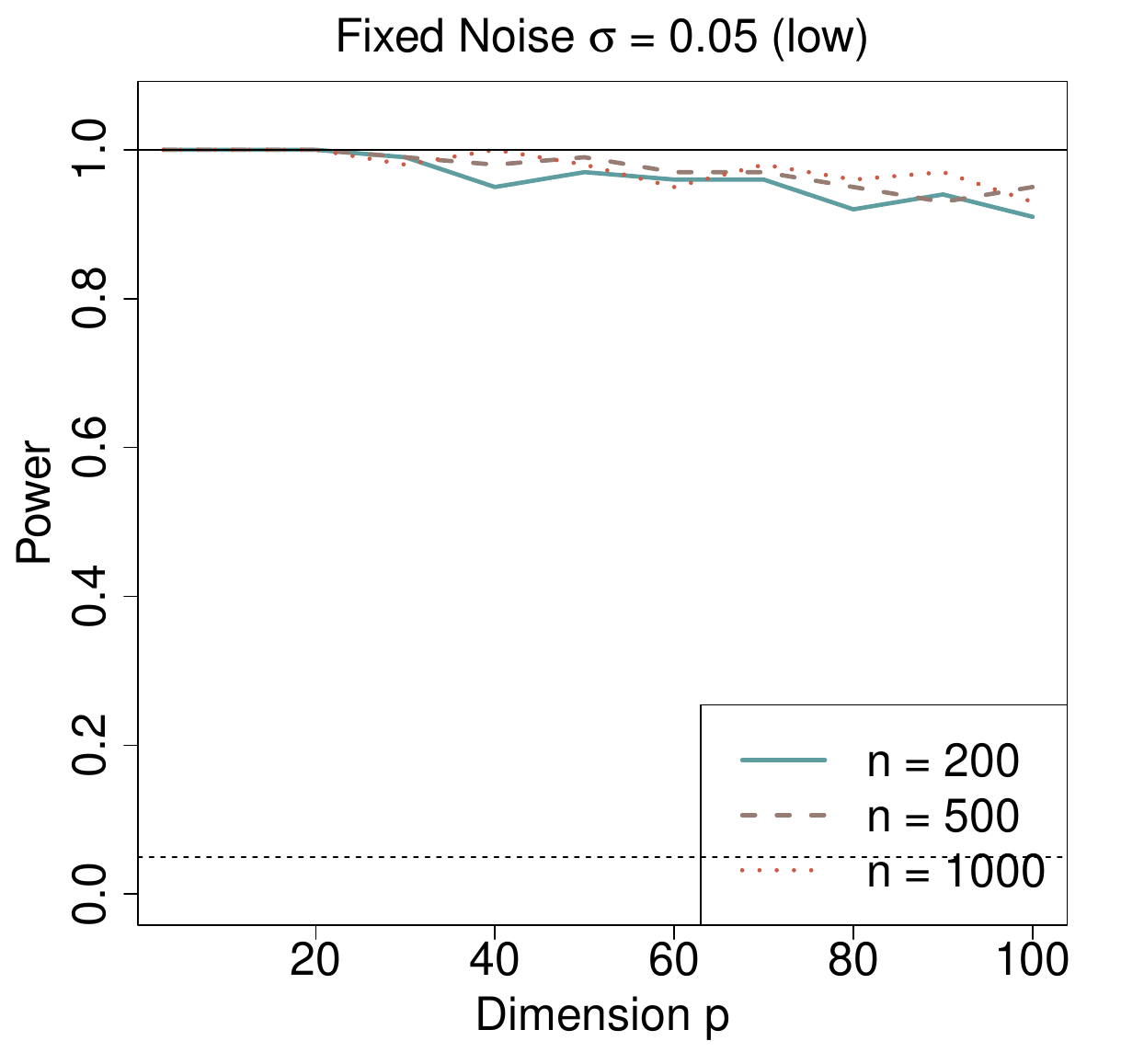} \includegraphics[width=0.45\linewidth]{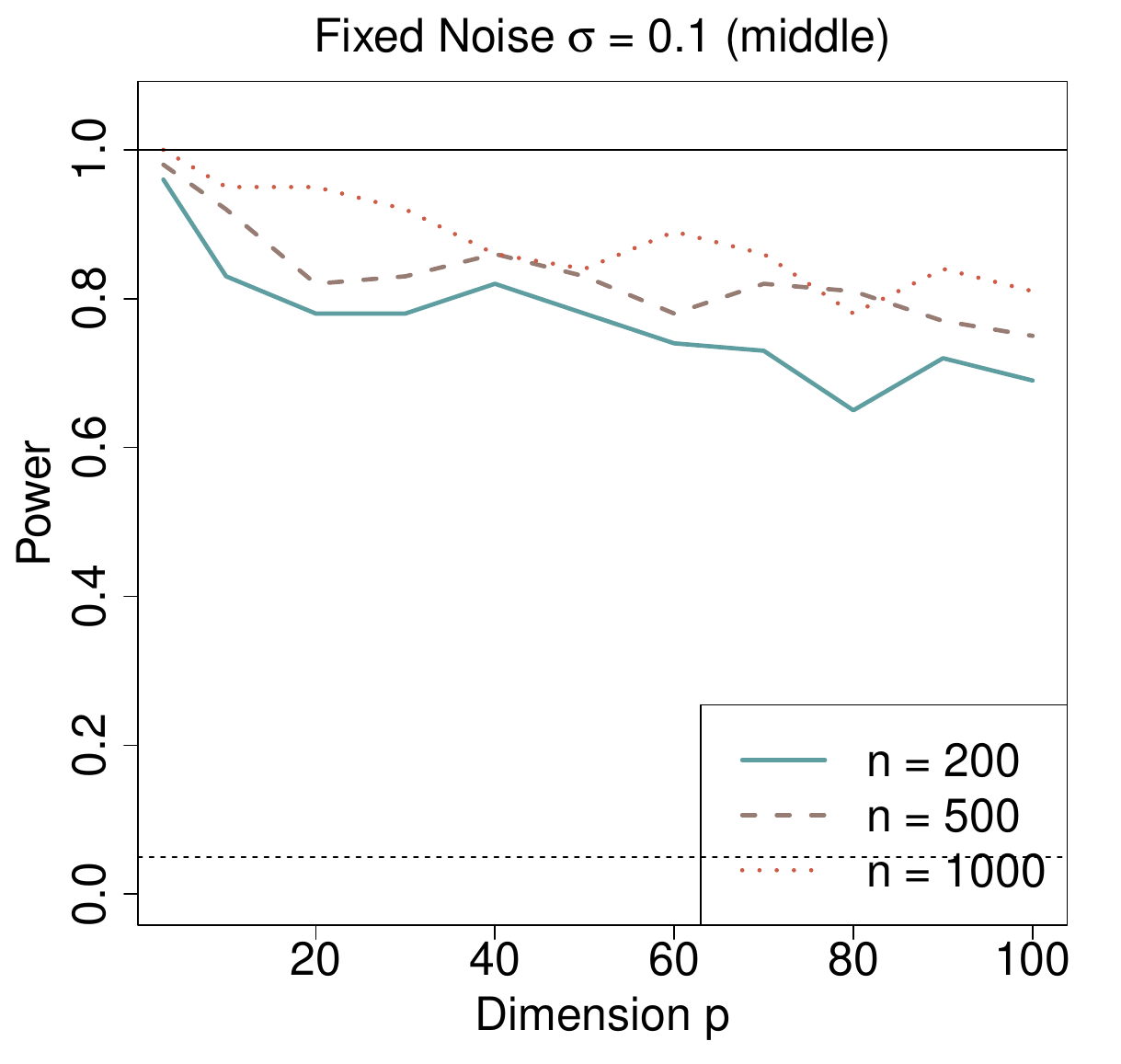}
        \includegraphics[width=0.45\linewidth]{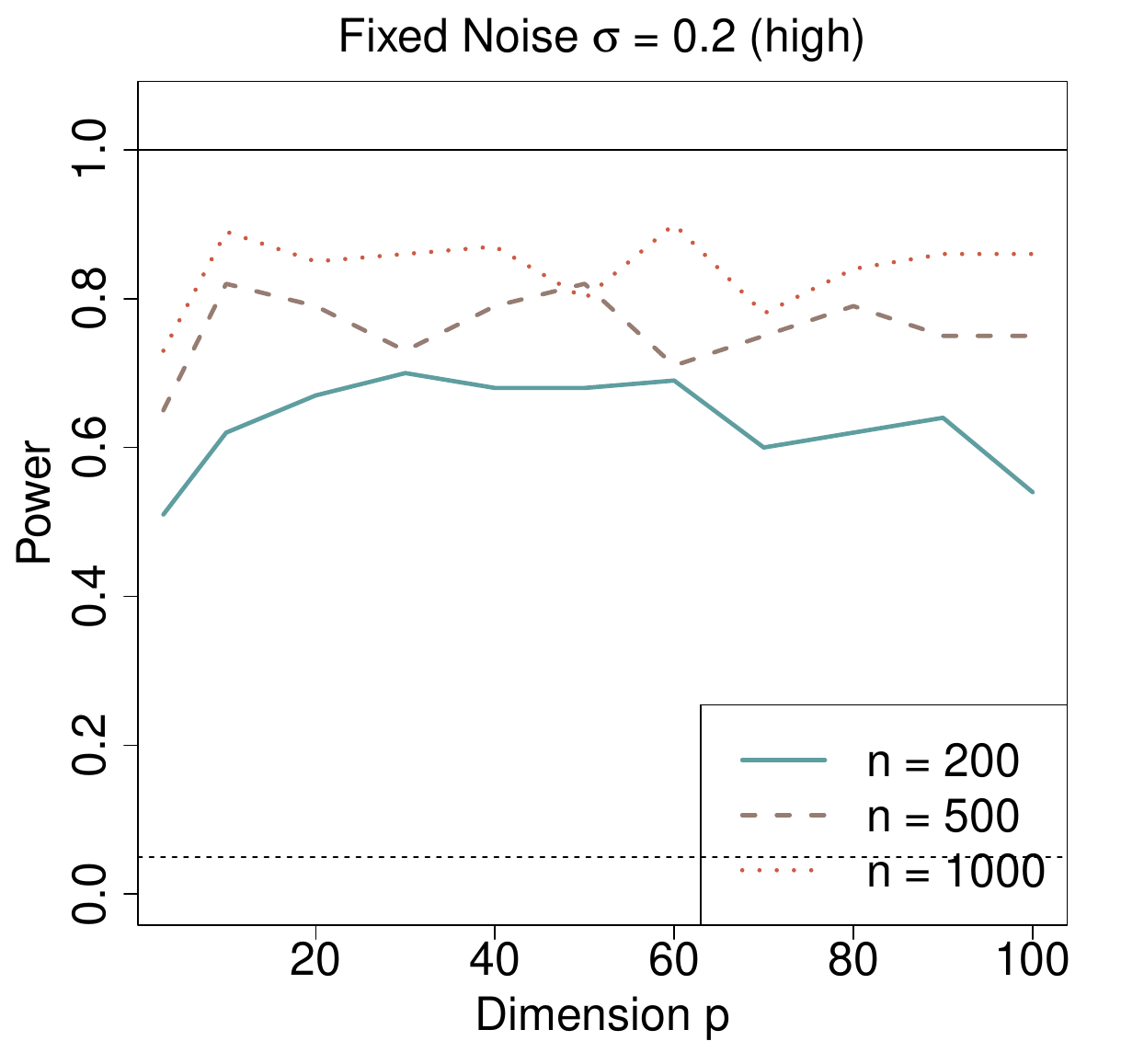}
        \includegraphics[width=0.45\linewidth]{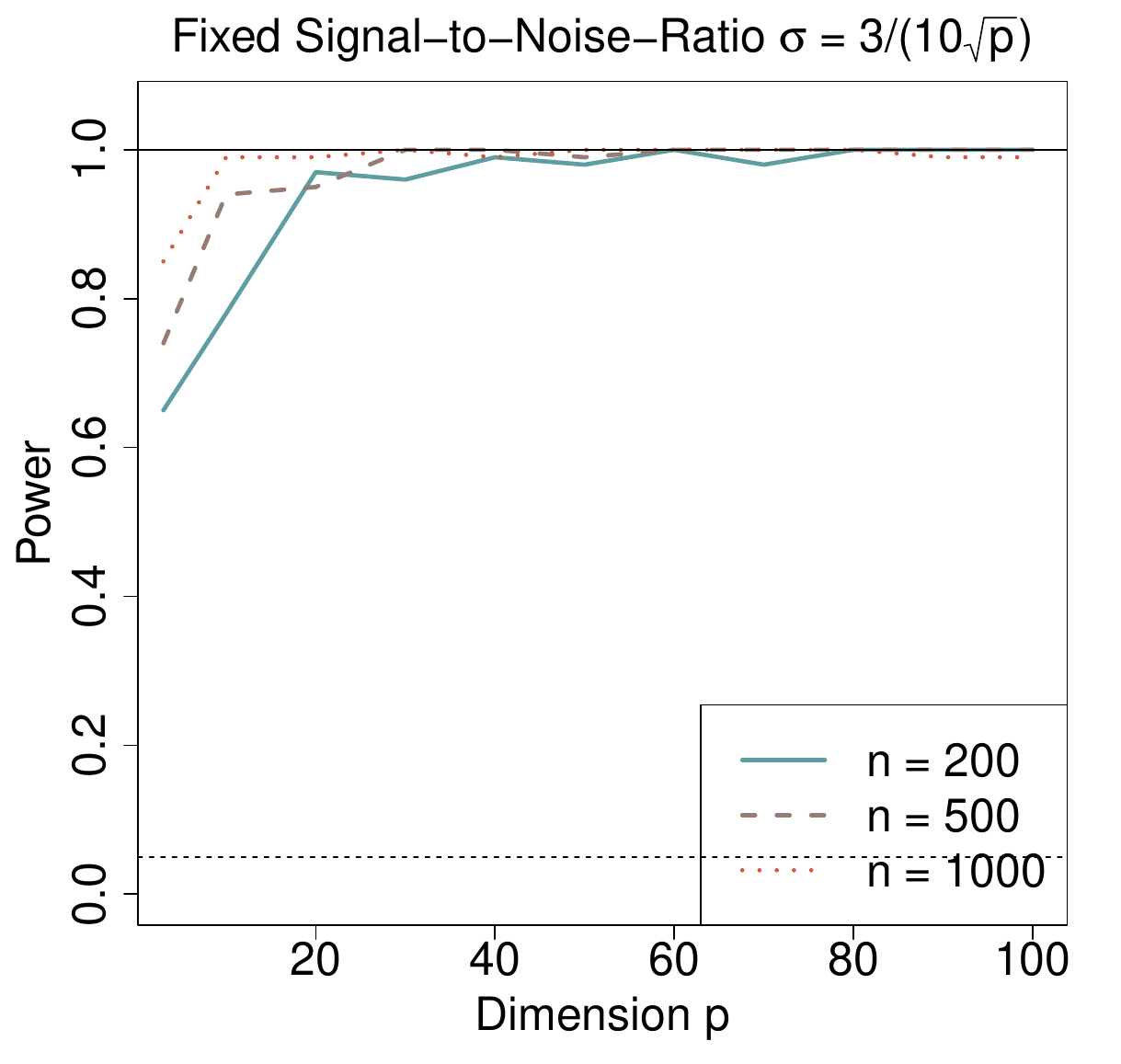}
        \caption{Power analysis for increasing ambient space dimension $p$ and  noise-contaminated random objects $\Tilde{X}_{1}, \ldots, \Tilde{X}_{n},$  with sample sizes $n= 200, 500, 1000.$ Each observation is modeled as $\Tilde{X}_{i} = X_{i} + \mathbf{\epsilon}_{i},$ where $X_{i}$ is uniformly distributed on an intrinsically two-dimensional positively curved space $\mathcal{E}_{p},$ and $\mathbf{\epsilon}_{i} \sim N_{p}(\mathbf{0}, \sigma^{2} \mathbf{I}_{p})$ represents additive Gaussian noise. The noise level $\sigma$ is considered under four settings: fixed noise levels $\sigma = 0.05$ (low), $ 0.1$ (middle), $ 0.2$ (high) and a fixed signal-to-noise ratio of $\frac{3}{10\sqrt{p}}.$}
         \label{fig:ref2:hdim_simul}
    \end{figure}

   In our second simulation, we consider an additive Gaussian noise model with noise level $\sigma$ varying with the ambient dimension $p.$ Figure \ref{fig:ref2:hdim_simul} illustrates the power analysis for increasing ambient space dimension $p$ for the case of  noise-contaminated random objects $\Tilde{X}_{1}, \ldots, \Tilde{X}_{n},$  with sample sizes $n= 200, 500, 1000.$ Each observation is modeled as $\Tilde{X}_{i} = X_{i} + \mathbf{\epsilon}_{i},$ where $X_{i}$ is uniformly distributed on an intrinsically two-dimensional positively curved space $\mathcal{E}_{p}$ and $\mathbf{\epsilon}_{i} \sim N_{p}(\mathbf{0}, \sigma^{2} \mathbf{I}_{p})$ represents additive Gaussian noise, with $\mathbf{0} = (0,\ldots,0)^{T} \in \real^{p}$ and $p \times p$ identity matrix $\mathbf{I}_{p}$. The noise level $\sigma$ is considered under four settings: fixed noise levels $\sigma = 0.05$ (low), $ 0.1$ (middle), $ 0.2$ (high), and a fixed signal-to-noise ratio of $\frac{3}{10\sqrt{p}}.$ 

    For the low and moderate noise settings $\sigma = 0.05$, $ 0.1$, as the ambient dimension $p$ increases, statistical power generally decreases. This decline occurs because the signal from $\mathcal{E}_{p},$ which has  intrinsic positive curvature, is increasingly overshadowed by the additive Gaussian noise, which follows a flat Euclidean structure. In the high noise case $\sigma = 0.2,$ the ambient dimension $p$ has a relatively minor effect on power. This is because even for the low-dimensional case $p = 3,$ the strong noise component dominates, weakening the signal from the positively curved intrinsic space $\mathcal{E}_{p}.$ Under the fixed signal-to-noise ratio setting, increasing the ambient dimension $p$ reduces the contribution of noise in each coordinate, proportional to $1/\sqrt{p}.$  This results in an increase in statistical power as $p$ grows. Overall, our proposed tests exhibit robustness to additive ambient noise, particularly when the ambient dimension substantially exceeds the intrinsic dimension of the underlying space.

\clearpage
\section{Additional real data analysis}
\label{sec:appen2}

\subsection{MNIST Digits}\label{sec:appen2:mnist}

The MNIST dataset is a widely used collection of handwritten digits, freely available for training various image processing systems. It comprises 70,000 images, each sized at $28 \times 28$ pixels, belonging to ten categories, with 60,000 images in the training set and 10,000 in the testing set. The dataset has a correct class label from a set of ten possible digits $\left\{0, \ldots, 9 \right\}.$

The intrinsic dimension of each digit image space might be significantly lower than the ambient space dimension of $784$ (since each image is $28 \times 28$ pixels), as images of the same digits often share similar patterns. While nonlinear dimension reduction methods, including  locally linear embedding, ISOMAP  and Laplacian eigenmaps, have been successfully applied to image data in recent years, the intrinsic curvature of the space has not been explored yet, as far as we know. In this section, we therefore use the 60,000 images from the MNIST training set to investigate the intrinsic curvature of each digit image space. 

We utilize the Euclidean distance as an ambient distance by flattening these 2D images into 1D vectors, and estimate the intrinsic curvature $\rho_{I}$ in Section 4.2. Table \ref{tab:digits} shows the intrinsic Confidence intervals $\mathcal{I}_{I,n}(1-\alpha)$ of each digit image space, at level $\alpha = 0.05$. We conclude that all digits are concentrated on a lower-dimensional set with positive curvature, where the curvature is most expressed for the digit 2, followed by 5 and 9,  and is least expressed for the digit 8, followed by 1 and 6. 

\begin{table}[h!]
    \caption{Confidence intervals $\mathcal{I}_{I,n}(1-\alpha)$ in (15) for $\rho_{I}$ for the MNIST digit data.}
        \centering
        \begin{tabular}{lcccccccccc}\toprule
             & 0 & 1 & 2 & 3 & 4 & 5 & 6 & 7 & 8 & 9 \\
             \midrule\midrule
        Num samples & 5923 & 6742 & 5958 & 6131 & 5842 & 5421 & 5918 & 6265 & 5851 & 5949 \\
        \midrule
        \begin{tabular}{@{}l@{}}Lower bound \\ $95\%$ C.I. $\mathcal{I}_{I,n}$ \end{tabular} & 0.09 & 0.04 & 0.15 & 0.13 & 0.11 & 0.12 & 0.05 & 0.10 & 0.003 & 0.12 \\
        \midrule
        \begin{tabular}{@{}l@{}}Upper bound \\ $95\%$ C.I. $\mathcal{I}_{I,n}$ \end{tabular} & 0.11 & 0.05 & 0.16  & 0.15 & 0.13 & 0.14 & 0.06 & 0.12 & 0.015 & 0.13 \\
        \midrule
        Curvature & + & + & +  & + & + & + & + & + & + & + \\
        \bottomrule
        \end{tabular}
        \label{tab:digits}
    \end{table}

    \subsection{US airport weather station temperature distributions}  \label{chap71}
    The need to measure local volatility in temperatures, particularly during seasons with extreme temperatures, has long been acknowledged; see, e.g. \cite{donadelli2021computing}. We utilize daily temperature data recorded at U.S. airport weather stations, which can be obtained from \url{https://www.ncdc.noaa.gov/cdo-web/search?datasetid=GHCND}. Specifically, we focus on daily maximum temperatures (in Fahrenheit) during the summer period (June 21 - Sep 20) and daily minimum temperatures during the winter period (Dec 22 - Mar 21) from 39 major airports spanning 63 years (1960.06 to 2023.03). 
    
    For each station, we construct the distribution of maximum temperatures in summer and minimum temperatures in winter for every year, resulting in 63 distributions for each season, respectively. 
    For these distributions we adopt the 2-Wasserstein  metric that is frequently used for  spaces  of one-dimensional distributions as we consider here \citep{bols:03,kant:06:1}. For the $i^{th}$ airport, we acquire the distributions for summer maximum temperatures $S^{(i)}_{j}$ and winter minimum temperatures $W^{(i)}_{j},$ for the $j^{th}$ year, respectively, where $i = 1, \ldots, 39,$ $j = 1, \ldots, 63.$ Subsequently, the metric variances are computed for each airport to quantify the variation of the  temperature distributions across all  63 years.  
    
\begin{figure}[!hbt]
      \centering
        \subfigure[]{\includegraphics[width=0.48\textwidth]{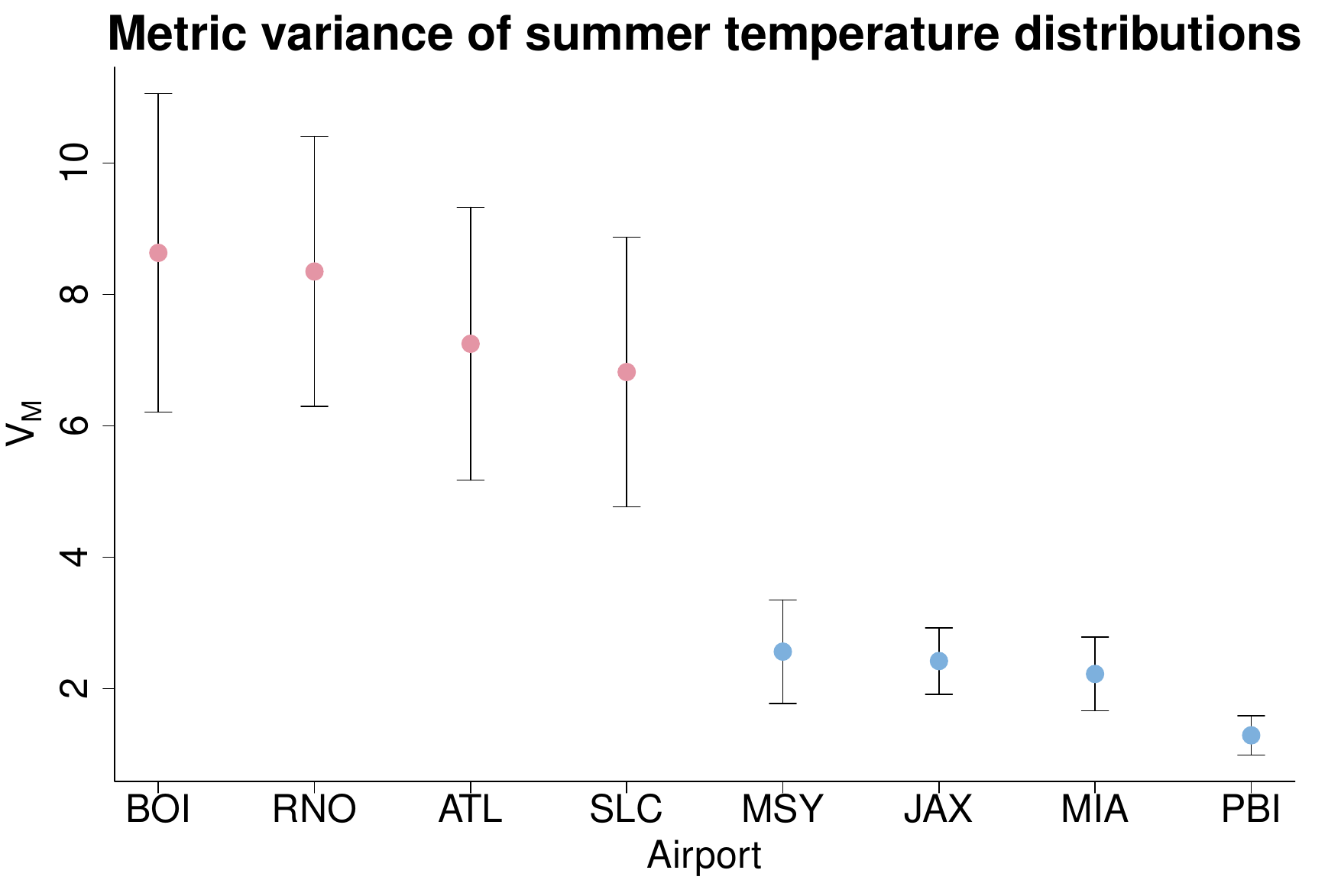}\label{fig:tempsub1}}
        \subfigure[]{\includegraphics[width=0.48\textwidth]{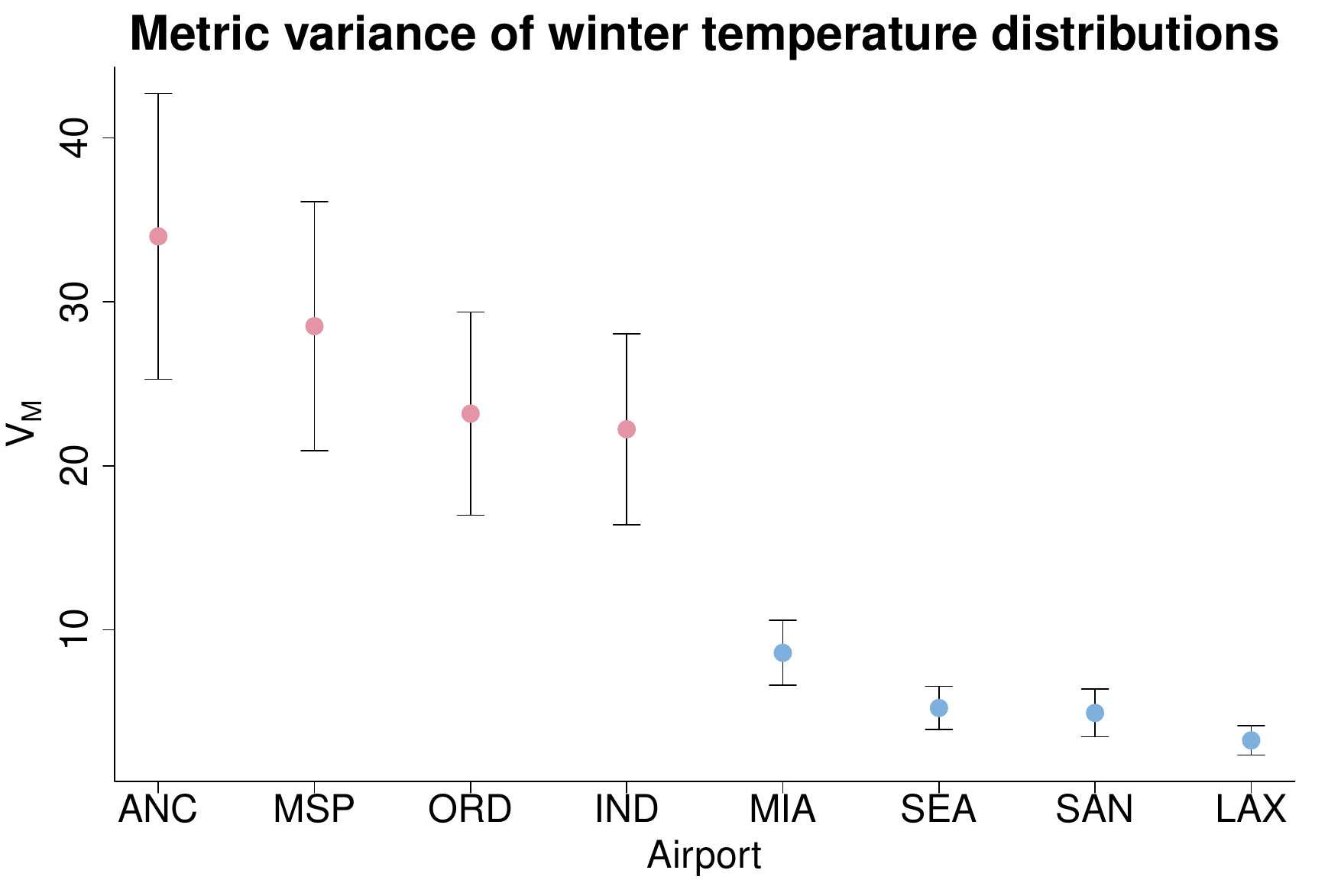}\label{fig:tempsub2}}
      \caption{The four airports with respectively highest  (red) and lowest (blue) metric variances  $\hat{V}_{M}$ in (5) with $95\%$ confidence intervals for summer maximum temperature probability distributions (left panel) and winter minimum temperature probability distributions (right panel), among 39 airports (BOI = Boise, RNO =  Reno, ATL = Atlanta, SLC =  Salt Lake City, MSY = New Orleans, JAX = Jacksonville, MIA = Miami,  PBI, West Palm Beach, ANC = Anchorage, MSP = Minneapolis-Saint Paul, ORD = Chicago, IND = Indianapolis, MIA = Miami, SEA = Seattle, SAN = San Diego, LAX = Los Angeles).}
      \label{fig:temperature}
    \end{figure}

    The estimated metric variances and  $95\%$ confidence intervals for the annual temperature distributions at selected airports are displayed in Figure \ref{fig:temperature}, where the airports with highest and lowest annual temperature volatility in summer and winter are highlighted.   The Boise airport has the highest and the Palm Beach airport the lowest summer temperature variation. Generally, airports located near the coastline have lower variation. Regarding winter temperatures,  
    Anchorage has the highest and Los Angeles  the lowest variation. 
    
\newpage
\section{Convergence of intrinsic distance estimation}
\label{suppl:sec:conv_intrin}

Let $\mA \sub \M$ be a compact set, where the ambient space $(\M, d)$ is a $p$-dimensional Euclidean space with Euclidean distance $d,$ and with  intrinsic dimension $k < p.$ The intrinsic space $\mA$ is said to be \textit{$K$-geodesically smooth} for some positive $K > 0,$ if for all $(x, y) \in \mA \times \mA,$ 1) there exists a geodesic $\gamma : [0, T] \rightarrow \mA$ from $x$ to $y,$ which is $C^{1}$-smooth, i.e., it has a continuous first derivatives, 2) there exists a real function $\beta$ such that $\lim\limits_{t \rightarrow 0} \beta(t) = 0$ and $l(\gamma) \leq \beta(d(x,y)),$ where $l(\cdot)$ is the length of the geodesic, and 3) the first derivative $\dot{\gamma}$ is $K$-Lipschitz continuous. A probability measure $\mP$ supported on $\mA$ is said to be \textit{$\epsilon$-standard with respect to a measure $\mu$} for $\epsilon > 0,$ if there exists $\Delta > 0$ such that $\mP(\mathcal{B}_{\delta}(x)) \geq \epsilon  \mu(\mathcal{B}_{\delta}(x))$ for all $x \in \mA$ and $\delta \in [0, \Delta].$

To provide the rate of convergence for the intrinsic distance estimation, we consider the following assumptions \citep{aaron2018convergence}:

\textit{(B1)} $\mA$ is a $k$-dimensional compact manifold that is $K$-geodesically smooth for $K > 0.$

\textit{(B2)} The probability measure $\mP$ supported on $\mA$ is  $\epsilon$-standard with respect to  the Lebesgue measure in the ambient space $\M.$ 

A compact manifold that is twice continuously differentiable with no boundaries satisfies  assumption \textit{(B1)}, and if   the probability distribution $\mP$ on $\mA$ has a continuous density that is bounded below by a positive constant $c_0$  assumption \textit{(B2)} \citep{aaron2018convergence} is satisfied. All proofs are provided in Section \ref{suppl:sec:prf}. Let $\tilde{\rho}_{I,n}$ denote the intrinsic metric curvature obtained by replacing the intrinsic distance $d_{I}$ with its estimated version $\hat{d}_{I}.$ 

\begin{Theorem}\label{thm:supp:conv}
    Suppose conditions \textit{(B1)}-\textit{(B2)} hold and that  the weighted graph in Algorithm 1 is constructed with ball sizes $r = r_{n}$  that satisfy  
    \begin{equation*}
        \left(A_{0}\frac{\log n}{n}\right)^{2/3k} \leq r_{n} \leq  \left(A_{1}\frac{\log n}{n}\right)^{2/3k}
    \end{equation*}
    for some $A_{0},$ $A_{1} > 0.$ Then 
    \begin{equation*}
    \max\limits_{i,j}\bigg\lvert \hat{d}_{I}(X_i, X_j) - d_{I}(X_{i}, X_{j})\bigg\rvert = O\left(\left(\frac{\log n}{n}\right)^{2/3k}\right)
\end{equation*} and \begin{equation*}
        \lvert\tilde{\rho}_{I, n} - {\rho}_{I, n}\rvert = O\left(\max\left\{ \left(\frac{\log n}{n}\right)^{2/3k}, \left(\frac{1}{n} \right)^{1/2} \right\}\right) \quad \text{a.s.}
    \end{equation*}
\end{Theorem}

To explicitly build a convenience data-based sequence of ball parameters  $r_n$ only from samples for  spaces that satisfy the additional assumption \textit{(B1)}-\textit{(B2)},  the following corollary may be helpful. 
    
\begin{Corollary}\label{cor:supp:conv}
    Suppose $\mA$ is a compact manifold that is twice differentiable with no boundaries  and the probability measure $\mP$ is concentrated on $\mA$ and has a continuous density that is bounded below by a positive constant $c_0.$ If the weighted graph in Algorithm 1 is constructed using  ball size $r = r_{n} = c(\max_{x \in \mathcal{X}_n}(\min_{y \neq x, y \in \mathcal{X}_n}d(x, y))^{2/3}$ for some constant $c > 0,$  then 
        \begin{equation*}
        \max\limits_{i,j}\bigg\lvert \hat{d}_{I}(X_i, X_j) - d_{I}(X_{i}, X_{j})\bigg\rvert = O\left(\left(\frac{\log n}{n}\right)^{2/3k}\right)
    \end{equation*} 
    and \begin{equation*}
        \lvert\tilde{\rho}_{I, n} - {\rho}_{I, n}\rvert = O\left( \max\left\{ \left(\frac{\log n}{n}\right)^{2/3k}, \left(\frac{1}{n} \right)^{1/2} \right\} \right) \quad \text{a.s.}
    \end{equation*}
\end{Corollary}

The convergence of the intrinsic distance estimation for more  general subspaces $\mA \sub \M$ remains an open problem. Currently, error bounds for intrinsic geodesic distances are available only under specific settings, where the ambient space $(\M, d)$ is a Euclidean space $\mathbb{R}^p$ for some $p>1$ and $\mA$ is a $k$-dimensional compact manifold that is $C^{q}$-smooth with a dense subset. Corollary S.\ref{cor:supp:conv} shows that the intrinsic distance estimator $\hat{d}_{I},$ obtained from Algorithm 1 in the main paper, has an error rate bound of \begin{equation*}
    \max\limits_{i,j}\bigg\lvert \hat{d}_{I}(X_i, X_j) - d_{I}(X_{i}, X_{j})\bigg\rvert = O\left(\left(\frac{\log n}{n}\right)^{2/3k}\right).
\end{equation*} 
Under the further assumption that $\mA$ is isometric to a convex domain, this convergence rate can be improved to $O\left( \left(\frac{\log n}{n} \right)^{1/k}\right)$  \citep{arias2020perturbation}. \cite{aamari2023optimal} established a new error bound of the order $O\left(\left(\frac{\log n}{n}\right)^{2/k}\right)$ by estimating the intrinsic distance using an updated algorithm with mesh construction via tangential Delaunay complexes \citep{boissonnat2014manifold}.

\newpage
\section{Proofs and technical results}
\label{suppl:sec:prf}

\no\textit{\underline{Proof of Proposition 1}}

Consider independent and identically distributed (i.i.d.) random objects, i.e., metric-space valued r.v.s  $X_{1}, \ldots, X_{n}.$ The sample metric variance can be expressed as 
\begin{align*}
    \hat{V}_{M} = \binom{n}{2}^{-1}\sum_{1\leq i < j \leq n} \psi(X_{i},X_{j}),
\end{align*}
which is the U-statistic with kernel function $\psi : \mathcal{M} \times \mathcal{M} \rightarrow \mathbb{R},$ \begin{align}
    \psi(x,y) = \frac{1}{2}d^{2}(x,y).
    \label{notpsi}
\end{align}

Define kernels \begin{align}
    \begin{split}
    h^{(1)}(X_{i}) &= \mathbb{E}_{X'|X_{i}}\left\{\psi(X_{i},X')|X_{i} \right\} - V_{M}, \\
    h^{(2)}(X_{i},X_{j}) &= \psi(X_{i},X_{j}) - h^{(1)}(X_{i}) - h^{(1)}(X_{j}) - V_{M},
    \end{split}
    \label{noth}
\end{align}
where $X'$ is an i.i.d. copy of $X_{1}, \ldots, X_{n}.$ 
Then
\begin{align}
    \begin{split}
        \hat{V}_{M} &= \frac{1}{\binom{n}{2}}\sum_{1\leq i < j \leq n} \psi(X_{i},X_{j}) \\
        &= \frac{1}{\binom{n}{2}}\sum_{1\leq i < j \leq n}  \left[V_{M} + h^{(1)}(X_{i}) + h^{(1)}(X_{j}) + h^{(2)}(X_{i},X_{j})\right]\\
        &= V_{M} + 2\binom{n}{1}^{-1}\sum_{i=1}^{n} h^{(1)}(X_{i}) + \binom{n}{2}^{-1}\sum_{1\leq i < j \leq n} h^{(2)}(X_{i},X_{j}) \\
        &= V_{M} + 2 H_{n}^{(1)}+ H_{n}^{(2)},
    \end{split}
    \label{Hdecomp}
\end{align}
where \begin{align}
    H_{n}^{(1)} = \binom{n}{1}^{-1}\sum_{i=1}^{n} h^{(1)}(X_{i})~\text{and}~ H_{n}^{(2)} = \binom{n}{2}^{-1}\sum_{1\leq i < j \leq n} h^{(2)}(X_{i},X_{j})
    \label{notH}
\end{align}
are uncorrelated. This is the H-decomposition of U-statistics \citep{hoeffding1961strong}. \\~

\no\textit{\underline{1) Proof of $\hat{V}_{M} \rightarrow V_{M}$ almost surely}}

We adopt the proofs from Section 3.4 in \cite{lee2019u}. Let $\hat{V}_{M}^{k}$ be a sample metric variance using $k$ random samples $X_{1}, \ldots, X_{k}.$ Then $\{\hat{V}_{M}^{n}\}_{n=1}^{\infty} $ is a reverse martingale adapted to the $\sigma-$fields $\left\{\mathcal{F}_{n} \right\}_{n=1}^{\infty},$ where $\mathcal{F}_{n} = \sigma\left(\hat{V}_{M}^{n},\hat{V}_{M}^{n+1},\ldots\right)$ and $\mathcal{F}_{\infty} = \cap_{n} \mathcal{F}_{n}.$ This is because \begin{enumerate}
    \item[(a)] $\{\mathcal{F}_{n}\}$ are decreasing sequences, i.e., $\mathcal{F}_{1} \supseteq \mathcal{F}_{2} \supseteq \cdots \supseteq \mathcal{F}_{\infty}.$

    \item[(b)] $\hat{V}_{M}^{n}$ is $\mathcal{F}_{n}$-measurable, for all $n \geq 1.$

    \item[(c)] $\mathbb{E}\lvert \hat{V}_{M}^{n} \rvert < \infty,$ for all $n \geq 1,$ under \textit{(M0)}.

    \item[(d)] $\mathbb{E}\{\hat{V}_{M}^{m}|\mathcal{F}_{n}\} = \mathbb{E}\left\{\mathbb{E}(\frac{1}{2}d^{2}(X_{1},X_{2})|\mathcal{F}_{m})|\mathcal{F}_{n}\right\} = \mathbb{E}\left\{\frac{1}{2}d^{2}(X_{1},X_{2})|\mathcal{F}_{n}\right\}=\hat{V}_{M}^{n},$ for all $1 \leq m \leq n,$ provided that $\hat{V}_{M}^{n} = \mathbb{E}\{\hat{V}_{M}^{n}|\mathcal{F}_{n}\} = \mathbb{E}\left\{\frac{1}{2}d^{2}(X_{1},X_{2})|\mathcal{F}_{n} \right\}$ and $\mathcal{F}_{n} \subseteq \mathcal{F}_{m}.$
\end{enumerate}

Then a property of reverse martingales implies that $\hat{V}_{M}^{n}$ converges to 

\no$\mathbb{E}\left\{\frac{1}{2}d^{2}(X_{1},X_{2})|\mathcal{F}_{\infty} \right\}$ almost surely \citep{shorack2009empirical}. Hence \begin{align*}
    \mathbb{E}\left\{\lim_{n \rightarrow \infty} \hat{V}_{M}^{n} \right\} = \mathbb{E}\left\{\mathbb{E}\left(\frac{1}{2}d^{2}(X_{1},X_{2})\bigg\rvert\mathcal{F}_{\infty} \right) \right\} = \mathbb{E}\left\{ \frac{1}{2}d^{2}(X_{1},X_{2})\right\} = V_{M}.
\end{align*} 

Let $\mathcal{G}_{n} = \sigma(X_{n},X_{n+1}, \ldots)$ and $\mathcal{G}_{\infty} = \cap_{n} \mathcal{G}_{n},$ which is a tail $\sigma-$field. The Kolmogorov Zero-One law implies that $\mathcal{G}_{\infty}$-measurable functions are almost surely constant. Therefore,   to establish almost surely convergence, it suffices to demonstrate that $\lim\limits_{n \to \infty} \hat{V}_{M}^{n}$ is $\mathcal{G}_{\infty}$-measurable, which implies that it must equal its expectation $V_{M}.$ It is sufficient to show that $\lim\limits_{n \to \infty} \hat{V}_{M}^{n}$ is $\mathcal{G}_{m}$-measurable, for arbitrary $m > 2.$

For any $n>m,$ let $\mathcal{A} = \{(i,j)~|~ m \leq i < j \leq n \},$ $\mathcal{B} = \{(i,j) \mid 1 \leq i < j < m\},$ and $\mathcal{C} = \{(i,j) \mid 1 \leq i < m \leq j\}.$ Then
\begin{align*}
    \hat{V}_{M}^{n} = \binom{n}{2}^{-1} A_{n} + \binom{n}{2}^{-1} B_{n} + \binom{n}{2}^{-1} C_{n},
\end{align*}
where $A_{n} = \sum\limits_{(i,j)\in \mathcal{A}} \frac{1}{2}d^{2}(X_{i},X_{j}),$  $B_{n} = \sum\limits_{(i,j)\in \mathcal{B}} \frac{1}{2}d^{2}(X_{i},X_{j}),$ and $C_{n} = \sum\limits_{(i,j)\in \mathcal{C}} \frac{1}{2}d^{2}(X_{i},X_{j}).$ 

The first term $\binom{n}{2}^{-1}A_{n}$ is $\mathcal{G}_{m}$-measurable for all $n \geq m,$ and thus $\lim\limits_{n} \binom{n}{2}^{-1}A_{n}$ is $\mathcal{G}_{m}$-measurable. 
For the second term,  $\lim\limits_{n}\binom{n}{2}^{-1}B_{n} = 0,$ which is $\mathcal{G}_{m}$-measurable, as $B_{n}$ is constant for all $n \geq m.$ 
Under \textit{(M0)}, for the third term $\binom{n}{2}^{-1}C_{n} = O(n^{-1}),$ because 
\begin{align*}
    \begin{split}
        \binom{n}{2}^{-1} C_{n} &= \binom{n}{2}^{-1} \sum_{i=1}^{m-1} \sum_{j = m}^{n} \frac{1}{2} d^{2}(X_{i},X_{j}) \\
        &=  \binom{n}{2}^{-1} (n-m+1) \sum_{i=1}^{m-1} \frac{1}{n-m+1} \sum_{j = m}^{n} \frac{1}{2} d^{2}(X_{i},X_{j}),
    \end{split}
\end{align*}
and $\frac{1}{n-m+1} \sum_{j = m}^{n} \frac{1}{2} d^{2}(X_{i},X_{j}) = O(1),$ given \textit{(M0)}. Then  $\lim\limits_{n} \binom{n}{2}^{-1}C_{n} = 0,$ which is $\mathcal{G}_{m}$-measurable.

Therefore  $\lim\limits_{n \to \infty} \hat{V}_{M}^{n}$ is $\mathcal{G}_{\infty}$-measurable and by the Kolmogorov Zero-One law, $ \hat{V}_{M}^{n}$ is constant $V_{M}$ with probability 1. \qed \\~

\no\textit{\underline{2) Proof of $\mathbb{E}\lvert \hat{V}_{M} - V_{M} \rvert^{2} = O(n^{-1})$}}

By the H-decomposition in \eqref{Hdecomp}, $\hat{V}_{M} - V_{M} = 2 H_{n}^{(1)}+ H_{n}^{(2)}.$ Therefore it is enough to show that $\mathbb{E}\lvert H_{n}^{(1)} \rvert^{2}$ and $\mathbb{E}\lvert H_{n}^{(2)} \rvert^{2}$ are of order $O(n^{-1}),$  then applying  the Minkowski inequality.
Both $\left\{\binom{n}{1} H^{(1)}_{n} \right\}_{n=1}^{\infty}$ and $\left\{\binom{n}{2} H^{(2)}_{n} \right\}_{n=2}^{\infty}$ are forward martingales adapted to the $\sigma-$fields $\left\{\mathcal{F}_{n} \right\}_{n=1}^{\infty},$ where $\mathcal{F}_{n} = \sigma\left(X_{1},\ldots, X_{n} \right)$ and $\mathcal{F}_{\infty} = \bigcup_{n} \mathcal{F}_{n}$ \citep{lee2019u}. This is because
\begin{enumerate}
    \item[(a)] $\{\mathcal{F}_{n}\}$ are increasing sequences, i.e., $\mathcal{F}_{1} \subseteq \mathcal{F}_{2} \subseteq \cdots \subseteq \mathcal{F}_{\infty}.$

    \item[(b)] $\binom{n}{c} H^{(c)}_{n}$ is $\mathcal{F}_{n}$-measurable, for all $n \geq 1,$ $c=1,2.$

    \item[(c)] $\mathbb{E} \lvert \binom{n}{c} H^{(c)}_{n} \rvert< \infty,$ for all $n \geq 1,$ $c=1,2,$ under \textit{(M0)}.

    \item[(d)] Let $\mathcal{C}_{1}(n) = \{i : 1\leq i \leq n\}, $ and $\mathcal{C}_{2}(n) = \{(i,j) : 1\leq i < j \leq n\}.$ For $c=1,2,$ \begin{align*}
        \mathbb{E}\left\{\binom{n+1}{c}H_{n+1}^{(c)}\bigg|\mathcal{F}_{n}\right\} = \sum\limits_{\mathcal{C}_{c}(n+1)} \mathbb{E}\{h^{(c)}(X_{i_{1}},\ldots, X_{i_{c}})|\mathcal{F}_{n}\},
    \end{align*}
    and $\mathbb{E}\{h^{(c)}(X_{i_{1}},\ldots, X_{i_{c}})|\mathcal{F}_{n}\}$ is zero if any $i_{j} = n + 1,$ and 
    
    \no$h^{(c)}(X_{i_{1}},\ldots, X_{i_{c}})$ otherwise. Hence
    \begin{align*}
        \mathbb{E}\left\{\binom{n+1}{c}H_{n+1}^{(c)}\bigg|\mathcal{F}_{n}\right\} = \sum\limits_{\mathcal{C}_{c}(n)} h^{(c)}(X_{i_{1}},\ldots, X_{i_{c}}) = \binom{n}{c}H_{n}^{(c)}.
    \end{align*}
\end{enumerate}

\textit{\underline{Step 1:} Bound for $\mathbb{E}\lvert H_{n}^{(1)} \rvert^{2}.$}

By Jensen's inequality, \begin{align}
    \mathbb{E}\lvert h^{(1)}(X_{1}) \rvert^{2} = \mathbb{E}\lvert \mathbb{E}_{X_{2}|X_{1}}\left\{\psi(X_{1},X_{2})- V_{M} |X_{1} \right\} \rvert^{2}
      \leq \mathbb{E}\lvert \psi(X_{1},X_{2}) - V_{M} \rvert^{2} = \alpha < \infty,
      \label{jeneq}
\end{align}
where $\alpha = \mathbb{E}\lvert \psi(X_{1},X_{2}) - V_{M} \rvert^{2}$ is finite under \textit{(M1)}. 

The martingale inequality in \cite{dharmadhikari1968bounds} shows that 
if $\left\{\Xi_{n} \right\}$ is a forward martingale with $\Xi_{0} = 0,$ then 
\begin{align}
	\mathbb{E}\lvert \Xi_{n} \rvert^{2} \leq C \beta_{n} n,
	\label{mtgleeq}
\end{align}
where $C$ is a constant and $\beta_{n} = \frac{1}{n} \sum_{j=1}^{n} \mathbb{E}\lvert \Xi_{j} - \Xi_{j-1} \rvert^{2}.$

By setting $\Xi_{n} = nH_{n}^{(1)} = \sum_{i=1}^{n} h^{(1)}(X_{i})$ in \eqref{mtgleeq}, there exists a constant $C_{1} >0$ with  
\begin{align*}
    \mathbb{E}\lvert nH^{(1)}_{n} \rvert^{2} \leq C_{1}\mathbb{E}\lvert h^{(1)}(X_{1}) \rvert^{2}n \leq  C_{1}\alpha n,
\end{align*}
whence 
\begin{align}
    \mathbb{E}\lvert H^{(1)}_{n} \rvert^{2} \leq  C_{1}\alpha n^{-1}.
    \label{suppeq1}
\end{align}

\textit{\underline{Step 2:} Bound for $\mathbb{E}\lvert H_{n}^{(2)} \rvert^{2}$}

Since 
\begin{align*}
    \binom{n}{2}H_{n}^{(2)} = \sum_{j=2}^{n} \sum_{i=1}^{j-1} h^{(2)}(X_{i},X_{j}) = \sum_{j=2}^{n} S_{j},
\end{align*}
where $S_{j} = \sum_{i=1}^{j-1} h^{(2)}(X_{i},X_{j}),$ $j=2,\ldots,n,$   
applying the martingale inequality in \eqref{mtgleeq} to  $\left\{\binom{n}{2}H^{(2)}_{n} \right\}_{n=2}^{\infty}$ implies that there exists a $C_{2} >0$ such that \begin{align}
    \mathbb{E}\bigg\lvert \binom{n}{2}H^{(2)}_{n} \bigg\rvert^{2} \leq C_{2}\left(\max\limits_{2 \leq j\leq n}\mathbb{E}\lvert S_{j} \rvert^{2}\right)n.
    \label{suppeq2}
\end{align}

For a fixed $j \geq 2,$ and $k=1,\ldots,j-1,$ we introduce $T_{kj} = \sum\limits_{i=1}^{k} h^{(2)}(X_{i},X_{j})$,  
which is a martingale adapted to $\sigma(X_{1},\ldots, X_{k},X_{j}).$  
By applying the martingale inequality in \eqref{mtgleeq} to $\{T_{kj} \},$ there exists  a global constant $C_{3} >0$  such that for $1 \le j \le n,$ 
\begin{align*}
\mathbb{E}\lvert S_{j} \rvert^{2} = \mathbb{E}\lvert T_{j-1,j} \rvert^{2} \leq C_{3} \frac{1}{j-1} \sum_{k=1}^{j-1} \mathbb{E}\lvert h^{(2)}(X_{k}, X_{j}) \rvert^{2} j = C_{3} \mathbb{E}\lvert h^{(2)}(X_{1}, X_{2}) \rvert^{2} j,
\end{align*} 
\bco
\begin{align*}
\mathbb{E}\lvert S_{j} \rvert^{2} 
    \leq C_{3} \mathbb{E}\lvert h^{(2)}(X_{1},X_{2}) \rvert^{2} j, \quad \text{for }  j=2,\ldots,n.
    \end{align*}
\fi
whence \begin{align}    
    \max_{2 \leq j \leq n}\mathbb{E}\lvert S_{j} \rvert^{2} \leq  C_{3} \mathbb{E}\lvert h^{(2)}(X_{1},X_{2}) \rvert^{2} n,
    \label{suppeq3}
\end{align}
and by the 
Minkowski inequality and \eqref{jeneq}  there exists a constant $C_{4} >0$ such that 
\begin{align}
    \begin{split}
        \mathbb{E}\lvert h^{(2)}(X_{1},X_{2}) \rvert^{2} &= \mathbb{E}\lvert \psi(X_{1},X_{2})-V_{M} - h^{(1)}(X_{1}) - h^{(1)}(X_{2}) \rvert^{2} \\
        & \leq C_{4} \left(\mathbb{E}\lvert \psi(X_{1},X_{2})-V_{M}\rvert^{2} + \mathbb{E}\lvert h^{(1)}(X_{1})\rvert^{2} + \mathbb{E}\lvert h^{(1)}(X_{2})\rvert^{2} \right) \\
        & \leq C_{4}( 3\alpha).
    \end{split}
    \label{suppeq4}
\end{align}

Combining \eqref{suppeq2}, \eqref{suppeq3}  and \eqref{suppeq4}, there exist a $C_{5} >0$ satisfying \begin{align}
    \mathbb{E}\lvert H_{n}^{(2)}\rvert^{2} \leq C_{5}\alpha n^{-2}.
    \label{suppeq5}
\end{align}
By \eqref{suppeq1} and \eqref{suppeq5}, we have $\mathbb{E}\lvert H_{n}^{(1)}\rvert^{2} = O(n^{-1})$ and $\mathbb{E}\lvert H_{n}^{(2)}\rvert^{2} = O(n^{-2}).$ By applying Minkowski inequality, $\mathbb{E}\lvert \hat{V}_{M} - V_{M} \rvert^{2} \leq \mathbb{E}\lvert  2 H_{n}^{(1)}\rvert^{2} + \mathbb{E}\lvert H_{n}^{(2)}\rvert^{2}.$ 
Thus, 
\begin{align*}
    \mathbb{E}\lvert \hat{V}_{M} - V_{M} \rvert^{2} = O(n^{-1}).
\end{align*} \qed

\no\textit{\underline{Proof of Theorem 1}}

We employ the same notation for the kernel functions $\psi : \mathcal{M} \times \mathcal{M} \rightarrow \mathbb{R},$ $h^{(1)} : \mathcal{M} \rightarrow \mathbb{R},$ and $h^{(2)} : \mathcal{M} \times \mathcal{M} \rightarrow \mathbb{R}$ as defined in \eqref{notpsi}, \eqref{noth} and for the H-decomposition terms $H_{n}^{(1)}$ and $H_{n}^{(2)}$ in \eqref{notH}.\\~

\textit{\underline{Step 1:} Sample metric variance $\hat{V}_{M}$}

Using the decomposition in \eqref{Hdecomp},
\begin{align*}
    \sqrt{n}(\hat{V}_{M}-V_{M}) = \sqrt{n}(2 H_{n}^{(1)}+ H_{n}^{(2)}) = \frac{2}{\sqrt{n}}\sum_{i=1}^{n}h^{(1)}(X_{i}) + \sqrt{n}H_{n}^{(2)}.
\end{align*}
The inequality \eqref{suppeq5} in the proof of Proposition 1 implies 
\begin{align}
    \text{Var}(\sqrt{n}H_{n}^{(2)}) = O(n^{-1}).
    \label{suppeq6}
\end{align}
Since $E(\sqrt{n}H_{n}^{(2)}) =0$ according to Step 3 of the proof of Proposition 1, it follows that \begin{align*}
    \sqrt{n}H_{n}^{(2)} = o_{P}(1).
\end{align*} 
Thus,
\begin{align*}
	\sqrt{n}(\hat{V}_{M}-V_{M}) = \sqrt{n} \left( \frac{1}{n} \sum_{i=1}^{n} 2h^{(1)}(X_{i}) \right)+ o_{p}(1).
\end{align*}

Note that $\mathbb{E}\left(2h^{(1)}(X_{1})\right) = 0.$ Applying the CLT for i.i.d. random variables and Slutsky's Theorem leads to
\begin{align}
	\sqrt{n}(\hat{V}_{M}-V_{M}) \rightarrow N\left(0, \sigma^{2}_{M} \right) \quad \textit{in distribution,}
	\label{VmCLT}
\end{align}
where $\sigma^{2}_{M} = \var\left(2h^{(1)}(X)\right) = \var_{X}\left[\mathbb{E}_{X'|X}\left\{d^{2}\left(X,X'\right)|X\right\}\right].$ \\~

\textit{\underline{Step 2:} Sample \F variance $\hat{V}_{F}$}

Following the proof of Theorem 1 in  \cite{dubey2019frechet},

\begin{align*}
\begin{split}
    \sqrt{n}(\hat{V}_{F} - V_{F}) &= \sqrt{n}\left[\frac{1}{n}\sum_{i=1}^{n}\left\{d^{2}(\hat{\mu}_{\oplus},X_{i}) - d^{2}(\mu_{\oplus},X_{i}) \right\}\right]\\
    &\qquad+ \sqrt{n}\left[\frac{1}{n}\sum_{i=1}^{n}\left\{d^{2}(\mu_{\oplus},X_{i}) - \mathbb{E}d^{2}(\mu_{\oplus},X) \right\}\right].
\end{split}
\end{align*}

Under \textit{(F0)-(F2)}, Proposition 1 in \cite{dubey2019frechet} shows that
\begin{align*}
    \sqrt{n}\frac{1}{n}\sum_{i=1}^{n}\left\{d^{2}(\hat{\mu}_{\oplus},X_{i}) - d^{2}(\mu_{\oplus},X_{i}) \right\} = o_{p}(1).
\end{align*}

Thus,
\begin{align}
	\sqrt{n}(\hat{V}_{F}-V_{F}) =  \sqrt{n}\left[\frac{1}{n}\sum_{i=1}^{n}\left\{d^{2}(\mu_{\oplus},X_{i}) - \mathbb{E}d^{2}(\mu_{\oplus},X) \right\}\right] + o_{p}(1).
\end{align}

Since $\mathbb{E}\left\{d^{2}(\mu_{\oplus},X_{1}) - \mathbb{E}d^{2}(\mu_{\oplus},X)\right\} = 0,$ applying the CLT for i.i.d. random variables and Slutsky's Theorem leads to
\begin{align}
	\sqrt{n}(\hat{V}_{F}-V_{F}) \rightarrow N\left(0, \sigma^{2}_{F} \right) \quad \textit{in distribution,}
	\label{VfCLT}
\end{align}
where $\sigma^{2}_{F} = \var\left(d^{2}(\mu_{\oplus},X)\right).$ \\~

\textit{\underline{Step 3:} Joint distribution of sample metric variance $\hat{V}_{M}$ and \F variance $\hat{V}_{F}$}

Following \eqref{VmCLT} in \textit{Step 1} and \eqref{VfCLT} in \textit{Step 2}, we aim at 
\begin{align*}
    &\sqrt{n}\left( \begin{pmatrix}
        \hat{V}_{M} \\
        \hat{V}_{F} 
    \end{pmatrix} - \begin{pmatrix}
        {V}_{M} \\
        {V}_{F} 
    \end{pmatrix}\right) \rightarrow N\left(\mathbf{0},\Sigma\right) \quad \textit{in distribution,}
\end{align*}
where $\Sigma = \begin{pmatrix}
            \sigma^{2}_{M} & \sigma_{FM}\\
            \sigma_{FM} & \sigma^{2}_{F}
        \end{pmatrix},$ and $\sigma_{FM}  = \cov\left( \mathbb{E}_{X'|X}\left\{d^{2}\left(X,X'\right)|X\right\}, d^{2}(\mu_{\oplus},X)\right)$.
        
Applying the Cram\'er-Wold device, it is enough to show that for all $\textbf{t} = (t_{1}, t_{2})^{T} \in \mathbb{R}^{2},$
\begin{align*}
    \sqrt{n}\left\{t_{1}\left(\hat{V}_{M} -  V_{M}\right) +  t_{2}\left(\hat{V}_{F} -  V_{F}\right) \right\} \rightarrow N\left(\mathbf{0},\textbf{t}^{T} \Sigma\textbf{t}\right) \quad \textit{in distribution.}
\end{align*}

For all $ (t_{1}, t_{2})^{T} \in \mathbb{R}^{2}, \quad t_1^2 + t_2^2>0$,
\begin{align*}
\begin{split}
    &\sqrt{n}\left\{t_{1}\left(\hat{V}_{M} -  V_{M}\right) +  t_{2}\left(\hat{V}_{F} -  V_{F}\right) \right\} \\
     &\quad =  \frac{1}{n}\sum_{i=1}^{n} \left[ t_{1} \left\{2h^{(1)}(X_{i})\right\} + t_{2}\left\{d^{2}(\mu_{\oplus},X_{i}) - \mathbb{E}d^{2}(\mu_{\oplus},X) \right\} \right] + o_{p}(1).
\end{split}
\end{align*}

By the CLT for i.i.d. random variables and Slutsky's Theorem,
\begin{align*}
    \sqrt{n}\left\{t_{1}\left(\hat{V}_{M} -  V_{M}\right) +  t_{2}\left(\hat{V}_{F} -  V_{F}\right) \right\} \rightarrow N\left(0, \lambda^{2}\right),
\end{align*}
where 
\begin{align*}
	\lambda^{2} = \var\left( t_{1} \left\{2h^{(1)}(X_{i})\right\} + t_{2}d^{2}(\mu_{\oplus},X_{i}) \right) = t^{2}_{1}\sigma^{2}_{M} + t^{2}_{2}\sigma^{2}_{F} + 2t_{1}t_{2} \sigma_{FM}=  \textbf{t}^{T} \Sigma\textbf{t}.
\end{align*}\\

\no\textit{\underline{Proof of Proposition 2}}

We employ the same notation for the kernel functions $\psi : \mathcal{M} \times \mathcal{M} \rightarrow \mathbb{R},$ $h^{(1)} : \mathcal{M} \rightarrow \mathbb{R},$ and $h^{(2)} : \mathcal{M} \times \mathcal{M} \rightarrow \mathbb{R}$ as defined in \eqref{notpsi}, \eqref{noth} and for the H-decomposition terms $H_{n}^{(1)}$ and $H_{n}^{(2)}$ in \eqref{notH}.

Using the decomposition in \eqref{Hdecomp},
\begin{align*}
    \sqrt{n}(\hat{V}_{M} - V_{M})/\sigma_{M} = 2\sqrt{n} H_{n}^{(1)}/\sigma_{M}+ \sqrt{n}H_{n}^{(2)}/\sigma_{M} := \mathcal{H}^{(1)} + \mathcal{H}^{(2)},
\end{align*}
where $\mathcal{H}^{(1)} = 2\sqrt{n} H_{n}^{(1)}/\sigma_{M}$ and $\mathcal{H}^{(2)} = \sqrt{n}H_{n}^{(2)}/\sigma_{M}.$

Under \textit{(M0)-(M3)}, Theorem 1 of Section 3.3.2. in \cite{lee2019u} leads to 
\begin{align*}
\begin{split}
 &\sup_{x}\lvert\mathbb{P}\left(\sqrt{n}(\hat{V}_{M}-V_{M})/\sigma_{M} \leq x \right) - \Phi(x)\rvert \\
  & \leq  \left(C_{1} \frac{ \mathbb{E}\lvert h^{(1)}(X_{1})\rvert^{3}}{\sigma^{3}_{M}} + C_{2} \frac{\mathbb{E}\lvert h^{(2)}(X_{1},X_{2})\rvert^{\frac{5}{3}}}{\sigma^{5/3}_{M}} + C_{3} \frac{\left(\mathbb{E}\lvert h^{(1)}(X_{1})\rvert^{3}\mathbb{E}\lvert h^{(2)}(X_{1},X_{2})\rvert^{\frac{3}{2}}\right)^{\frac{2}{3}}}{\sigma^{3}_{M}}  \right) n^{-\frac{1}{2}},
\end{split}
\end{align*}
for some positive constants $C_{1}, C_{2},$ and $C_{3}.$ Under \textit{(M0)-(M3)}, the terms $ \mathbb{E}\lvert h^{(1)}(X_{1})\rvert^{3},$ $\mathbb{E}\lvert h^{(2)}(X_{1},X_{2})\rvert^{\frac{5}{3}}$ and $\mathbb{E}\lvert h^{(2)}(X_{1},X_{2})\rvert^{\frac{3}{2}}$ are finite. Thus,
\begin{align*}
    \sup_{x}\left|\mathbb{P}\left(\sqrt{n}(\hat{V}_{M}-V_{M})/\sigma_{M} \leq x \right) - \Phi(x)\right| = O(n^{-\frac{1}{2}}).
\end{align*}\qed 

\no\textit{\underline{Proof of Proposition 3}}

\no\textit{\underline{1) Proof of $\hat{\sigma}^{2}_{M} \rightarrow \sigma^{2}_{M}$ in probability}}

Note that
 \begin{align*}
 \begin{split}
	\hat{\sigma}^{2}_{M} &= \frac{1}{n}\sum\limits_{i=1}^{n}\left\{\frac{1}{n-1} \sum\limits_{j \neq i}^{n} d^{2}(X_{i},X_{j})\right\}^{2} - \left\{\frac{2}{n(n-1)}\sum\limits_{1 \leq i < j \leq n} d^{2}(X_{i},X_{j} )\right\}^{2},
\end{split}
\end{align*} and
 \begin{align*}
	\sigma^{2}_{M} = \mathbb{E} \left[\mathbb{E}_{X'|X}\left\{d^{2}\left(X,X'\right)|X\right\}\right]^{2} - \left(\mathbb{E}d^{2}(X,X')\right)^{2}.
\end{align*}

According to Theorem 1 and the continuous mapping theorem, the second term on the r.h.s. of  $\hat{\sigma}_{M}^{2}$ converges to the second term on the r.h.s. of  ${\sigma}_{M}^{2},$ given by 
\begin{align*}
	\left\{\frac{2}{n(n-1)}\sum\limits_{1 \leq i < j \leq n} d^{2}(X_{i},X_{j} )\right\}^{2} = (2 \hat{V}_{M})^{2} \rightarrow (2V_{M})^{2} = \left(\mathbb{E}d^{2}(X,X')\right)^{2} \quad \text{in probability.}
\end{align*}
Thus it remains to show the convergence of the first terms, 
\begin{align*}
	\frac{1}{n}\sum\limits_{i=1}^{n}\left\{\frac{1}{n-1} \sum\limits_{j \neq i}^{n} d^{2}(X_{i},X_{j})\right\}^{2} \rightarrow  \mathbb{E}_{X} \left[\mathbb{E}_{X'|X}\left\{d^{2}\left(X,X'\right)|X\right\}\right]^{2}  \quad \text{in probability}.
\end{align*}
We decompose 
\begin{align}
\begin{split}
    \frac{1}{n}\sum_{i=1}^{n}\left(\frac{1}{n-1} \sum_{j \neq i}^{n} d^{2}(X_{i},X_{j}) \right)^{2} - \mathbb{E}_{X}\left[\mathbb{E}_{X'|X}\left\{d^{2}(X,X')|X \right\} \right]^{2} = A_{n} + B_{n},
    \end{split}
    \label{asymv}
\end{align}
where $A_{n} =  \left[\frac{1}{n}\sum_{i=1}^{n}\left(\frac{1}{n-1} \sum_{j \neq i}^{n} d^{2}(X_{i},X_{j}) \right)^{2} - \frac{1}{n}\sum_{i=1}^{n}\left(\mathbb{E}_{X'|X_{i}}\left\{d^{2}(X_{i},X')|X_{i} \right\}\right)^{2} \right]$ and $B_{n} =  \left[\frac{1}{n}\sum_{i=1}^{n}\left(\mathbb{E}_{X'|X_{i}}\left\{d^{2}(X_{i},X')|X_{i} \right\}\right)^{2} - \mathbb{E}_{X}\left[\mathbb{E}_{X'|X}\left\{d^{2}(X,X')|X \right\} \right]^{2}\right].$ Here $X'$ is an i.i.d. copy of the   $X_{1},\ldots, X_{n}.$

Observe that the $A_{n}$ in \eqref{asymv} is
\begin{align}
\begin{split}
	A_{n} &=\frac{1}{n}\sum_{i=1}^{n}\left(\frac{1}{n-1} \sum_{j \neq i}^{n} d^{2}(X_{i},X_{j}) \right)^{2} - \frac{1}{n}\sum_{i=1}^{n}\left(\mathbb{E}_{X'|X_{i}}\left\{d^{2}(X_{i},X')|X_{i} \right\}\right)^{2}\\
	&\qquad =  \binom{n}{3}^{-1}\sum_{1\leq i < j < k \leq n} d^{2}(X_{i},X_{j}) d^{2}(X_{i},X_{k}) \\
	&\qquad\qquad -  \frac{1}{n}\sum_{i=1}^{n}\mathbb{E}_{X', X''|X_{i}}\left\{d^{2}(X_{i},X')d^{2}(X_{i},X'')|X_{i} \right\} + a_{n},
\end{split}
\label{subasymv2}
\end{align}
where $a_{n} = \frac{1}{n(n-1)^2}\sum_{i=1}^{n}\sum_{j \neq i} d^{4}(X_{i}, X_{j}) - \frac{1}{(n-1)}  \binom{n}{3}^{-1}\sum_{1\leq i < j < k \leq n} d^{2}(X_{i},X_{j}) d^{2}(X_{i},X_{k}).$

By the consistency of U-statistics and the WLLN one has 
\begin{align*}
\begin{split}
     &\binom{n}{3}^{-1}\sum_{1\leq i < j < k \leq n} d^{2}(X_{i},X_{j}) d^{2}(X_{i},X_{k})  -  \frac{1}{n}\sum_{i=1}^{n}\mathbb{E}_{X', X''|X_{i}}\left\{d^{2}(X_{i},X')d^{2}(X_{i},X'')|X_{i} \right\}\\
    &\quad=\left[\binom{n}{3}^{-1}\sum_{1\leq i < j < k \leq n} d^{2}(X_{i},X_{j}) d^{2}(X_{i},X_{k}) - \mathbb{E}\left(d^{2}(X,X')d^{2}(X,X'') \right)\right]\\
    & + \left[\mathbb{E}\left(d^{2}(X,X')d^{2}(X,X'') \right) -  \frac{1}{n}\sum_{i=1}^{n}\mathbb{E}_{X', X''|X_{i}}\left\{d^{2}(X_{i},X')d^{2}(X_{i},X'')|X_{i} \right\}\right] = o_{p}(1). 
\end{split}
\end{align*}

Also, under \textit{(M0)},
\begin{align*}
	&a_{n} = \frac{1}{n(n-1)^2}\sum_{i=1}^{n}\sum_{j \neq i} d^{4}(X_{i}, X_{j})\\
	 &\qquad\qquad- \frac{1}{(n-1)}  \binom{n}{3}^{-1}\sum_{1\leq i < j < k \leq n} d^{2}(X_{i},X_{j}) d^{2}(X_{i},X_{k})= O(n^{-1})  = o_{p}(1).
\end{align*}

\bco
Also, by the consistency of U-statistics and WLLN, we have
\begin{align*}
\begin{split}
     &\binom{n}{3}^{-1}\sum_{1\leq i < j < k \leq n} d^{2}(X_{i},X_{j}) d^{2}(X_{i},X_{k})  -  \frac{1}{n}\sum_{i=1}^{n}\mathbb{E}_{X', X''|X_{i}}\left\{d^{2}(X_{i},X')d^{2}(X_{i},X'')|X_{i} \right\}\\
    &\quad=\left[\binom{n}{3}^{-1}\sum_{1\leq i < j < k \leq n} d^{2}(X_{i},X_{j}) d^{2}(X_{i},X_{k}) - \mathbb{E}\left(d^{2}(X,X')d^{2}(X,X'') \right)\right]\\
    & + \left[\mathbb{E}\left(d^{2}(X,X')d^{2}(X,X'') \right) -  \frac{1}{n}\sum_{i=1}^{n}\mathbb{E}_{X', X''|X_{i}}\left\{d^{2}(X_{i},X')d^{2}(X_{i},X'')|X_{i} \right\}\right] = o_{p}(1). 
\end{split}
\end{align*}
\fi

Then \eqref{subasymv2} becomes
\begin{align*}
\begin{split}
    A_{n} = \left[\frac{1}{n}\sum_{i=1}^{n}\left(\frac{1}{n-1} \sum_{j \neq i}^{n} d^{2}(X_{i},X_{j}) \right)^{2} - \frac{1}{n}\sum_{i=1}^{n}\left(\mathbb{E}_{X'|X_{i}}\left\{d^{2}(X_{i},X')|X_{i} \right\}\right)^{2} \right] = o_{p}(1)
\end{split}
\end{align*}

Applying the weak law of large number (WLLN)  gives for the $B_{n}$ in \eqref{asymv} 
\begin{align}
    B_{n} = \left[\frac{1}{n}\sum_{i=1}^{n}\left(\mathbb{E}_{X'|X_{i}}\left\{d^{2}(X_{i},X')|X_{i} \right\}\right)^{2} - \mathbb{E}_{X}\left[\mathbb{E}_{X'|X}\left\{d^{2}(X,X')|X \right\} \right]^{2}\right] = o_{p}(1).
    \label{subasymv1}
\end{align}

Combining \eqref{subasymv2} and \eqref{subasymv1},  \eqref{asymv} becomes 
\begin{align*}
	\frac{1}{n}\sum\limits_{i=1}^{n}\left\{\frac{1}{n-1} \sum\limits_{j \neq i}^{n} d^{2}(X_{i},X_{j})\right\}^{2} -  \mathbb{E} \left[\mathbb{E}_{X'|X}\left\{d^{2}\left(X,X'\right)|X\right\}\right]^{2} = A_{n} + B_{n} = o_{p}(1).
\end{align*}

Thus, $\hat{\sigma}^{2}_{M} \rightarrow \sigma^{2}_{M}$ in probability. \qed

\vspace{2cm} 
\no\textit{\underline{2) Proof of $\hat{\sigma}^{2}_{F} \rightarrow \sigma^{2}_{F}$ in probability}}

The consistency of $\hat{\sigma}^{2}_{F}$ was established in Proposition 2 of \cite{dubey2019frechet}. \qed \vspace{2in} 

\no\textit{\underline{3) Proof of $\hat{\sigma}_{FM} \rightarrow \sigma_{FM}$ in probability}}

Note that
\begin{align*}
	\hat{\sigma}_{FM} &= \frac{1}{n}\sum\limits_{i=1}^{n} \left[\frac{d^{2}(\hat{\mu}_{\oplus},X_{i})}{n-1} \{\sum\limits_{j \neq i}^{n} d^{2}(X_{i},X_{j})\} \right] \\
	&\qquad - \left\{\frac{1}{n}\sum\limits_{i=1}^{n}d^{2}(\hat{\mu}_{\oplus},X_{i})\right\} \left\{\frac{2}{n(n-1)}\sum\limits_{1 \leq i < j \leq n} d^{2}(X_{i},X_{j} )\right\}
\end{align*}
and
\begin{align*}
	\sigma_{FM} = \mathbb{E}\{d^{2}(\mu_{\oplus}, X)d^{2}(X,X')\}  - \mathbb{E}d^{2}(\mu_{\oplus},X)\mathbb{E}d^{2}(X,X').
\end{align*}
By the consistency of $\hat{V}_{F}$ and $\hat{V}_{M},$ the second terms converge in probability, 
\begin{align}
	\left\{\frac{1}{n}\sum\limits_{i=1}^{n}d^{2}(\hat{\mu}_{\oplus},X_{i})\right\} \left\{\frac{2}{n(n-1)}\sum\limits_{1 \leq i < j \leq n} d^{2}(X_{i},X_{j} )\right\} - \mathbb{E}d^{2}(\mu_{\oplus},X)\mathbb{E}d^{2}(X,X') = o_{p}(1).
	\label{prop3second}
\end{align}
Thus, it remains to show the convergence in probability of the first terms, given by
\begin{align}
\begin{split}
	 &\frac{1}{n}\sum\limits_{i=1}^{n} \left[\frac{d^{2}(\hat{\mu}_{\oplus},X_{i})}{n-1} \{\sum\limits_{j \neq i}^{n} d^{2}(X_{i},X_{j})\} \right] -  \mathbb{E}\{d^{2}(\mu_{\oplus}, X)d^{2}(X,X')\} \\
	 &\qquad = C_{n} + D_{n} =  o_{p}(1),
\end{split}
	 \label{prop3first}
\end{align}
where $C_{n} = \frac{1}{n}\sum\limits_{i=1}^{n} \left[\frac{d^{2}(\hat{\mu}_{\oplus},X_{i})}{n-1} \{\sum\limits_{j \neq i}^{n} d^{2}(X_{i},X_{j})\} \right] - \frac{1}{n}\sum\limits_{i=1}^{n} \left[\frac{d^{2}(\mu_{\oplus},X_{i})}{n-1} \{\sum\limits_{j \neq i}^{n} d^{2}(X_{i},X_{j})\} \right],$ and $D_{n} = \frac{1}{n}\sum\limits_{i=1}^{n} \left[\frac{d^{2}(\mu_{\oplus},X_{i})}{n-1} \{\sum\limits_{j \neq i}^{n} d^{2}(X_{i},X_{j})\} \right] - \mathbb{E}\{d^{2}(\mu_{\oplus}, X)d^{2}(X,X')\}.$

Under \textit{(F1)}, Theorem 1 of \cite{pete:19} leads to  $d(\hat{\mu}_{\oplus}, \mu_{\oplus}) = o_{p}(1),$ whence 
\begin{align*}
	 \lvert C_{n} \rvert &= \Bigg\lvert\frac{1}{n}\sum\limits_{i=1}^{n} \left[\frac{d^{2}(\hat{\mu}_{\oplus},X_{i})}{n-1} \{\sum\limits_{j \neq i}^{n} d^{2}(X_{i},X_{j})\} \right] - \frac{1}{n}\sum\limits_{i=1}^{n} \left[\frac{d^{2}(\mu_{\oplus},X_{i})}{n-1} \{\sum\limits_{j \neq i}^{n} d^{2}(X_{i},X_{j})\} \right]\Bigg\rvert \\
	 &= \Bigg\lvert  \frac{1}{n}\sum\limits_{i=1}^{n} \left[\frac{d^{2}(\hat{\mu}_{\oplus},X_{i})-d^{2}(\mu_{\oplus},X_{i})}{n-1} \{\sum\limits_{j \neq i}^{n} d^{2}(X_{i},X_{j})\} \right]  \Bigg\rvert \\
	 & \leq 2 \text{diam}(\mathcal{M}) \frac{d(\hat{\mu}_{\oplus}, \mu_{\oplus})  }{n(n-1)}\sum\limits_{i=1}^{n} \sum\limits_{j \neq i}^{n} d^{2}(X_{i},X_{j})
	     = o_{p}(1),
\end{align*}
because $\frac{1}{n(n-1)}\sum\limits_{i=1}^{n} \sum\limits_{j \neq i}^{n} d^{2}(X_{i},X_{j}) = O(1)$ under \textit{(F0)}.
Also, the first term of $D_{n}$ in \eqref{prop3first} is represented as
\begin{align*}
	 &\frac{1}{n}\sum\limits_{i=1}^{n} \left[\frac{d^{2}(\mu_{\oplus},X_{i})}{n-1} \{\sum\limits_{j \neq i}^{n} d^{2}(X_{i},X_{j})\} \right]\\
	 &\quad =  \frac{1}{n(n-1)}\sum\limits_{i=1}^{n}\sum\limits_{j \neq i}^{n} \left[\left\{\frac{d^{2}(\mu_{\oplus},X_{i}) + d^{2}(\mu_{\oplus},X_{j})}{2} \right\} d^{2}(X_{i},X_{j})\right],
\end{align*}
which is the U-statistic with the kernel $\phi(x,y) = \left\{ \frac{d^{2}(\mu_{\oplus},x)  + d^{2}(\mu_{\oplus},y) }{2} \right\}d^{2}(x,y).$ 
The mean of this  U-statistic is 
\begin{align*}
	\mathbb{E}\left\{\frac{d^{2}(\mu_{\oplus}, X)}{2}d^{2}(X,X')\right\} +  \mathbb{E}\left\{\frac{d^{2}(\mu_{\oplus}, X')}{2}d^{2}(X,X')\right\} =   \mathbb{E}\{d^{2}(\mu_{\oplus}, X)d^{2}(X,X')\} .
\end{align*}
Then by the consistency of U-statistics,
\begin{align*}
	 D_{n} = \frac{1}{n}\sum\limits_{i=1}^{n} \left[\frac{d^{2}(\mu_{\oplus},X_{i})}{n-1} \{\sum\limits_{j \neq i}^{n} d^{2}(X_{i},X_{j})\} \right] - \mathbb{E}\{d^{2}(X,X')d^{2}(\mu_{\oplus}, X)\} = o_{p}(1).
\end{align*}

Thus, by  combining \eqref{prop3second} and \eqref{prop3first},  $\hat{\sigma}^{2}_{FM} \rightarrow \sigma^{2}_{FM}$ in probability. \qed \\~

\no\textit{\underline{Lemma 1}  \citep{aleksandrov1951theorem}}
\textit{Let $\kappa$ be a real number. For all vertices $a, b, c$ on the space $(\mathcal{M}, d),$ satisfying $d(a,b) + d(a,c) + d(b,c) < 2 D_{\kappa},$ where $D_{\kappa} = \infty$ for $\kappa \leq 0,$ and $D_{\kappa} = \frac{\pi}{\sqrt{\kappa}}$ for $\kappa >0,$ there exists points $\tilde{a},\tilde{b},$ and $\tilde{c}$ on model space $(\M_{\kappa}, d_{\kappa})$ such that $d(a,b) = d_{\kappa}(\tilde{a}, \tilde{b}),$ $d(b,c) = d_{\kappa}(\tilde{b}, \tilde{c}),$ and $d(a,c) = d_{\kappa}(\tilde{a}, \tilde{c}).$} \\

\noindent  For a more detailed explanation, see Lemma 2.14 in Chapter I.2 of \cite{bridson2013metric}. \\

\no\textit{\underline{Proof of Theorem 2}}

We first define an (Alexandrov) angle between two geodesics $\gamma_{1}$ and $\gamma_{2}$ emanating from $a$ in a geodesic space $\left(\mathcal{M},d\right)$ as
\begin{align*}
	\angle_{a}(\gamma_{1},\gamma_{2}) = \limsup_{s,t \rightarrow 0} \bar{\angle}_{a}\left(\gamma_{1}(s), \gamma_{2}(t) \right),
\end{align*}
where the comparison angle between $b$ and $c$ at $a$ is $\bar{\angle}_{a}\left(b,c\right) = \arccos \frac{d^{2}\left(a,b\right)+d^{2}\left(a,c\right)-d^{2}\left(b,c\right)}{2d\left(a,b\right)d\left(a,c\right)}.$  For a uniquely geodesic space $\left(\mathcal{M},d\right)$, $\angle_{a}\left(\gamma_{1},\gamma_{2}\right)$ does not depend on the lengths of $\gamma_{1}$ or $\gamma_{2}.$ For $a,b,c \in \mathcal{M},$ we can define the angle $\angle_{a}\left(b,c\right) = \angle_{a}\left(\gamma_{ab},\gamma_{ac}\right).$

By Lemma 1 and properties of bounded curvature spaces (see, e.g., Proposition 1.7 in Chapter II.1 of \cite{bridson2013metric} or Section 2 in \cite{lin2021total}), if the curvature of $(\M, d)$ is upper (lower, respectively) bounded by $\kappa,$ for vertices $a, b, c \in (\M, d)$ satisfying $d(a,b) + d(a,c) + d(b,c) < 2D_{\kappa},$ there exist points $\tilde{a}, \tilde{b}, \tilde{c} \in (\M_{\kappa}, d_{\kappa})$ such that $d(a,b) = d_{\kappa}(\tilde{a}, \tilde{b}),$ $d(a,c) = d_{\kappa}(\tilde{a}, \tilde{c}),$ and $\angle_{a}(b,c) = \angle_{\tilde{a}}(\tilde{b},\tilde{c}),$ and $d(b,c) \geq d_{\kappa}(\tilde{b}, \tilde{c})$ ($d(b,c) \leq d_{\kappa}(\tilde{b}, \tilde{c})$, respectively). Also, under the conditions of Theorem 2, the metric space $\mathcal{M}$ is a unique geodesic space \citep{stur:03} and has curvature upper bounded by $0.$ 
       
Then for $a = \mu_{\oplus},$ $b = X,$ and $c = X',$ we have 
\begin{align*}
	d(X, X') \geq d_{0}(\tilde{X}, \tilde{X'}),
\end{align*}
where $\tilde{\mu}_{\oplus}, \tilde{X}, \tilde{X'} \in (\M_{0}, d_{0}),$ form a comparison triangle on the Euclidean space such that $d(\mu_{\oplus}, X) = d_{0}(\tilde{\mu}_{\oplus}, \tilde{X}),$ $d(\mu_{\oplus}, X') = d_{0}(\tilde{\mu}_{\oplus}, \tilde{X'}),$ and $\angle_{\mu_{\oplus}}(X,X') = \angle_{\tilde{\mu}_{\oplus}}(\tilde{X},\tilde{X'}).$ When $\M$ is a curved space, one may have $\mathbb{E}\tilde{X}  \neq \tilde{\mu}_{\oplus}.$

By the law of cosines on the Euclidean space, 
\begin{align}
\begin{split}
	d^{2}(X, X') &\geq d^{2}_{0}(\tilde{X}, \tilde{X'}) \\
	& = d^{2}_{0}(\tilde{\mu}_{\oplus}, \tilde{X}) + d^{2}_{0}(\tilde{\mu}_{\oplus}, \tilde{X'}) - 2 d_{0}(\tilde{\mu}_{\oplus}, \tilde{X})d_{0}(\tilde{\mu}_{\oplus}, \tilde{X'})\cos\angle_{\tilde{\mu}_{\oplus}}(\tilde{X},\tilde{X'}) \\
	&=  d^{2}(\mu_{\oplus}, X) + d^{2}(\mu_{\oplus}, X') - 2d(\mu_{\oplus}, X)d(\mu_{\oplus}, X')\cos\angle_{\mu_{\oplus}}(X,X').
\end{split}
\label{negcosine}
\end{align}
The expectation of the cross-product term in \eqref{negcosine} is given by
\begin{align}
\begin{split}
	&\mathbb{E}\left[d(\mu_{\oplus},X)d(\mu_{\oplus},X')\cos_{\mu_{\oplus}}(X,X')\right] \\
	&\qquad = \mathbb{E}_{X'}\left[d(\mu_{\oplus},X') \mathbb{E}_{X|X'}\left\{d(\mu_{\oplus},X)\cos\angle_{\mu_{\oplus}}(X,X') \right\}   \right].
\end{split}
\label{neglemma}
\end{align}
By Lemma S.7. in \cite{lin2021total}, \begin{align*}
    \mathbb{E}\left\{d(\mu_{\oplus},X)\cos \angle_{\mu_{\oplus}}(X, \xi) \right\} \leq 0
\end{align*} for $\xi \in \mathcal{M}.$ Then since $d(\cdot, \cdot) \geq 0,$ for the term in  \eqref{neglemma} 
\begin{align}
	\mathbb{E}\left[d(\mu_{\oplus},X)d(\mu_{\oplus},X')\cos_{\mu_{\oplus}}(X,X')\right]  \leq 0.
	\label{neglemmares}
\end{align}
Taking expectations in \eqref{negcosine},
\begin{align*}
    \begin{split}
        2V_{M} &= \mathbb{E}d^{2}(X,X') \\
        &\geq  \mathbb{E}d^{2}(\mu_{\oplus}, X) +  \mathbb{E}d^{2}(\mu_{\oplus}, X') - 2 \mathbb{E}d(\mu_{\oplus}, X)d(\mu_{\oplus}, X')\cos\angle_{\mu_{\oplus}}(X,X') \\
        &= 2V_{F} - 2 \mathbb{E}d(\mu_{\oplus}, X)d(\mu_{\oplus}, X')\cos\angle_{\mu_{\oplus}}(X,X') \\
        &\geq 2V_{F}.
    \end{split}
\end{align*}
The last inequality is implied by \eqref{neglemmares}.
Thus, $V_{M} \geq V_{F}.$

If $(\mathcal{M},d)$ is a space with \textit{strictly negative curvature} satisfying the conditions of Theorem 2, we can find $\delta < 0$ such that $(\mathcal{M},d)$ is upper bounded by $\delta.$ Then for $a = \mu_{\oplus},$ $b = X$ and $c = X',$  
\begin{align*}
	d(X, X') \geq d_{\delta}(\check{X}, \check{X'}) > d_{0}(\tilde{X}, \tilde{X'}),
\end{align*}
where $\check{\mu}_{\oplus}, \check{X}, \check{X'} \in (\M_{\delta}, d_{\delta})$ and $\tilde{\mu}_{\oplus}, \tilde{X}, \tilde{X'} \in (\M_{0}, d_{0})$ such that $d(\mu_{\oplus}, X)= d_{\delta}(\check{\mu}_{\oplus}, \check{X}) = d_{0}(\tilde{\mu}_{\oplus}, \tilde{X}),$ $d(\mu_{\oplus}, X')= d_{\delta}(\check{\mu}_{\oplus}, \check{X'}) = d_{0}(\tilde{\mu}_{\oplus}, \tilde{X'})$ and  \newline $\angle_{\mu_{\oplus}}(X,X') = \angle_{\check{\mu}_{\oplus}}(\check{X},\check{X'})= \angle_{\tilde{\mu}_{\oplus}}(\tilde{X},\tilde{X'}).$
Following the same argument as above, 
\begin{align*}
    d^{2}(X,X') > d^{2}(\mu_{\oplus},X)+ d^{2}(\mu_{\oplus},X') -2d(\mu_{\oplus},X)d(\mu_{\oplus},X')\cos\angle_{\mu_{\oplus}}(X,X') .
\end{align*}
Finally, under the condition \textit{(S1)},
\begin{align*}
    2V_{M} >  2V_{F} -2\mathbb{E}_{X'}\left[d(\mu_{\oplus},X')\mathbb{E}_{X|X'}\left\{d(\mu_{\oplus},X)\cos\angle_{\mu_{\oplus}}(X,X')\right\}\right] \geq 2V_{F}.
\end{align*} 
Thus, $V_{M} > V_{F}.$\qed \\

\no\textit{\underline{Proof of Theorem 3}}

We follow the proof of Theorem 2. Under the conditions of Theorem 3, the unique geodesic metric space $\mathcal{M}$ has curvature lower bounded by $0.$ Then by Lemma 1 and properties of bounded curvature space (e.g., Proposition 1.7 in Chapter II.1 of \cite{bridson2013metric} or Section 2 in \cite{lin2021total}), for $a = \mu_{\oplus},$ $b = X$ and $c = X',$ 
\begin{align*}
	d(X, X') \leq d_{0}(\tilde{X}, \tilde{X'}),
\end{align*}
where $\tilde{\mu}_{\oplus}, \tilde{X}, \tilde{X'} \in (\M_{0}, d_{0}),$ form a comparison triangle on the Euclidean space such that $d(\mu_{\oplus}, X) = d_{0}(\tilde{\mu}_{\oplus}, \tilde{X}),$ $d(\mu_{\oplus}, X') = d_{0}(\tilde{\mu}_{\oplus}, \tilde{X'}),$ and $\angle_{\mu_{\oplus}}(X,X') = \angle_{\tilde{\mu}_{\oplus}}(\tilde{X},\tilde{X'}).$ 

By the law of cosines on the Euclidean space, 
\begin{align}
\begin{split}
	d^{2}(X, X') &\leq d^{2}_{0}(\tilde{X}, \tilde{X'}) \\
	& = d^{2}_{0}(\tilde{\mu}_{\oplus}, \tilde{X}) + d^{2}_{0}(\tilde{\mu}_{\oplus}, \tilde{X'}) - 2 d_{0}(\tilde{\mu}_{\oplus}, \tilde{X})d_{0}(\tilde{\mu}_{\oplus}, \tilde{X'})\cos\angle_{\tilde{\mu}_{\oplus}}(\tilde{X},\tilde{X'}) \\
	&=  d^{2}(\mu_{\oplus}, X) + d^{2}(\mu_{\oplus}, X') - 2d(\mu_{\oplus}, X)d(\mu_{\oplus}, X')\cos\angle_{\mu_{\oplus}}(X,X').
\end{split}
\label{poscosine}
\end{align}
The expectation of the cross-product term in \eqref{poscosine} is given by
\begin{align}
\begin{split}
	&\mathbb{E}\left[d(\mu_{\oplus},X)d(\mu_{\oplus},X')\cos_{\mu_{\oplus}}(X,X')\right]\\
	& \quad= \mathbb{E} \left[d_{0}(\tilde{\mu}_{\oplus}, \tilde{X})d_{0}(\tilde{\mu}_{\oplus}, \tilde{X'})\cos\angle_{\tilde{\mu}_{\oplus}}(\tilde{X},\tilde{X'}) \right] =   \mathbb{E} \langle \tilde{X} - \tilde{\mu}_{\oplus}, \tilde{X'} - \tilde{\mu}_{\oplus}\rangle  \\
	& \quad=  \mathbb{E} \langle (\tilde{X} - \mathbb{E}\tilde{X})  + ( \mathbb{E}\tilde{X} - \tilde{\mu}_{\oplus} ), (\tilde{X'} - \mathbb{E}\tilde{X'})  +( \mathbb{E}\tilde{X'} - \tilde{\mu}_{\oplus}) \rangle\\
	&\quad = ( \mathbb{E}\tilde{X} - \tilde{\mu}_{\oplus} )^{2} \geq 0,
\end{split}
\label{poslemma}
\end{align}
where $\langle \cdot, \cdot \rangle$ is the inner product in Euclidean space $(\M_{0}, d_{0}).$
Taking expectations in \eqref{poscosine},
\begin{align*}
    \begin{split}
        2V_{M} &= \mathbb{E}d^{2}(X,X') \\
        &\leq  \mathbb{E}d^{2}(\mu_{\oplus}, X) +  \mathbb{E}d^{2}(\mu_{\oplus}, X') - 2 \mathbb{E}d(\mu_{\oplus}, X)d(\mu_{\oplus}, X')\cos\angle_{\mu_{\oplus}}(X,X') \\
        &= 2V_{F} - 2 \mathbb{E}d(\mu_{\oplus}, X)d(\mu_{\oplus}, X')\cos\angle_{\mu_{\oplus}}(X,X') \\
        &\leq 2V_{F}.
    \end{split}
\end{align*}
The last inequality is implied by \eqref{poslemma}.
Thus, $V_{M} \leq V_{F}.$

If $(\mathcal{M},d)$ is a \textit{strictly positive curvature} space satisfying the conditions of Theorem 3, we can find $\epsilon > 0$ such that $(\mathcal{M},d)$ is lower bounded by $\epsilon.$ Then for $a = \mu_{\oplus},$ $b = X$ and $c = X',$ 
\begin{align*}
	d(X, X') \leq d_{\epsilon}(\check{X}, \check{X'}) < d_{0}(\tilde{X}, \tilde{X'}),
\end{align*}
where $\check{\mu}_{\oplus}, \check{X}, \check{X'} \in (\M_{\epsilon}, d_{\epsilon})$ and $\tilde{\mu}_{\oplus}, \tilde{X}, \tilde{X'} \in (\M_{0}, d_{0})$ such that $d(\mu_{\oplus}, X)= d_{\epsilon}(\check{\mu}_{\oplus}, \check{X}) = d_{0}(\tilde{\mu}_{\oplus}, \tilde{X}),$ $d(\mu_{\oplus}, X')= d_{\epsilon}(\check{\mu}_{\oplus}, \check{X'}) = d_{0}(\tilde{\mu}_{\oplus}, \tilde{X'})$ and \newline  $\angle_{\mu_{\oplus}}(X,X') = \angle_{\check{\mu}_{\oplus}}(\check{X},\check{X'})= \angle_{\tilde{\mu}_{\oplus}}(\tilde{X},\tilde{X'}).$
Following the same argument above, 
\begin{align*}
    d^{2}(X,X') < d^{2}(\mu_{\oplus},X)+ d^{2}(\mu_{\oplus},X') -2d(\mu_{\oplus},X)d(\mu_{\oplus},X')\cos\angle_{\mu_{\oplus}}(X,X') .
\end{align*}
Finally, under the condition \textit{(S1)},
\begin{align*}
    2V_{M} <  2V_{F} -2\mathbb{E}_{X'}\left[d(\mu_{\oplus},X')\mathbb{E}_{X|X'}\left\{d(\mu_{\oplus},X)\cos\angle_{\mu_{\oplus}}(X,X')\right\}\right] \leq 2V_{F}.
\end{align*} 
Thus, $V_{M} < V_{F}.$\qed \\

\bco
\textit{\underline{Proof of Proposition \ref{prop4}}} \\

The $1 \Rightarrow 4$ is straightforward. The equivalence $3 \Leftrightarrow 4$ was initially established by \cite{schoenberg1938metricpositive}. The equivalence $2 \Leftrightarrow 3$ was also discovered by \cite{schoenberg1938metricmonotone}. Finally, the $2 \Rightarrow 1$ was addressed in Theorem 1 of \cite{feragen2015geodesic}.\qed\\

\fi
\no\textit{\underline{Proof of Theorem 4}}

Observing
\begin{align*}
    \hat{\rho} = g(\hat{V}_{M}, \hat{V}_{F}),
\end{align*}
where $g(x_{1}, x_{2}) = \frac{x_{2}}{x_{1}} - 1$ is a differentiable function with gradient function $\nabla g= (-\frac{x_{2}}{x^{2}_{1}}, \frac{1}{x_{1}})^{T},$  a  simple application of the $\delta$-method establishes the asymptotic normality of $\hat{\rho},$ given by
\begin{align*}
    \sqrt{n}\left( \hat{\rho} - \rho \right) \rightarrow N\left(0,\sigma^{2}\right) \quad \text{in distribution,}
\end{align*}
where the asymptotic variance is $\sigma^{2} = a^{T}\Sigma a$ and $a$ is the gradient function $\nabla g$ evaluated at $(V_{M}, V_{F}).$ 

Proposition 3 leads to a consistent estimator for $\sigma^{2},$ given by 
\begin{align*}
    \hat{\sigma}^{2} = \hat{a}^{T} \hat{\Sigma} \hat{a},
\end{align*}
where $\hat{a} = \left(-\frac{\hat{V}_{F}}{\hat{V}^{2}_{M}}, \frac{1}{\hat{V}_{M}}\right)^{T}$ and the estimated asymptotic covariance matrix is  $$\hat{\Sigma} = \begin{pmatrix}
        \hat{\sigma}^{2}_{M} & \hat{\sigma}_{MF}\\
        \hat{\sigma}_{FM} & \hat{\sigma}^{2}_{F}
    \end{pmatrix}.$$
By consistency of asymptotic variance $\hat{\sigma}^{2},$ we obtain the  asymptotic normality 
{\begin{equation*}
    \sqrt{n}(\hat{\rho} - \rho)/\hat{\sigma} \rightarrow N(0, 1) \quad\text{in distribution.}
\end{equation*} Thus, under $H_0: \rho = 0,$ $T_{n} \rightarrow N(0,1)$ in distribution.}
\qed\\

\no\textit{\underline{Proof of Theorem 5}}

Following the same argument in case of the sample \F mean $\hat{\mu}_{\oplus},$ under the conditions \textit{(A0)-(A4)}, the sample intrinsic \F mean $\hat{\mu}_{I,\oplus}$ is consistent for 
the population intrinsic \F mean $\mu_{I,\oplus}$ (Theorem 1 of \cite{pete:19}).
Under conditions \textit{(A0)-(A4)}, Theorem 1 leads to 
    \begin{align*}
    \sqrt{n}\left((\hat{V}_{I, M},\hat{V}_{I,F})^{T}
    - (V_{I,M}, V_{I,F})^{T} \right) \rightarrow N\left(\mathbf{0},\Sigma_{I}\right) \quad \text{in distribution},
\end{align*} where $\Sigma_{I} = \begin{pmatrix}
        \sigma^{2}_{I,M} & \sigma_{I,FM}\\
        \sigma_{I,FM} & \sigma^{2}_{I, F}
    \end{pmatrix}.$ 
Theorem 4 can then be directly applied under  conditions \textit{(A0)-(A4)}, {we have \begin{equation*}
    \sqrt{n}(\hat{\rho}_I - \rho_I)/\hat{\sigma}_I \rightarrow N(0, 1) \quad \text{in distribution. }
\end{equation*} Thus, under $H_0: \rho_I = 0,$ $T_{I, n} \rightarrow N(0,1)$ in distribution.}
\qed

\no\textit{\underline{Proof of Theorem S.\ref{thm:supp:conv}}} 

Following Corollary 2.1 of \cite{aaron2018convergence}, under conditions \textit{(B1)}-\textit{(B2)}, if the weighted graph in Algorithm 1 is constructed using a ball size $r = r_{n}$ such that 
    \begin{equation*}
        \left(A_{0}\frac{\log n}{n}\right)^{2/3k} \leq r_{n} \leq  \left(A_{1}\frac{\log n}{n}\right)^{2/3k},
    \end{equation*}
    for some $A_{0},$ $A_{1} > 0,$  it holds that  
    \begin{equation}
    \max\limits_{i,j}\bigg\lvert \hat{d}_{I}(X_i, X_j) - d_{I}(X_{i}, X_{j})\bigg\rvert = O\left(\left(\frac{\log n}{n}\right)^{2/3k}\right).
    \label{conv_dist}
\end{equation} 

By definition, 
\begin{equation*}
    \rho_{I} = \frac{V_{I,F}}{V_{I,M}} - 1,\quad \hat{\rho}_{I} = \frac{\hat{V}_{I,F}}{\hat{V}_{I,M}}-1,\quad \text{and}\quad\tilde{\rho}_{I,n} = \frac{\tilde{V}_{I,F}}{\tilde{V}_{I,M}} - 1,
\end{equation*}
where
$V_{I,F} = \mathbb{E}d_{I}^{2}(\mu_{I,\oplus}, X),$ $\hat{V}_{I,F} = \frac{1}{n}\sum\limits_{i=1}^{n} d^{2}_{I}(\hat{\mu}_{I, \oplus}, X_{i}),$ $\tilde{V}_{I,F} = \frac{1}{n}\sum\limits_{i=1}^{n} \hat{d}^{2}_{I}(\hat{\mu}_{I, \oplus}, X_{i}),$ \newline $V_{I,M} = \frac{1}{2}\mathbb{E}d_{I}^{2}(X, X'),$ $\hat{V}_{I,M} = \frac{1}{n(n-1)}\sum\limits_{1 \leq i < j \leq n} d^{2}_{I}(X_{i}, X_{j})$ and $\tilde{V}_{I,M} = \frac{1}{n(n-1)}\sum\limits_{1 \leq i < j \leq n} \hat{d}^{2}_{I}(X_{i}, X_{j}).$
In the following, the  notation $\lesssim$ refers to an inequality modulo a  multiplicative constant. We observe 
\begin{align*}
\begin{split}
    \lvert \tilde{V}_{I,F} - \hat{V}_{I,F} \rvert &\leq \frac{1}{n} \sum_{i=1}^{n} \lvert\hat{d}_{I}(\hat{\mu}_{I, \oplus}, X_{i}) - d_{I}(\hat{\mu}_{I, \oplus}, X_{i})\rvert \lvert\hat{d}_{I}(\hat{\mu}_{I, \oplus}, X_{i}) + d_{I}(\hat{\mu}_{I, \oplus}, X_{i})\rvert \\
    &\lesssim \frac{1}{n} \sum_{i=1}^{n} \lvert\hat{d}_{I}(\hat{\mu}_{I, \oplus}, X_{i}) - d_{I}(\hat{\mu}_{I, \oplus}, X_{i})\rvert = O\left(\left(\frac{\log n}{n}\right)^{2/3k}\right), 
\end{split}
\end{align*}
where the second $\lesssim $ follows from the boundedness of $\mathcal{A},$ the uniform convergence of \newline $\max\limits_{i,j}\bigg\lvert \hat{d}_{I}(X_i, X_j) - d_{I}(X_{i}, X_{j})\bigg\rvert$ and the fact that $\hat{\mu}_{\oplus} \in \mathcal{X}_{n} := \{X_{1}, \ldots, X_{n} \}$. The last equality is a consequence of \eqref{conv_dist}. Similarly,
\begin{align*}
\begin{split}
    \lvert \tilde{V}_{I,M} - \hat{V}_{I,M} \rvert &\leq \frac{1}{n(n-1)} \sum_{1\leq i< j \leq n} \lvert\hat{d}_{I}(X_{i}, X_{j}) - d_{I}(X_{i}, X_{j})\rvert \lvert\hat{d}_{I}(X_{i}, X_{j}) + d_{I}(X_{i}, X_{j})\rvert \\
    &\lesssim \frac{1}{n(n-1)} \sum_{1\leq i< j \leq n} \lvert\hat{d}_{I}(X_{i}, X_{j}) - d_{I}(X_{i}, X_{j})\rvert = O\left(\left(\frac{\log n}{n}\right)^{2/3k}\right).
\end{split}
\end{align*}
Additionally, \begin{equation*}
    \hat{V}_{I,F} = O(1) \quad \text{a.s.} \quad \text{and} \quad  \hat{V}_{I,M} = O(1) \quad \text{a.s.}
\end{equation*} 
Next, 
\begin{align*}
    \begin{split}
        \lvert \tilde{\rho}_{I,n} - \rho_{I,n}\rvert \leq \lvert \tilde{\rho}_{I,n} - \hat{\rho}_{I,n}\rvert + \lvert\hat{\rho}_{I,n} - \rho_{I,n}\rvert.
    \end{split}
\end{align*} For the first term $\tilde{\rho}_{I,n} - \hat{\rho}_{I,n}$, we have
\begin{align*}
    \begin{split}
        \lvert \tilde{\rho}_{I,n} - \hat{\rho}_{I,n} \rvert &= \Bigg \lvert \frac{\tilde{V}_{I,F}}{\tilde{V}_{I,M}} - \frac{\hat{V}_{I,F}}{\hat{V}_{I,M}} \Bigg \rvert \\
        &= \Bigg \lvert \frac{(\tilde{V}_{I,F} - \hat{V}_{I,F})\hat{V}_{I,M} - (\tilde{V}_{I,M} - \hat{V}_{I,M})\hat{V}_{I,F} }{(\tilde{V}_{I,M}-\hat{V}_{I,M})\hat{V}_{I,M} + \hat{V}^{2}_{I,M}}  \Bigg \rvert \\
        &\lesssim \lvert (\tilde{V}_{I,F} - \hat{V}_{I,F})\hat{V}_{I,M} - (\tilde{V}_{I,M} - \hat{V}_{I,M})\hat{V}_{I,F} \rvert \\
        &\leq \lvert (\tilde{V}_{I,F} - \hat{V}_{I,F})\hat{V}_{I,M}\rvert + \lvert (\tilde{V}_{I,M} - \hat{V}_{I,M})\hat{V}_{I,F} \rvert \\
        &= O\left(\left(\frac{\log n}{n}\right)^{2/3k}\right).
    \end{split}
\end{align*}
By Theorem 5, we can obtain the convergence rate of the second term \begin{equation*}
    \lvert \hat{\rho}_{I,n} - \rho_{I,n} \rvert = O(n^{-1/2})\quad \text{a.s.}
\end{equation*}
and thus  \begin{align*}
    \begin{split}
        \lvert \tilde{\rho}_{I,n} - \rho_{I,n}\rvert &\leq \lvert \tilde{\rho}_{I,n} - \hat{\rho}_{I,n}\rvert + \lvert\hat{\rho}_{I,n} - \rho_{I,n}\rvert = O\left(\max \left\{\left(\frac{\log n}{n}\right)^{2/3k}, \left(\frac{1}{n}\right)^{1/2} \right\}\right) ~\text{a.s.}
    \end{split}
\end{align*}    

\no\textit{\underline{Proof of Corollary S.\ref{cor:supp:conv}}}

The result is a direct  consequence of Corollary 2.2 of \cite{aaron2018convergence} and Theorem S.\ref{thm:supp:conv}. 

\clearpage

\section{Verification of conditions for simulations and real data analysis}
\label{suppl:sec:veri_cond}

\no 1. \textbf{Example 1 (Section 5.1)}: We consider a subset of the bivariate Wasserstein space, $\mathcal{D}\subset \mathcal{W}_2(\real^2),$ for  \begin{equation*}
    \mathcal{D} = \left\{ N(0, \Lambda(\theta)) \mid \Lambda(\theta) = R(\theta) \Lambda_{0} R(\theta)^{T} \in \real^{2 \times 2}, ~ \theta \in [0,1] \right\},
    \end{equation*} with a  rotation matrix $R(\theta) = \begin{pmatrix}
    \cos(\frac{\pi}{2}\theta) & - \sin(\frac{\pi}{2}\theta) \\
    \sin(\frac{\pi}{2}\theta) & \cos(\frac{\pi}{2}\theta)
\end{pmatrix}$ and $\Lambda_{0} = \text{diag}(\lambda_1, \lambda_2),$ where  $\lambda_{1} = 4$ and $\lambda_{2} = 1.$ Specifically, the space $\mathcal{D}$ forms a location-scatter family, as defined by \cite{ahidar2020convergence}. This follows from the fact that $\mathcal{D}$ is a subset of the multivariate mean zero Gaussian distribution space with covariance matrices constrained by bounded eigenvalues. Formally, \begin{align*}
        \mathcal{D} \sub \left\{ N_{p}(0, \Lambda) \mid \lambda_{1}I_p \leq \Lambda \leq \lambda_{2}I_p \right\},
    \end{align*} where $\lambda_{1} = 1,$ $\lambda_{2} = 4,$ and $p = 2.$
    
    We first show condition $(F2).$ Example 2.4 in \cite{ahidar2020convergence} implies  that $N\left(\frac{\delta \epsilon}{2}, B_{\delta}(\xi), d \right) \leq (\frac{C}{\epsilon})^{D},$ for a constant $C > ,$ and $D = p + p(p+1)/2,$  where  $p = 2$ in the bivariate case we consider here. Then
        \begin{align*}
            \int_{0}^{1}&\sqrt{1 + \log N(\delta\epsilon/2,B_{\delta}(\xi),d)} d\epsilon \leq  \int_{0}^{1} \sqrt{1 + D\log C - D \log \epsilon}~d\epsilon \\
            &\leq 1+  D\log C + \sqrt{D}\int_{0}^{1} \sqrt{-\log \epsilon} ~ d\epsilon \\
            &\leq 1+  D\log C + \sqrt{D}\int_{1}^{\infty} \sqrt{y} \exp(-y) ~ dy < \infty.
        \end{align*}
        The second inequality comes from $\sqrt{a + x} \leq a + \sqrt{x},$ for $a \geq 1,$ and $x > 0.$ The third inequality is obtained by substituting  $y = - \log \epsilon.$ Then \begin{equation*}
            \delta \int_{0}^{1}\sqrt{1 + \log N(\delta\epsilon/2,B_{\delta}(\xi),d)} d\epsilon  \rightarrow 0  \text{ as } \delta \rightarrow 0.
        \end{equation*}
    The total boundedness condition  \textit{(F0)} holds because $\mathcal{D}$ is closed and bounded with respect to the metric, and it is a subset of a $D$-dimensional Euclidean space \citep{ahidar2020convergence}. By the Heine-Borel theorem, $\mathcal{D}$ is compact, hence  totally bounded. The uniqueness of the population \F mean \textit{(F1)} is established by \cite{le2017existence}.

    For assumptions \textit{(M0)}-\textit{(M3)}, we consider a continuous distribution over $\mathcal{D}$ that has no atoms. The left hand inequalities in \textit{(M0)} and \textit{(M2)} correspond to
    \begin{align*}
         \mathbb{E}d^{2}(X,X') > 0, \text{ and } \text{Var}_{X}\left[\mathbb{E}_{X'|X}\left\{d^{2}\left(X,X'\right)|X\right\}\right] > 0.
    \end{align*} 
    Additionally, the total boundedness in \textit{(F0)} ensures boundedness, which guarantees the right hand inequality in \textit{(M0)}, i.e.,  $\mathbb{E} d^{2}\left(X,X'\right) < \infty,$ as well as the inequalities  in \textit{(M1)} and \textit{(M3)}, $\mathbb{E}d^{4}(X,X')  < \infty$ and   $\mathbb{E}_{X}\lvert \mathbb{E}_{X|X'} \left\{d^{2}(X,X')|X \right\} \rvert^{3} < \infty,$ for i.i.d. random objects $X, X' \in \mathcal{M}.$

\begin{figure}[!ht]
        \centering
        \includegraphics[width=0.5\linewidth]{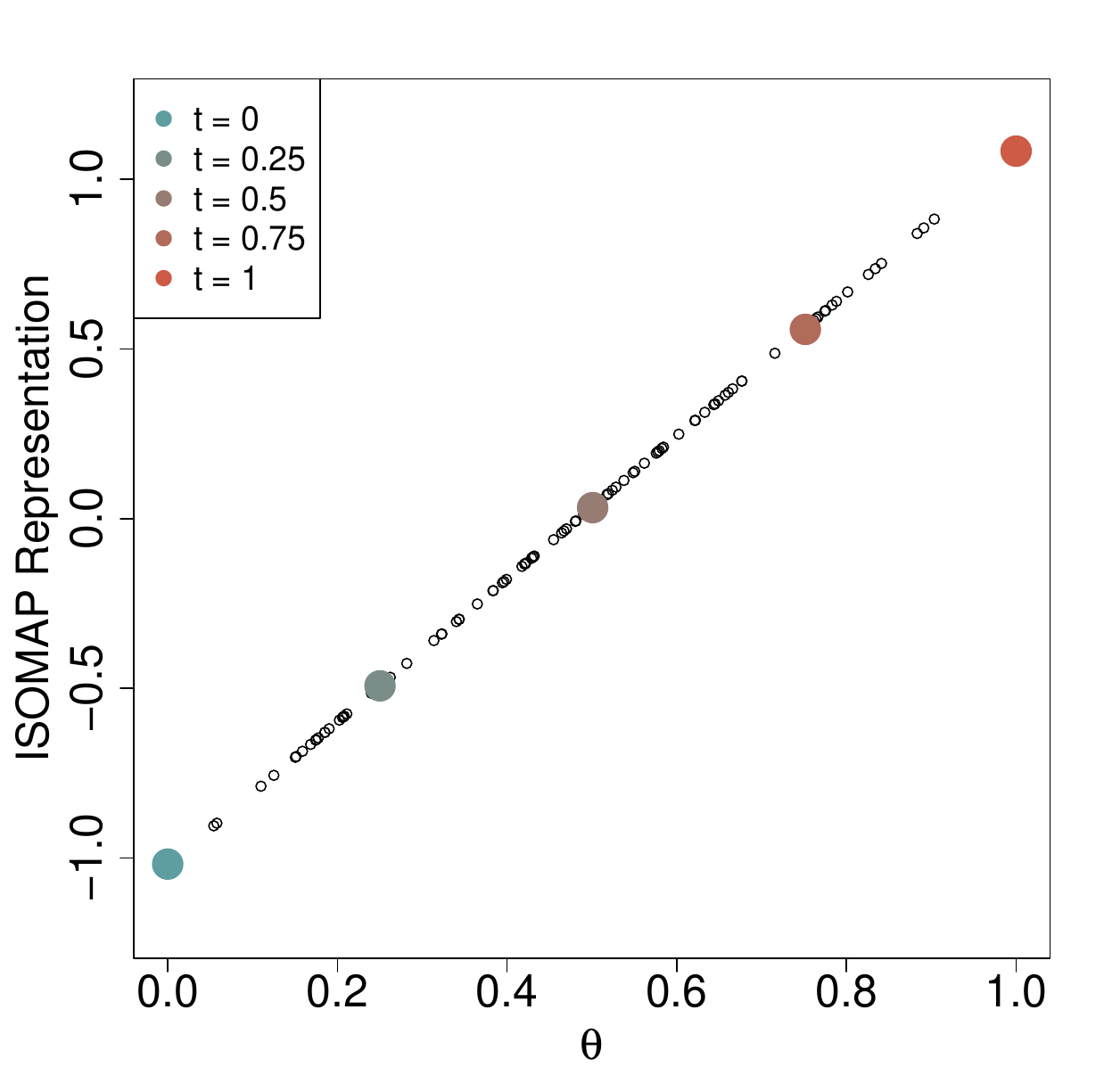}
        \caption{ISOMAP representation interpolation $s(t) = (1-t)\hat{\psi}(x) + t\hat{\psi}(y),$ where $\hat{\psi} : \mathcal{D} \rightarrow \real^{k}$ is the representation map of ISOMAP, obtained by random bivariate normal distributions, $X_{1}, \ldots, X_{100}$ (black dots), generated from the intrinsic space $\mathcal{D}$ as a function of the normalized rotation angle $0 \leq \theta \leq 1.$}
        \label{fig:ref1_fig1}
    \end{figure}

    The intrinsic space $(\mathcal{D}, d_{I})$ here is  a one-dimensional compact manifold that is smooth with boundary, such as a closed interval $[m, M] \in \real.$ This is because the space depends only on the one-dimensional normalized rotation angle $\theta \in [0, 1].$ Figure \ref{fig:ref1_fig1} provides empirical evidence, showing that the representation of random distributions $X_{1}, \ldots, X_{100}$ aligns with a one-dimensional structure proportional to the angle $\theta$. This one-dimensional compact manifold with boundary satisfies assumptions \textit{(A0)} through \textit{(A4)}.\\

    2. \textbf{Gait Synchronization Analysis (Section 6.1)}: In this  real data analysis, the gait synchronization data are represented as  a sample of  $2 \times 2$ symmetric positive definite (SPD) matrices, equipped with the Bures-Wasserstein metric, which defines the ambient metric space. Under the assumption of bounded eigenvalues of the SPD space, this space has the same geometric structure as the space of mean-zero bivariate Gaussian distributions equipped with the 2-Wasserstein metric \citep{bhatia2019bures}. Accordingly, the  same arguments as in the previous subsection can be applied to verify assumptions $(F0)-(F2)$, $(M0)-(M3)$, and $(A0)-(A4).$ \\

    3. \textbf{Example 2 (Section 5.2)}:  \bco Implementing your request that simulations and data scenarios should be in accordance with the assumptions, we have made a minor adjustment in the set-up for this simulation so that indeed all assumptions are satisfied.  Since we only consider the intrinsic geodesic space $\mathcal{B}_{1},$  $\mathcal{B}_{2},$  $\mathcal{B}_{3}$ of the 3-dimensional Euclidean space $\real^{3}$, our focus is to verify \textit{(A0)}-\textit{(A4)}.

    Specifically, the second simulation covers three different point cloud data with \begin{align*}
        \begin{split}
            \mathcal{B}_{1} &= \left\{(x,y,z) \in \mathbb{R}^{3} ~ | ~ x^{2} + y^{2} + z^{2} = 1, ~ {z\geq 0} \right\},\\
            \mathcal{B}_{2} &= \left\{(x,y,z) \in \mathbb{R}^{3} ~ | ~ x^{2} - y^{2} - z^{2} = -1, ~ {0 \leq z \leq 4} \right\},\\
            \mathcal{B}_{3} &= \left\{(x,y,0) \in \mathbb{R}^{3} ~ | ~ 0 \leq x,y \leq 1 \right\},
        \end{split}
    \end{align*}
    where $\mathcal{B}_{1}$ represents the upper hemisphere,  $\mathcal{B}_{2}$ is a hyperboloid obtained by rotating a hyperbolic graph around the $z$-axis, and $\mathcal{B}_{3}$ is a plane. These spaces are positive curved, negative curved and flat spaces, respectively. 

    Our point cloud data generation is as follows: For $i = 1, \ldots, n,$
    \begin{align*}
        \begin{split}
            (x_{1i}, y_{1i}, z_{1i}) &= (\cos\theta_{i}\sin\psi_{i}, \sin\theta_{i}\sin\psi_{i}, \cos\psi_{i} ) \in \mathcal{B}_{1}, ~\psi_{i} \sim {\text{U}[0, \pi/2]}, \theta_{i} \sim {\text{U}[0, 2\pi ]},\\
            (x_{2i}, y_{2i}, z_{2i}) &= (\upsilon_{i}, \sqrt{1+\upsilon_{i}^{2}}\cos\theta_{i}, \sqrt{1+\upsilon_{i}^{2}}\sin\theta_{i})\in \mathcal{B}_{2}, ~\upsilon_{i} \sim {TN_{[-r,r]}(0, 1)}, \theta_{i} \sim \text{U}[0, \pi], \\
            (x_{3i}, y_{3i}, z_{3i}) &= \left(\mu_{i}, \nu_{i}, 0 \right) \in \mathcal{B}_{3}, ~ \mu_{i}, \nu_{i} \sim {\text{U}[0,1]},
        \end{split}
    \end{align*}
    where $\text{U}[a, b]$ denotes a uniform distribution over the interval $[a, b],$ and $TN_{[-r, r]}(0, 1)$ is a one-dimensional truncated normal distribution with mean 0 and variance 1, restricted to the interval $[-r, r].$ We set $r = \sqrt{15}.$  
    
    \fi 
    
    Based on the generation of random samples in each space   $\mathcal{B}_{1}$,  $\mathcal{B}_{2}$  $\mathcal{B}_{3}$,     the assumptions of strictly positiveness $(A0),$ non-zero metric variance $V_{M},$ and non-degenerate asymptotic variance $(A4)$ hold. 
    For $(A1)$,  the space $\mathcal{B}_{1}$ is a closed upper hemisphere of radius 1, making it a compact Euclidean rectangle with uniform distribution, which is also totally bounded. While  hyperbolic spaces are not generally bounded, imposing the constraint $0 \leq z \leq 4$ ensures boundedness. This is because for any $X_{a} = (x_{a}, y_{a}, z_{a})$ and $X_{b} = (x_{b}, y_{b}, z_{b})$ in $\mathcal{B}_{3},$  defining $X_{0} = (0, 0, 1),$ the hyperbolic geodesic distance satisfies:
    \begin{align*}
        d_{\text{hyper}}(X_{a}, X_{b}) &\leq d_{\text{hyper}}(X_{a}, X_{0}) + d_{\text{hyper}}(X_{b}, X_{0})\\
        &= \text{arccosh}(z_{a}) +  \text{arccosh}(z_{b}) \\
        &\leq 2\log(4 + \sqrt{15}) < \infty.
    \end{align*} The space $\mathcal{B}_{3}$ is closed and forms a subset of the 3-dimensional Euclidean space. By the Heine-Borel theorem, it is compact, which implies total boundedness.

    For \textit{(A2)}, it is easy to see that the  \F mean of $\mathcal{B}_{1}$ and of $\mathcal{B}_{2}$ is $(0, 0, 1)$ and the \F mean of $\mathcal{B}_{3}$ is $(1/2, 1/2, 0).$ Then ${(A3)}$ is satisfied based on Proposition 3 in the supplementary materials of \cite{pete:19}, which shows that a bounded Riemannian manifold of finite dimension $r$ equipped with the geodesic distance satisfies the entropy condition 
    \begin{equation*}
            \int_{0}^{1}\sqrt{1 + \log N(\delta\epsilon/2,B_{\delta}(\xi),d)} d\epsilon = O(1)
    \end{equation*}
    as $\delta \rightarrow 0.$

    In real-world scenarios, data can often be contaminated with noise, which may cause certain assumptions to be violated depending on the noise structure. We consider error-contaminated settings with relatively small noise to assess the robustness of our test, and our simulation results show that our method can still detect the intrinsic positive, flat, and negative curvature even in  the presence of noise. \\

    4. \textbf{Energy data (Section 6.2)}: The second real data analysis, which uses energy data, considers random objects on the positive quadrant of the intrinsic spherical space $\mathbb{S}_{+}^{2},$ equipped with the geodesic distance. The ambient space is the $3$-dimensional Euclidean  space. This setting aligns with $\mathcal{B}_{1}$ in Example 2, as discussed in the previous paragraph, and the same arguments apply to verify the assumptions.\\

\clearpage
\section{Sensitivity analysis for the choice of input distance $d$}
\label{suppl:sec:sens_anal}

The selection of an appropriate metric $d$ is an important and open research question that has received limited attention. Recent discussion on this topic can be found in Section 8 of \cite{mull:24}.  In cases where data lie on Riemannian manifolds embedded in Euclidean ambient spaces, such as spheres or hyperbolic spaces, the Euclidean distance is commonly used as the ambient metric. Therefore in our point cloud simulation (Section 5.2) and energy data analysis (Section 6.2) we have  adopted the Euclidean distance as the ambient metric. 
    
    For our distributional data simulation (Section 5.1), we use the Wasserstein distance as the metric $d$. This choice is motivated by the fact that key Riemannian manifold concepts, such as the tangent space, exponential map, and logarithm map, can be naturally extended to Wasserstein space \citep{ambrosio2008gradient, panaretos2019statistical}. An alternative choice for distributional data is the Fisher-Rao metric  \citep{dai2022statistical}, which, while not as directly connected to optimal transport theory as the Wasserstein metric \citep{villani2003topics}, has computational advantages. Specifically, it serves as the geodesic distance for the square root of probability densities, 
    \begin{equation*}
        d_{FR}(\xi_{1}, \xi_{2}) = \arccos\left(\int \sqrt{f_{\xi_{1}}(x)f_{\xi_{2}}(x) } dx\right),
    \end{equation*}
    where $\xi_{1}, \xi_{2} \in \mathcal{W}_{2}(\mathbb{R}^{p})$ with densities $f_{\xi_{1}}(x),$ $f_{\xi_{2}}(x)$. 
    
    For the gait synchronization data analysis (Section 6.1), we employ the Bures-Wasserstein metric to measure distances between SPD matrices. This metric is equivalent to the Wasserstein metric for the space of non-degenerate Gaussian distributions \citep{bhatia2019bures}. In the space of SPD matrices, feature preservation along geodesics is often desirable, as the choice of metric determines the paths connecting data objects. The classical Frobenius distance, while commonly used, suffers from the swelling effect, which can distort statistical interpretations. This issue can be mitigated using the Cholesky distance \citep{dryden2009non, lin2019riemannian}, which provides a more stable representation.

\begin{figure}[!h]
      \centering
        \includegraphics[width=0.75\linewidth]{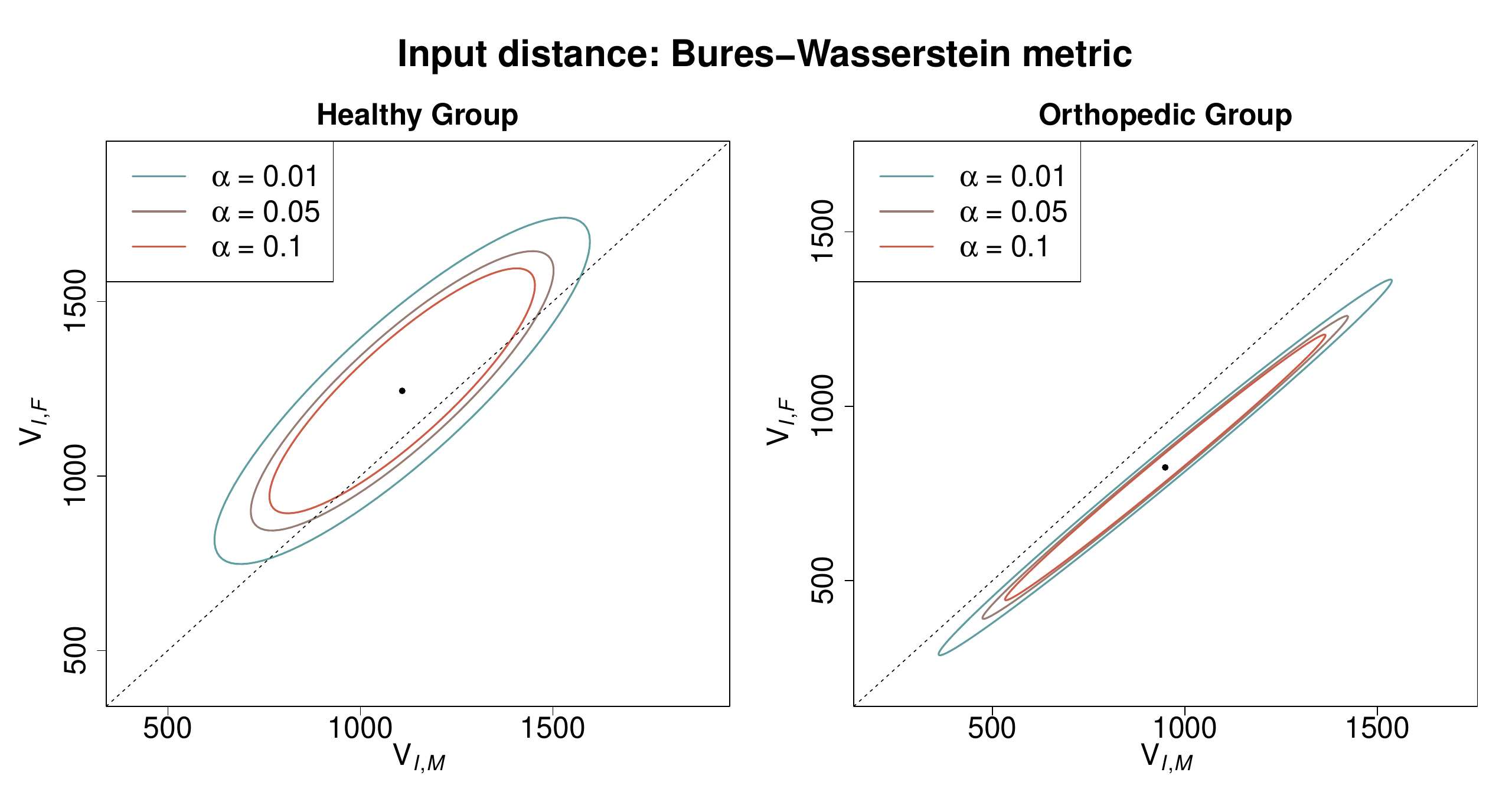}
        \includegraphics[width=0.75\linewidth]{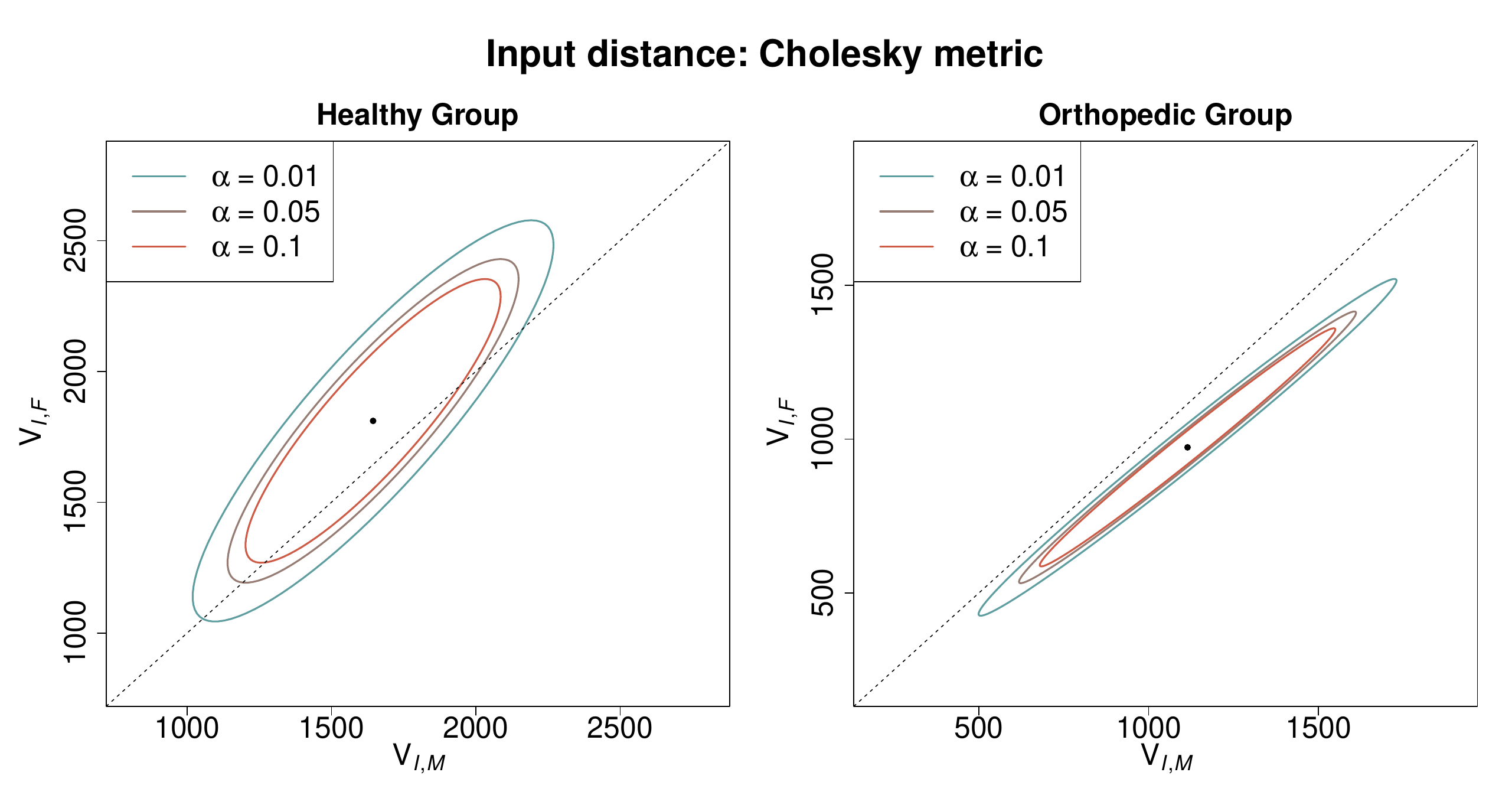}
        \includegraphics[width=0.75\linewidth]{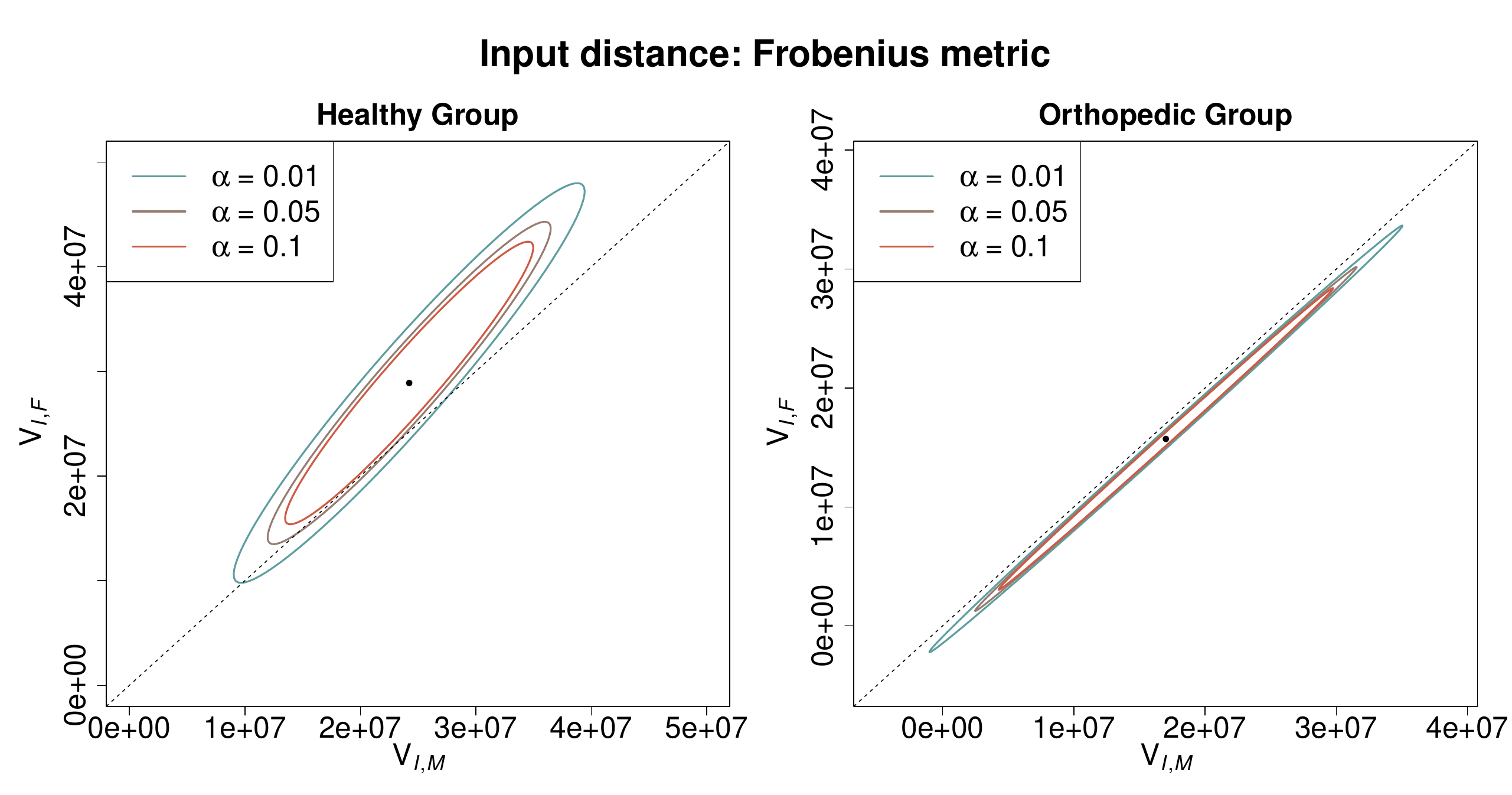}
      \caption{Comparison of confidence regions $\mathcal{C}_{I,n}(1-\alpha)$ in (16) for $({V}_{I,M}, {V}_{I,F})$ representing gait synchronization SPD matrices for healthy and orthopedic disorder groups, with $\alpha = 0.01, 0.05, 0.1$. The first row corresponds to the Bures-Wasserstein input metric, the second row to the Cholesky input metric, and the third row to the Frobenius input metric, all yielding consistent curvature test results.}
      \label{fig:ref2:gait}
    \end{figure}

    For the gait synchronization data analysis (Section 6.1), we compare our curvature test results using three different input metrics $d$: 1) the Bures-Wasserstein metric, 2) Frobenius metric, and 3) Cholesky metric.
To infer the intrinsic curvature of the space of gait synchronization SPD matrices we compare the intrinsic \F variance $\hat{V}_{I,F}$ and metric variance $\hat{V}_{I,M}$. The confidence regions $\mathcal{C}_{I, n}(1-\alpha)$ in (16) for $(V_{I, M},V_{I,F})^{T}$ are illustrated in Figure \ref{fig:ref2:gait} for the healthy and orthopedic disorder groups, with significance levels $\alpha = 0.01, 0.05, 0.1$. Specifically, the first row corresponds to the Bures-Wasserstein input metric, the second row to the Cholesky input metric, and the third row to the Frobenius input metric. For the healthy group, regardless of the choice of input distance, all confidence regions intersect with the set $\left\{({V}_{I,M},{V}_{I,F})~|~{V}_{I,M} ={V}_{I,F} \right\},$ but with $\hat{V}_{I,F}$ consistently greater than $\hat{V}_{I,M},$ which may indicate a slight positive curvature. In contrast, for the orthopedic disorder group, the confidence regions are consistently contained in the set $\left\{({V}_{I,M},{V}_{I,F})~|~{V}_{I,M} > {V}_{I,F} \right\}$ at level $\alpha=0.01$, indicating the presence of negative curvature. 
    
    For the distributional simulation scenario in Section 5.1, we compare our curvature test results using two different input metrics $d$: 1) the 2-Wasserstein metric and 2) the Fisher-Rao metric.

    \begin{figure}[!h]
        \centering
        \includegraphics[width=0.8\linewidth]{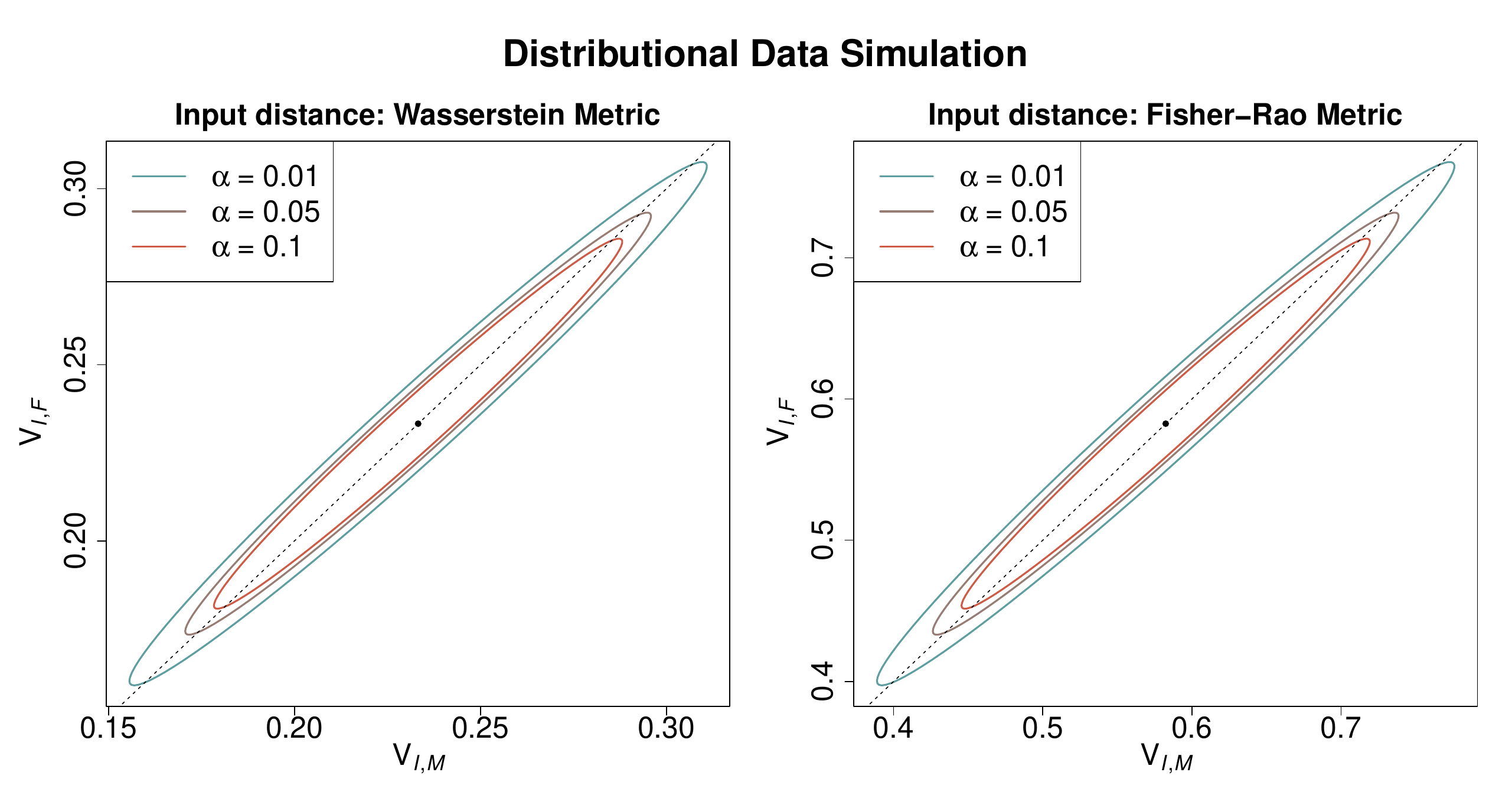}
        \caption{The proposed intrinsic curvature test results using the confidence regions $\mathcal{C}_{I,n}(1-\alpha)$ in (16) for $({V}_{I,M}, {V}_{I,F})$ with $\alpha = 0.01, 0.05, 0.1$. The first column corresponds to the 2-Wasserstein input metric, and the second row to the Fisher-Rao metric, both indicating a consistent result of intrinsic flat curvature.}
        \label{fig:ref2:dist_simul}
    \end{figure}

    In Figure \ref{fig:ref2:dist_simul}, we compare the intrinsic \F variance $\hat{V}_{I,F}$ and the metric variance $\hat{V}_{I,M}$. The confidence regions $\mathcal{C}_{I, n}(1-\alpha)$ in (16) for $(V_{I, M},V_{I,F})^{T}$ are displayed. Regardless of whether the Wasserstein or Fisher-Rao metric is used, all confidence regions intersect with the set $\left\{({V}_{I,M},{V}_{I,F})~|~{V}_{I,M} = {V}_{I,F} \right\}$  both indicating a consistent result that a flat intrinsic  curvature is compatible with the data. 

    In conclusion, our experiment shows that at least for the scenario considered, the choice of input distance $d$ does not significantly affect the results, indicating a degree of robustness. However, in general, the selection of an appropriate input distance requires further research, as briefly discussed in  Section 7.

\clearpage
\section{Relation between metric curvature and Alexandrov curvature}
\label{suppl:sec:rel_curv}

    By definition, the metric curvature $\rho$ is determined by the ratio of the \textit{expectation} of $d^{2}(\mu_{\oplus},X)$ and $\frac{1}{2}d^{2}(X,X'),$ 
    \begin{equation*}
        \rho = \frac{V_{F}}{V_{M}} - 1 = \frac{2\mathbb{E}d^{2}(\mu_{\oplus},X)}{\mathbb{E}d^{2}(X,X')} - 1,
    \end{equation*}
    where $X'$ represents an i.i.d. copy of $X.$ Thus, the actual value of $\rho$ depends on the probability measure $\mP$ of the random objects $X.$  As a  consequence, for    a fixed model space $(\M_{\kappa}, d_{\kappa})$ with the curvature parameter $\kappa$,   while the sign of $\kappa$ determines the sign of $\rho$ (as stated in Proposition 4), the actual value of $\rho$ can vary depending on the probability measure $\mathcal{P}.$

\begin{figure}[!ht]
        \centering
        \includegraphics[width=0.5\linewidth]{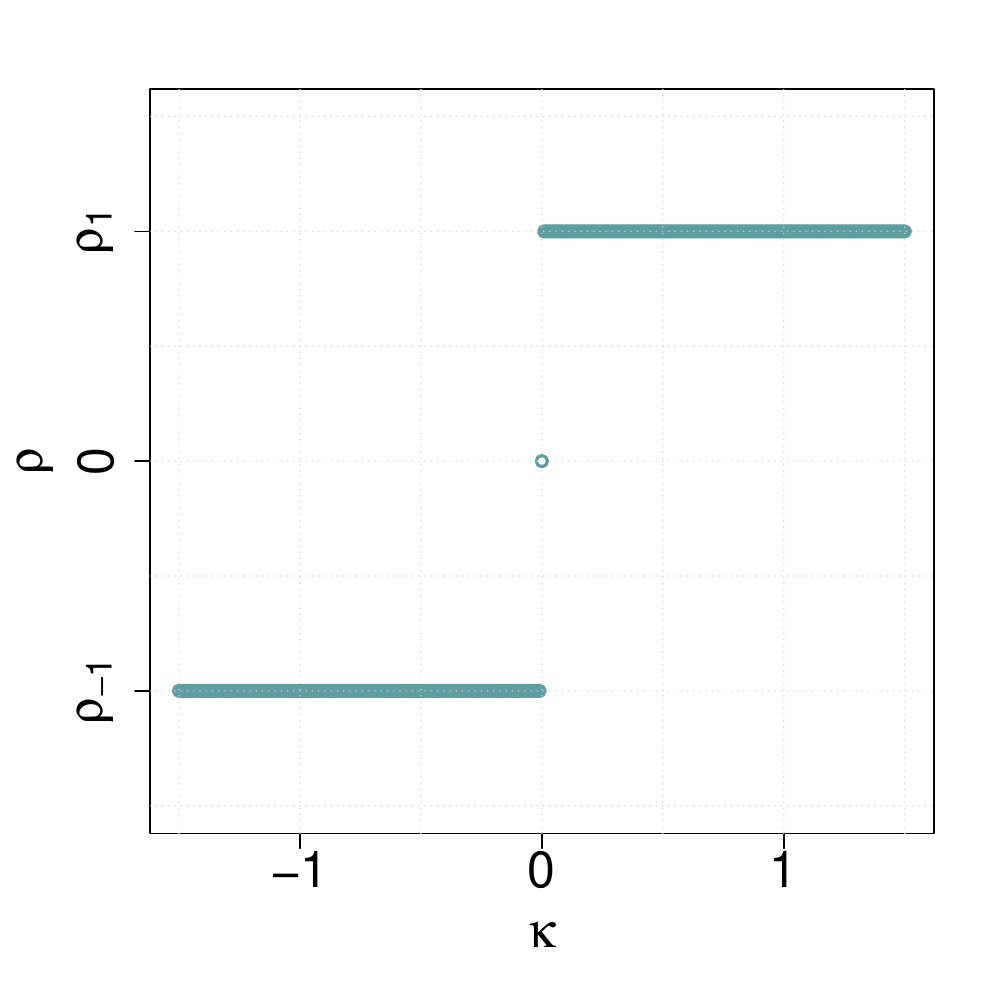}
        \caption{Relationship between the metric curvature $\rho$ and the Alexandrov curvature $\kappa$ for  the special case where the probability measure $\mP$  is fixed and the \F mean is independent of $\kappa.$} 
        \label{fig:ref3:step}
    \end{figure}

    In some special cases, we can establish a somewhat stronger relationship. We consider flat, positively curved, and negatively curved model spaces, i.e., $\M_{\kappa},$ for $\kappa = 0,$ $\kappa > 0,$ and $\kappa <0.$ 
    \begin{itemize}
        \item  For $\kappa = 0,$ the model space is the Euclidean plane $\M_{\kappa} = \real^{2}$ with the Euclidean distance. Then irrespective of the  probability measure $\mP$ of $X,$ as long as it is positive, Proposition 4 implies that $\rho = \kappa = 0.$

        \item For $\kappa > 0,$ the model space is the sphere \begin{equation*}
            \M_{\kappa} = \mathbb{S}^{2} = \left\{(x,y,z) \mid x^2 + y^2 + z^2 = 1 \right\},
        \end{equation*} with the angular distance $d_{\kappa}(a,b) = \arccos(x_{a}x_{b} + y_{a}y_{b} + z_{a}z_{b})/\sqrt{\kappa},$ where $a = (x_{a}, y_{a}, z_{a}) $ and $b = (x_{b}, y_{b}, z_{b}).$  Consider  a positive probability measure $\mP$ that generates $X$.   The metric curvature associated with $\M_{\kappa}$ is 
        \begin{equation*}
            \rho_{\kappa} = \frac{2\mathbb{E}d_{\kappa}^{2}(\mu_{\oplus},X)}{\mathbb{E}d_{\kappa}^{2}(X,X')} - 1,
        \end{equation*} where $X = (x_{1}, y_{1}, z_{1}),$ $X' = (x_{2}, y_{2}, z_{2}),$ and $\mu_{\oplus} = (x_{\oplus}, y_{\oplus}, z_{\oplus}).$ We assume that the unique \F mean remains the same  for all $\kappa > 0,$ i.e., \begin{equation*}
            \mu_{\oplus} = \argmin_{\xi \in \M_{\kappa}} \mathbb{E}d_{\kappa}^2(\xi,X).
        \end{equation*} If we consider a measure $\mP$ that has a symmetric density with respect to the north pole $(0, 0, 1) \in \mathbb{S}^{2},$ the \F mean can be fixed as $\mu_{\oplus} = (0,0,1)$ regardless of $d_{\kappa}.$ 
        

        Then we obtain
        \begin{align*}
            \rho_{\kappa} &= \frac{2\mathbb{E}\{\arccos^{2}(x_{1}x_{\oplus}+y_{1}y_{\oplus} + z_{1}z_{\oplus})\}/\kappa }{\mathbb{E}\{\arccos^{2}(x_{1}x_{2}+y_{1}y_{2} + z_{1}z_{2})\}/\kappa} - 1 \\
            &= \rho_{1} = C_{1} > 0,
        \end{align*}
        for all $\kappa > 0$ and some constant $C_{1}$ that does not depend on $\kappa.$ The second equality follows from the cancellation of $1/\kappa,$ and $C_{1} > 0$ according to  Proposition 4. 

        \item For $\kappa < 0,$ the model space is the hyperbolic space \begin{equation*}
            \M_{\kappa} = \mathbb{H}^{2} = \left\{(x,y,z) \mid x^2 + y^2 - z^2 = -1 \right\},
        \end{equation*} with the hyperbolic distance $d_{\kappa}(a,b) = \text{arccosh}(z_{a}z_{b} - x_{a}x_{b} - y_{a}y_{b})/\sqrt{-\kappa}$ for $\kappa < 0.$ Similarly, under the assumption that the \F mean does not depend on $\kappa,$ for all $\kappa < 0,$ we can obtain $\rho_{\kappa} = \rho_{-1} = C_{2}< 0,$ for some constant $C_{2}$ that does not depend on $\kappa.$ 
    \end{itemize}
    For these special cases, the dependence of the  metric curvature $\rho$  is a monotone increasing step function of the Alexandrov curvature $\kappa$,  as illustrated in Figure \ref{fig:ref3:step}.

    If one desires a linear  relationship, one may instead consider an alternative measure $\rho',$ defined as
    \begin{equation*}
        \rho' = \frac{1}{V_{M}} - \frac{1}{V_{F}}.
    \end{equation*}
    By leveraging the relationship between \F variance $V_{F}$ and metric variance $V_{M}$ in Theorem 2, Theorem 3, and Corollary 1 in Section 3.2, this alternative measure $\rho'$ still satisfies  Proposition  4, i.e.,    
    if  $\left(\M, d\right)$ has  strictly negative curvature, then $\rho' < 0$,  if  $\left(\M, d\right)$ has strictly positive curvature and condition \textit{(F1)} holds, then $\rho' > 0, $ and  if $\left(\mathcal{M}, d\right)$ is a flat space, then $\rho' = 0.$
Again invoking the above assumptions that the positive probability measure $\mP$ is fixed and the unique \F mean is independent of $\kappa$ for $\kappa > 0,$ one can derive the  relationship of $\rho'$ and $\kappa$ in spaces  $\M_{\kappa}.$  Denoting the dependency of $\rho'$ on $\kappa$ by  $\rho'_{\kappa},$ it emerges that 
      \begin{align*}
        \rho'_{\kappa} &= \frac{2}{\mathbb{E}d_{\kappa}^{2}(X,X')} - \frac{1}{\mathbb{E}d_{\kappa}^{2}(\mu_{\oplus},X)} \\
        &= \kappa\left(\frac{2}{\mathbb{E}\{\arccos^{2}(x_{1}x_{2}+y_{1}y_{2} + z_{1}z_{2})\}} - \frac{1}{\mathbb{E}\{\arccos^{2}(x_{1}x_{2}+y_{1}y_{2} + z_{1}z_{2})\}}\right) \\
        &= \kappa\rho'_{1} = \kappa C_{3} > 0,
    \end{align*}
    for some constant $C_{3}.$  Similarly, it holds that  $\rho'_{\kappa} = (-\kappa)\rho'_{-1} = \kappa C_{4} <0$ for $\kappa <0$ and some constant $C_{4} = (-\rho'_{-1}) > 0.$ 
    
\begin{figure}[!ht]
        \centering
        \includegraphics[width=0.5\linewidth]{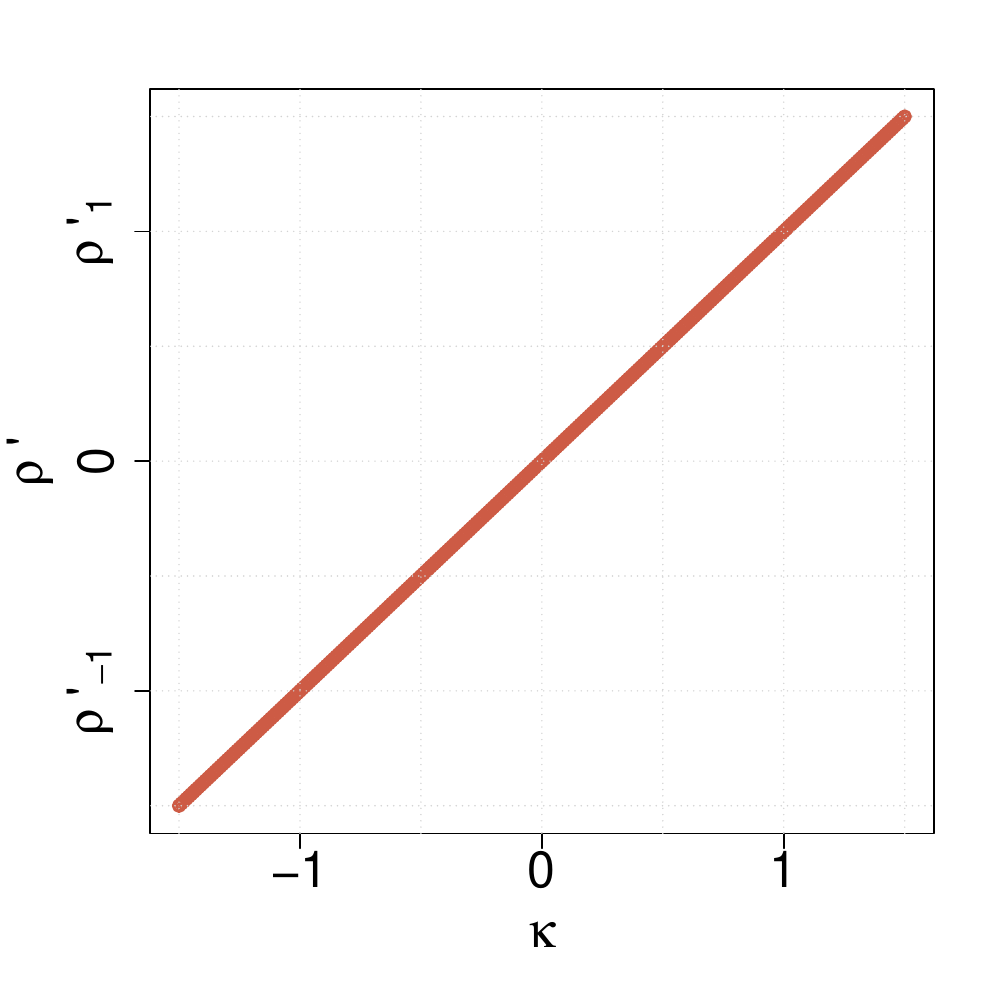}
        \caption{Relationship between the alternative metric curvature $\rho'$ and the Alexandrov curvature $\kappa$ under the special case where the probability measure $\mP$  is fixed and the \F mean is independent of $\kappa,$ for $\kappa >0,$ $\kappa = 0,$ and $\kappa < 0.$}
        \label{fig:ref3:linear}
    \end{figure}

    Figure \ref{fig:ref3:linear} illustrates the relationship between  $\rho'$ and the Alexandrov curvature $\kappa$ under these  additional assumptions. Furthermore, estimation of  $\hat{\rho}'$ and the  asymptotic normality of estimates is completely analogous to the corresponding results for $\rho$. 

\begin{singlespace}
\bibliography{DC}
\end{singlespace}

\end{document}